\documentclass[aps,prb,twocolumn,a4paper]{revtex4-1}
\usepackage[dvipdfmx]{graphicx}

\usepackage{amsmath,amsthm,amssymb}
\usepackage{amsmath,bm}
\usepackage{color}
\usepackage{ifthen}

\usepackage{multirow,bigdelim}

\newcount \commentout
\newcommand{\1}{\mbox{1}\hspace{-0.25em}\mbox{l}}

\newcommand{\sgn}{\mathrm{sgn}\!}

\newcommand{\mytitle}{
Fate of exceptional points under interactions: Reduction of topological classifications
\\
}

\newlength{\figwidth}
\newlength{\figlarge}
\begin{document}
\title{
\mytitle
}
\author{
Tsuneya Yoshida
}
\affiliation{
Department of Physics, Kyoto University, Kyoto 606-8502, Japan
}
\author{
Yasuhiro Hatsugai
}
\affiliation{
Department of Physics, University of Tsukuba, Ibaraki 305-8571, Japan
}
\date{\today}
\begin{abstract}
Despite recent extensive studies of the non-Hermitian topology, understanding interaction effects is left as a crucial question. 
In this paper, we address interaction effects on exceptional points which are protected by the non-trivial point-gap topology unique to non-Hermitian systems. 
Our analysis in a two-dimensional parameter space elucidates the existence of exceptional points and symmetry-protected exceptional rings fragile against interactions; they are topologically protected only in non-interacting cases.
This fragility of exceptional points and symmetry-protected exceptional rings arises from the reduction of non-Hermitian topological classifications, which is elucidated by introducing topological invariants of the second-quantized Hamiltonian for both non-interacting and interacting cases. 
These topological invariants are also available to analyze the reduction phenomena of gapped systems.
The above results strongly suggest similar reduction phenomena of exceptional points in generic cases and open up a new direction of research in the non-Hermitian topology.
\end{abstract}
\maketitle


\section{
Introduction
}

Extensive efforts have been devoted to understanding effects of interactions on topological insulators/superconductors.
In particular, it has turned out that interplay between the topology and interactions triggers off exotic phenomena such as the emergence of fractional topological insulators\cite{Tsui_FQHEExp_PRL82,Laughlin_FQHE_PRL83,Tang_FChern_PRL11,Sun_FChern_PRL11,Neupert_FChern_PRL11,Regnalt_FChen_PRX11,Sheng_FChern_NComm12,Bergholtz_FChern_IntJModPhysB13} and topological Mott insulators~\cite{Pesin_TMI_NatPhys2010}.
Furthermore, it has been elucidated that interactions change the topological classification of free fermions~\cite{Schnyder_classification_free_2008,Kitaev_classification_free_2009,Ryu_classification_free_2010} which provides systematic understanding of topological states and serves as the corner stone of the material searching.
For instance, interactions change the $\mathbb{Z}$-classification to the $\mathbb{Z}_8$-classification for one-dimensional topological superconductors with time-reversal symmetry~\cite{Z_to_Zn_Fidkowski_PRB10}.
This fact indicates that the number of possible topological states is reduced by interactions; there exist an infinite number of topologically distinct states in non-interacting cases while there exist eight topologically distinct states in interacting cases.
Further extensive works have elucidated the ubiquity of such reduction of topological classifications~\cite{entanglement_Pollmann10,Turner_ZtoZ8_PRB11,Fidkowski_1Dclassificatin_PRB11,Chen_classification_1D_1,Chen_classification_1D_2,Lu_CS_2011,YaoRyu_Z_to_Z8_2013,Ryu_Z_to_Z8_2013,Qi_Z_to_Z8_2013,Levin_CS_2012,Hsieh_CS_CPT_2014,Isobe_Fu2015,chen_cohomology_3D,gu_supercohomology,kapustin_bosonic_cobordisms2014_1,kapustin_bosonic_cobordisms2014_2,kapustin_fermionic_cobordisms2014,Fidkowski_Z162013,Wang_Potter_Senthil2014,Metlitski_3Dinteraction2014,Wang_Senthil2014,You_Cenke2014,Morimoto_2015,Superlattice_Yoshida17}. 
Namely, the reduction phenomena occur for arbitrary dimensions and symmetry classes. 
In addition, they occur even in parameter spaces~\cite{CMJian_ZtoZnSynthetic_PRX18}.

In parallel with the above significant developments, in these years, a topological aspect of non-Hermitian systems attracts as one of hot topics in condensed matter physics~\cite{Hu_nH_PRB11,Esaki_nH_PRB11,Sato_nHPTEP12,Diehl_DissCher_NatPhys11,Bardyn_DissCher_NJP2013,Budich_DissCher_PRA15,TELeePRL16_Half_quantized,ZPGong_PRL17,Lieu_nHSSH_PRB2018,Gong_class_PRX18,Kawabata_gapped_PRX19,Lieu_Liouclass_PRL20,Yang_nHJPoly_PRL20,Chang_nHES_PRR20,Kumer_nHTopo_PRB22,Arouca_nHTopo_arXiv22,Cayao_nHTopo_arXiv22,Bergholtz_Review19,Ashida_nHReview_AdvPhys20,Yoshida_nHReview_PTEP20}.
For such systems, extensive works of the non-interacting non-Hermitian topology have discovered a variety of novel phenomena induced by the point-gap topology unique to non-Hermitian systems, such as non-Hermitian skin effects which result in extreme sensitivity to the presence/absence of boundaries~\cite{Alvarez_nHSkin_PRB18,SYao_nHSkin-1D_PRL18,KFlore_nHSkin_PRL18,Lee_Skin19,Lee_nHSkin_PRL19,Borgnia_ptGapPRL2020,Zhang_BECskin19,Okuma_BECskin19,Hofmann_ExpRecipSkin_19,Yoshida_MSkinPRR20,Bessho_nHNN_PRL21,Kawabata_TQFTSkin_PRL21,Okuma_nHSkinRev_arXiv22}.
Furthermore, the non-Hermiticity induces a new type of topological degeneracies dubbed exceptional points (EPs) which are protected by the point-gap topology~\cite{HShen2017_non-Hermi,YXuPRL17_exceptional_ring,Hassan_EP_PRL17,Zhou_ObEP_Science18,VKozii_nH_arXiv17,Yoshida_EP_DMFT_PRB18,Wojcik_DiscEP_PRB20,Yang_DiscEP_PRL21}.
This new type of topological degeneracies is further enriched by symmetry, which results in the emergence of symmetry-protected exceptional rings (SPERs) and symmetry-protected exceptional surfaces in two and three dimensions, respectively~\cite{Budich_SPERs_PRB19,Okugawa_SPERs_PRB19,Yoshida_SPERs_PRB19,Zhou_SPERs_Optica19,Kawabata_gapless_PRL19,Delplace_Resul_PRL21,Mandal_EP3_PRL21}.
The EPs and their symmetry-protected variants in non-interacting systems attract interdisciplinary interests because they are reported for a wide variety of systems~\cite{Jose_DissSuperEP_SciRep16,Zhou_ObEP_Science18,Ozawa_TopoPhoto_RMP19,VKozii_nH_arXiv17,Yoshida_EP_DMFT_PRB18,Yoshida_SPERs_mech19,Hofmann_ExpRecipSkin_19,LiSciencePT19,Partanen_EPQbit_PRB2019,Naghiloo_EPQbit_NatPhys2019,Yoshida_nHgame_SciRep2022}.

The above two progresses lead us the following issues to be addressed; effects of interactions on the non-Hermitian topology.
Although several works addressed this issue~\cite{Luitz_EPcorrPRR2019,Yoshida_nHFQH19,Yoshida_nHFQHJ_PRR20,Guo_nHToric_PRB20,Matsumoto_nHToric_PRL20,Zhang_nHTMI_PRB20,Liu_nHTMI_RPB20,Xu_nHBM_PRB20,Pan_PTHubb_oQS_PRA20,Mu_MbdySkin_PRB20,Yoshida_PtGpZtoZ2_PRB21,Yang_EPKitaev_PRL21,Shen_CorrSkin_arXiv21,Lee_MbdySkin_PRB21,Zhang_CorrSkin_arXiv22,Kawabata_CorrSkin_PRB22,Schafers_EPcorr_arXiv2022,Orito_CorrSkin_PRB22,Tsubota_CorrInv_PRB22,Gaugno_corrnHSkin_arXiv2022,Yoshida_reduction1Dptgp_arXiv2022,Qin_CorrPolarons_arXiv22}, fate of EPs under interactions remains highly crucial question.
The significance of this question is further enhanced by recent experimental progresses in cold atoms~\cite{Tomita_Zeno_SciAdv17,Tomita_2BdyLoss_PRA19,Takasu_nHPTcoldAtom_PTEP2020} and quantum circuits~\cite{Ma_LossQuantumCircuits_Nature2019}.

We hereby analyze effects of interactions on an EP and an SPER in a two-dimensional parameter space which are protected by symmetry.
In particular, we elucidate that interactions may destroy the EP and the SPER without breaking relevant symmetry.
The above fragility of EPs against interactions arises from the reduction of the non-Hermitian topological classification, which is obtained by comparing topological invariants of the second-quantized Hamiltonian for both non-interacting and interacting cases.
Specifically, our analysis elucidates that the reduction $\mathbb{Z}^{(N+P'+1)/2}\to\mathbb{Z}$ ($\mathbb{Z}\to\mathbb{Z}_2$) results in the fragility of EPs (SPERs) for systems with charge $\mathrm{U(1)}$ symmetry and spin-parity symmetry (chiral symmetry).
Here, we have focused on the Fock space with the particle number $N$. 
For even (odd) $N$, $P'$ takes $\pm 1$ ($0$).

The above topological invariants are also applicable to the reduction for gapped systems~\cite{Yoshida_PtGpZtoZ2_PRB21,Yoshida_reduction1Dptgp_arXiv2022}.
For gapped systems with charge $\mathrm{U(1)}$ symmetry and spin-parity symmetry, our topological invariants indicate the reduction of one-dimensional point-gap topology: $\mathbb{Z}^{(N+P'+1)/2}\to \mathbb{Z}$.
For gapped systems with chiral symmetry, our topological invariants indicate the reduction of zero-dimensional point-gap topology: $\mathbb{Z}\to \mathbb{Z}_2$.

The rest of this paper is organized as follows. 
Section~\ref{sec: spin-parity} is devoted to clarifying the fragility of EPs against interactions in systems with charge $\mathrm{U(1)}$ symmetry and spin-parity symmetry. 
Section~\ref{sec: chiral} elucidates the fragility of SPERs with chiral symmetry.  In Sec.~\ref{sec: summary}, a brief summary is provided.
In Appendix~\ref{sec: num Fock app}, we count the number of the subspaces for a given Fock space.
In Appendix~\ref{sec: EP for s=-1 app}, we demonstrate that there also exist EPs robust against interactions.
In Appendices~\ref{sec: gapped 1D app}~and~\ref{sec: gapped 0D app}, we address the reduction phenomena for gapped systems~\cite{Yoshida_PtGpZtoZ2_PRB21,Yoshida_reduction1Dptgp_arXiv2022} by computing the above topological invariants.

\section{
Exceptional points with charge $\mathrm{U}(1)$ symmetry and spin-parity symmetry
}
\label{sec: spin-parity}

There exist EPs protected by the point-gap topology only when the second-quantized Hamiltonian is quadratic. 
In order to see this, let us analyze interaction effects on EPs for the two-dimensional parameter space in the presence of charge $\mathrm{U(1)}$ symmetry and spin-parity symmetry. 
The Hamiltonian reads,
\begin{subequations}
\label{eq: generic H}
\begin{eqnarray}
\hat{H}&=& \hat{H}_0+\hat{H}_{\mathrm{int}},
\end{eqnarray}
\begin{eqnarray}
\label{eq: H0}
\hat{H}_0&=& \hat{\Psi}^\dagger_{\alpha} h_{\alpha\beta}(x,y) \hat{\Psi}_{\beta}.
\end{eqnarray}
\end{subequations}
Here, summation is assumed over the repeated indices. 
The first-quantized Hamiltonian is denoted by $h(x,y)$.
Real variables $x$ and $y$ describe the two-dimensional parameter space.
Operator $\hat{\Psi}^\dagger$ ($\hat{\Psi} $) denotes a set of creation operators $\hat{c}^\dagger_{\alpha}$ (annihilation operators $\hat{c} _{\alpha}$) of fermions. The subscripts $\alpha$ and $\beta$ label the internal degrees of freedom such as orbital and spin.
Here, one might consider that the above setup is somewhat artificial. However, EPs in such a parameter space have been reported for quantum circuits~\cite{Partanen_EPQbit_PRB2019,Naghiloo_EPQbit_NatPhys2019}.

In this section, we consider a system of fermions with spin-1/2 whose Hamiltonian preserves charge $\mathrm{U(1)}$ symmetry and spin-parity symmetry;
\begin{subequations}
\begin{eqnarray}
{}[\hat{H},\hat{N}]_{\mathrm{c}}&=& 0, \label{eq: U1 symm mbdyH} \\ 
{}[\hat{H},e^{i\pi \hat{S}_z}]_{\mathrm{c}}&=& 0,  \label{eq: spin-parity symm mbdyH}
\end{eqnarray}
\end{subequations}
with  $\hat{N}=\hat{N}_\uparrow+\hat{N}_\downarrow$ and $\hat{S}_z=(\hat{N}_\uparrow-\hat{N}_\downarrow)/2$.
Here $\hat{N}_\sigma$ denotes the operator of the total number of fermions in spin-state $\sigma=\uparrow,\downarrow$. 
The commutation relation denoted by square brackets $[\hat{A},\hat{B}]_{\mathrm{c}}:=\hat{A}\hat{B}-\hat{B}\hat{A}$.
The above equations indicates that the second-quantized Hamiltonian $\hat{H}$ can be block-diagonalized with respect to $\hat{N}$ and $\hat{P}=(-1)^{\hat{N}_\uparrow}=e^{i\pi \hat{N}/2}e^{i\pi \hat{S}_z}$.
We denote eigenvalues of $\hat{N}$, $\hat{S}_z$, $\hat{N}_\sigma$ and $\hat{P}$ by $N$, $S_{z}$, $N_\sigma$ and $P$, respectively.

In the rest of this section, we introduce topological invariants and demonstrate the presence of EPs which are fragile against interactions.

\subsection{
Topological invariants
}
\label{sec: spin-parity topoinv}
For the Fock space with $[N,P]$, the number $(N+P'+1)/2$ of $\mathbb{Z}$-invariants are introduced in the non-interacting cases, while the number of $\mathbb{Z}$-invariants is reduced to one in the presence of interactions. Here, $P'$ takes $P$ ($0$) for even (odd) $N$.
This fact indicates the reduction of the topological classification of $\hat{H}$: $\mathbb{Z}^{(N+P'+1)/2} \to \mathbb{Z}$ (for application to gapped systems, see Appendix~\ref{sec: gapped 1D app}).
In other words, there exist EPs which are destroyed by interactions without breaking relevant symmetry.
The key ingredient is the additional symmetry imposed on the quadratic Hamiltonian $\hat{H}_0$ [see Eq.~(\ref{eq: spin-parity symm mbdyH free})].

\subsubsection{
Non-interacting case
}
In the presence of the spin-parity symmetry~(\ref{eq: spin-parity symm mbdyH}), the number $(N+P'+1)/2$ of $\mathbb{Z}$-invariants can be introduced when the second-quantized Hamiltonian is quadratic.

Firstly, we note that the spin-parity symmetry imposes the following constraint on the first-quantized Hamiltonian
\begin{eqnarray}
\label{eq: comm h and s}
{}[h, s_z]_{\mathrm{c}} &=& 0,
\end{eqnarray}
with $s_z$ being the $z$-component of the first-quantized spin operator.
This commutation relation can be seen by noting the relation $ e^{i\pi \hat{S}_z} \hat{\Psi}^\dagger_{\alpha} e^{-i\pi \hat{S}_z}= e^{i\pi (s_z)_{\alpha\beta}} \hat{\Psi}^\dagger_{\beta}$.
The above constraint indicates that at the non-interacting level, the second-quantized Hamiltonian $\hat{H}_0$ satisfies
\begin{eqnarray}
\label{eq: spin-parity symm mbdyH free}
{}[\hat{H}_0,\hat{S}_z]_{\mathrm{c}}&=& 0,  
\end{eqnarray}
meaning that the $\hat{H}_0$ can be block-diagonalized with respect to $\hat{S}_z$.

Thus, the Fock space with $[N,P]$ can be divided into subspaces with $(N_\uparrow,N_\downarrow)$.
For each subspace with $(N_\uparrow,N_\downarrow)$, the following winding number can be introduced
\begin{eqnarray}
\label{eq: W (N Sz) free}
W_{(N_\uparrow,N_\downarrow)} &=& \oint \frac{d\bm{\lambda}}{2\pi i} \cdot \bm{\partial}_{\bm{\lambda}} \log \mathrm{det}[\hat{H}_{(N_\uparrow,N_\downarrow)}(\bm{\lambda})-E_{\mathrm{ref}}\1], \nonumber \\
\end{eqnarray}
with the block-diagonalized Hamiltonian $\hat{H}_{(N_\uparrow,N_\downarrow)}$, the reference energy $E_{\mathrm{ref}} \in \mathbb{C}$, and the identity matrix $\1$.
The integral is taken over a closed path parameterized by $\bm{\lambda}=(x,y)$ in the two-dimensional parameter space.
Here, we have supposed that along the path, the point-gap opens at the reference energy $E_{\mathrm{ref}}$; $\mathrm{det}[\hat{H}_{(N_\uparrow, N_\downarrow)}(\bm{\lambda})-E_{\mathrm{ref}}\1]\neq 0$ holds for $\bm{\lambda}$ parameterizing the path.
Here, we note that for a given set of $[N,P]$, the number $(N+P'+1)/2$ of the sets $(N_\uparrow,N_\downarrow)$ are allowed where $P'$ takes $P$ $(0)$ for even (odd) $N$.
The detailed derivation is provided in Appendix~\ref{sec: num Fock app}.

Therefore, it is concluded that the number $(N+P'+1)/2$ of $\mathbb{Z}$-invariants are introduced in the non-interacting cases.

\subsubsection{
Interacting case
}

For correlated systems, the second-quantized Hamiltonian can be block-diagonalized not with $\hat{S}_z$ but with $\hat{P}=(-1)^{\hat{N}_\uparrow}$ due to spin-parity symmetry.

Thus, the point-gap topology is characterized by
\begin{eqnarray}
\label{eq: W [N,P] }
W_{[N,P]} &=& \oint \frac{d\bm{\lambda}}{2\pi i} \cdot \bm{\partial}_{\bm{\lambda}} \log \mathrm{det}[\hat{H}_{[N,P]}( \bm{\lambda} )-E_{\mathrm{ref}}\1].
\end{eqnarray}
Here, $\hat{H}_{[N,P]}(\bm{\lambda})$ denotes the second-quantized Hamiltonian for the Fock space with $[N,P]$.

In the non-interacting case, the above winding numbers satisfy
\begin{eqnarray}
\label{eq: Wparity=Wsz+Wsz}
W_{[N,P]} &=&  \displaystyle{ \sum_{(N_\uparrow,N_\downarrow)}}\!\!\!\!\!' }\,\, W_{(N_\uparrow,N_\downarrow),
\end{eqnarray}
where the summation is taken over sets of  $N_\uparrow$ and $N_\downarrow$ satisfying $N_\uparrow+N_\downarrow=N$ and $(-1)^{N_\uparrow}=P$ for given $N$ and $P$.

Equation~(\ref{eq: Wparity=Wsz+Wsz}) indicates that for the Fock space with $[N,P]$, 
the point-gap topological states form the $\mathbb{Z}^{(N+P'+1)/2}$ group in the non-interacting case while the point-gap topological states form its subgroup $\mathbb{Z}$ in correlated cases.
In particular, it indicates that interactions may destroy EPs without breaking charge $\mathrm{U(1)}$ symmetry and spin-parity symmetry if they are characterized by vanishing $W_{[N,P]}$ and finite $W_{(N_\uparrow,N_\downarrow)}$.

\subsection{
Analysis of a toy model
}
\label{sec: spin-parity toy}

EPs can be fragile against interactions due to the reduction of the non-Hermitian topological classification for systems with charge $\mathrm{U(1)}$ symmetry and spin-parity symmetry.
In order to demonstrate this fact, let us analyze a three-orbital system described by the second-quantized Hamiltonian~(\ref{eq: generic H}) with $\hat{\Psi}=(\hat{c}_{a\uparrow},\hat{c}_{b\uparrow},\hat{c}_{c\uparrow},\hat{c}_{a\downarrow},\hat{c}_{b\downarrow},\hat{c}_{c\downarrow})^{T}$,
\begin{eqnarray}
\label{eq: spin-parity h}
h(x,y) &=&
\left(
\begin{array}{ccc|ccc}
0                            & x+iy                         & 0                              &                                &   &  \\
1                            & 0                            & 0                              &                                &   &  \\
0                            & 0                            & 0                              &                                &   &  \\\hline
                             &                              &                                & 0                              &          1                     & 0   \\
                             &                              &                                & x+is_\downarrow y              & 0                              & 0   \\
                             &                              &                                &   0                            & 0                              & 0   \\
\end{array}
\right),
\end{eqnarray}
and
\begin{eqnarray}
\label{eq: parity intVonly}
\hat{H}_{\mathrm{int}} &=& iV [(\hat{S}^+_a+\hat{S}^+_b)\hat{S}^+_c +\mathrm{h.c.}].
\end{eqnarray}
Here, a fermion in orbital $l=a,b,c$ and spin-state $\sigma=\uparrow,\downarrow$ is created (annihilated) by applying operator $\hat{c}^\dagger_{l\sigma}$ ($\hat{c}_{l\sigma}$).
The parameter $s_\downarrow$ takes $1$ or $-1$. Unless otherwise noted, we set $s_\downarrow=-1$ in the main text.
Operator $\hat{S}^+_l$ ($\hat{S}^-_l$) is defined as $\hat{S}^+_l=c^\dagger_{l\uparrow}c_{l\downarrow}$ ($\hat{S}^-_l=c^\dagger_{l\downarrow}c_{l\uparrow}$).
The prefactor $V$ is real. 
The above Hamiltonian satisfies $[\hat{H},\hat{n}_{c\sigma}]_{\mathrm{c}}=0$ for $\sigma=\uparrow,\downarrow$, and thus, we suppose that~\cite{U1-nc_ftnt} a fermion occupies orbital $c$. 
In this section, we focus on the Fock space with $[N,P]=[2,1]$ because there is no topologically protected EPs in the other subspace~\cite{CommentN2P-1_ftnt} with $N=2$.

Let us start with the non-interacting case. 
For the Fock space specified by $[N,P]=[2,1]$, the second-quantized Hamiltonian $\hat{H}_{0[2,1]}$ is written as
\begin{subequations}
\begin{eqnarray}
\label{eq: parity mbdyH0 4x4}
\hat{H}_{0[2,1]}&=& 
\left(
\begin{array}{c|c}
\hat{H}_{0(2,0)} & 0 \\\hline
0 & \hat{H}_{0(0,2)}
\end{array}
\right),
\end{eqnarray}
with
\begin{eqnarray}
\hat{H}_{0(2,0)}
&=&
\left(
\begin{array}{cc}
0                                & x+iy  \\
1                                & 0     \\
\end{array}
\right), \\
\hat{H}_{0(0,2)}
&=&
\left(
\begin{array}{cc}
 0                               & 1                          \\
 x+is_\downarrow y      & 0
\end{array}
\right).
\end{eqnarray}
\end{subequations}
Here we have chosen the basis as
\begin{eqnarray}
\label{eq: basis (N,P)=(2,1)}
(c^\dagger_{a\uparrow}c^\dagger_{c\uparrow}|0\rangle, c^\dagger_{b\uparrow}c^\dagger_{c\uparrow}|0\rangle, c^\dagger_{a\downarrow}c^\dagger_{c\downarrow}|0\rangle, c^\dagger_{b\downarrow}c^\dagger_{c\downarrow}|0\rangle). 
\end{eqnarray}
The matrix $\hat{H}_{0(2,0)}$ [$\hat{H}_{0(0,2)}$] is the Hamiltonian for the subspace with $(N_\uparrow,N_\downarrow)=(2,0)$ [$(0,2)$].
The above equation is consistent with the fact that the non-interacting Hamiltonian $\hat{H}_0$ can be block-diagonalized with respect to $\hat{S}_z$ in the presence of spin-parity symmetry [see Eq.~(\ref{eq: spin-parity symm mbdyH free})].

The Hamiltonian $\hat{H}_{[2,1]}$ exhibits EPs for $V=0$, which can be seen by diagonalizing $\hat{H}_{0(2,0)}$ and $\hat{H}_{0(0,2)}$.
Figures~\ref{fig: MbdyEfermi free (N,P)=(2,1)}(a)~and~\ref{fig: MbdyEfermi free (N,P)=(2,1)}(b) display eigenvalues of $\hat{H}_{0[2,1]}$ against $x$ and $y$. 
In these figures, EPs emerge at zero energy $E=0$ and $(x,y)=(0,0)$ which are denoted by red dots.
\begin{figure}[!h]
\begin{minipage}{1\hsize}
\begin{center}
\includegraphics[width=1\hsize,clip]{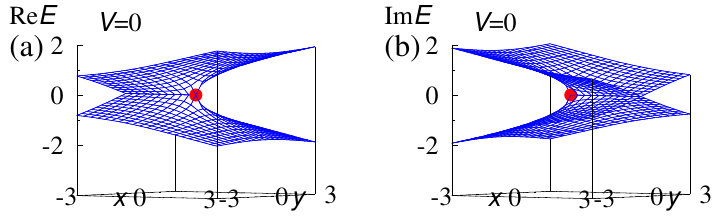}
\end{center}
\end{minipage}
\begin{minipage}{1\hsize}
\begin{center}
\includegraphics[width=1\hsize,clip]{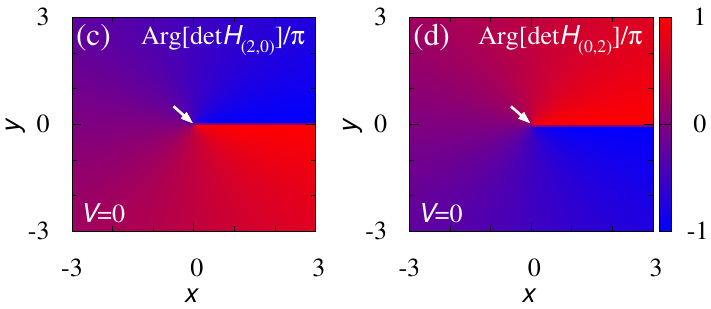}
\end{center}
\end{minipage}
\caption{
(a) [(b)] The real- [imaginary-] part of eigenvalues of $\hat{H}_{[2,1]}$ for $V=0$.
(c) [(d)] The argument of $\mathrm{det}\hat{H}_{0(2,0)}$ [$\mathrm{det}\hat{H}_{0(0,2)}$]. 
We recall that $\hat{H}_{[N,P]}$ and $\hat{H}_{(N_{\uparrow},N_{\downarrow})}$ denote the Hamiltonian for the Fock space with $[N,P]$ and $(N_{\uparrow},N_{\downarrow})$.
The data are obtained for $s_\downarrow=-1$.
}
\label{fig: MbdyEfermi free (N,P)=(2,1)}
\end{figure}
The point-gap topology protecting these EPs is characterized by the winding numbers.
For computation of $W_{(2,0)}$ and $W_{(0,2)}$, we plot $\mathrm{det}[\hat{H}_{(N_\uparrow,N_\downarrow)}]$ for $(N_\uparrow,N_\downarrow)=(2,0)$ and $(0,2)$ in Figs.~\ref{fig: MbdyEfermi free (N,P)=(2,1)}(c)~and~\ref{fig: MbdyEfermi free (N,P)=(2,1)}(d), respectively.
From these figure, we can extract the winding numbers $\left( W_{(2,0)} ,W_{(0,2)} \right)=(1,-1)$ for $E_{\mathrm{ref}}=0$ computed along a path enclosing the origin $(x,y)=(0,0)$.
Here, the path is taken so that it winds the origin in the counterclockwise direction.
Therefore, the EPs [see Figs.~\ref{fig: MbdyEfermi free (N,P)=(2,1)}(a)~and~\ref{fig: MbdyEfermi free (N,P)=(2,1)}(b)] are robust against perturbations at the non-interacting level because they are protected by the non-trivial point-gap topology.

In the presence of interactions, however, the above EPs are no longer protected by the topology, implying that they can be destroyed by interactions.
This is because the subspaces with $(N_\uparrow,N_\downarrow)=(2,0)$ and $(0,2)$ are unified in the presence of interactions.

Specifically, Eq.~(\ref{eq: Wparity=Wsz+Wsz}) elucidates that the point-gap topology is trivial in the presence of interactions; $W_{[2,1]}=W_{(2,0)}+W_{(0,2)}=0$ for $E_{\mathrm{ref}}=0$.
Correspondingly, introducing the interaction~(\ref{eq: parity intVonly}) destroys EPs. For the Fock space with $[N,P]=[2,1]$,
the second-quantized Hamiltonian is written as
\begin{subequations}
\begin{eqnarray}
\hat{H}_{[2,1]}&=& \hat{H}_{0[2,1]}+\hat{H}_{\mathrm{int}[2,1]},
\end{eqnarray}
\begin{eqnarray}
\hat{H}_{\mathrm{int}[2,1]} &=&
iV
\left(
\begin{array}{cc|cc}
0 &0 & 1 & 0  \\
0 &0  & 0 & 1 \\ \hline
1 &0 & 0 & 0 \\
0 &1 & 0 & 0
\end{array}
\right),
\end{eqnarray}
\end{subequations}
with the basis defined in Eq.~(\ref{eq: basis (N,P)=(2,1)}).
Figures~\ref{fig: MbdyEfermi corr (N,P)=(2,1)} displays eigenvalues of $\hat{H}_{[2,1]}$ for $V=1$.
As observed in this figure, EPs are destroyed by interaction $V$ which mixes the subspaces with $(N_{\uparrow}, N_{\downarrow})=(2,0)$ and $(0,2)$.

Putting the above results [Figs.~\ref{fig: MbdyEfermi free (N,P)=(2,1)}~and~\ref{fig: MbdyEfermi corr (N,P)=(2,1)} and Eq.~(\ref{eq: Wparity=Wsz+Wsz})] together, we end up with the conclusion that the reduction $\mathbb{Z}^2\to\mathbb{Z}$ results in the fragility of EPs against interactions for the Fock space with $[N,P]=[2,1]$.

We finish this section with two remarks. 
Firstly, we note that if the winding number $W_{[N,P]}$ is finite, the EPs are robust against interactions, which can be seen in the case for $s_\downarrow=1$ (see Appendix~\ref{sec: EP for s=-1 app}).

Secondly, we point out that for the Fock space with $[N,P]=[2,1]$, the winding numbers can be analytically computed along the path specified by $x^2+ y^2=1$.
Along this path the Hamiltonian is written as
\begin{eqnarray}
\label{eq: spin H x^2+y^2=1}
\hat{H}_{[2,1]}&=& 
\left(
\begin{array}{cc|cc}
0   & e^{i\theta} & iV           & 0 \\
1   & 0           & 0            & iV \\ \hline
iV  &   0         &    0         & 1 \\
 0  &  iV         & e^{i s_\downarrow \theta} & 0
\end{array}
\right),
\end{eqnarray}
with $(x,y)=(\cos\theta,\sin\theta)$ and $0  \leq \theta < 2\pi$.
Here, we have chosen the basis~(\ref{eq: basis (N,P)=(2,1)}).
Diagonalizing the above Hamiltonian, we have eigenvalues~\cite{4x4EPdiag_ftnt} for $s_\downarrow=-1$
\begin{subequations}
\label{eq: Ep Em (N,P)=(2,1) s=-1}
\begin{eqnarray}
E_{\mathrm{p}\pm}(\theta)&=& \left[\cos \frac{\theta}{2} \pm i\sqrt{\sin^2\frac{\theta}{2}+V^2} \right], \\
E_{\mathrm{m}\pm}(\theta)&=& -\left[\cos \frac{\theta}{2} \pm i\sqrt{\sin^2\frac{\theta}{2}+V^2} \right].
\end{eqnarray}
\end{subequations}
For $V=0$, eigenvalues for the subspace with $(N_\uparrow,N_\downarrow)=(2,0)$ are given by $(E_{\mathrm{p}+},E_{\mathrm{m}+})=(e^{i\frac{\theta}{2}},-e^{i\frac{\theta}{2}})$, which indicates $W_{(2,0)}=1$ for $E_{\mathrm{ref}}=0$.
In a simiar way, we obtain $W_{(0,2)}=-1$ along the loop.
For $V>0$, Eq.~(\ref{eq: Ep Em (N,P)=(2,1) s=-1}) indicates that the imaginary-part of all eigenvalues are finite for $0  \leq \theta < 2\pi$. 
This fact results in $W_{[2,1]}=0$ because no eigenvalue winds the origin.
The above analysis is consistent with Eq.~(\ref{eq: Wparity=Wsz+Wsz}).

\begin{figure}[!h]
\begin{minipage}{1\hsize}
\begin{center}
\includegraphics[width=1\hsize,clip]{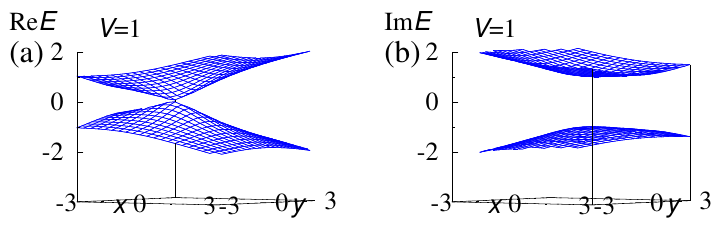}
\end{center}
\end{minipage}
\caption{
(a) [(b)] The real- [imaginary-] part of eigenvalues of $\hat{H}_{[2,1]}$ for $V=1$.
}
\label{fig: MbdyEfermi corr (N,P)=(2,1)}
\end{figure}

\section{
Symmetry-protected exceptional ring with chiral symmetry
}
\label{sec: chiral}
As is the case of EPs, there exist SPERs protected by the point-gap topology only when the Hamiltonian is quadratic. 
In order to see this, let us analyze interaction effects on SPERs for the two-dimensional parameter space in the presence of chiral symmetry. 
Consider the Hamiltonian~(\ref{eq: generic H}) preserving chiral symmetry
\begin{eqnarray}
\label{eq: chiral mbdy}
[\hat{H},\hat{\Xi}]_{\mathrm{c}}&=& 0,
\end{eqnarray}
with anti-unitary operator $\hat{\Xi}$ which is a product of time-reversal and charge conjugation operators.
In the rest of this section, we introduce zero-dimensional topological invariants and demonstrate the presence of SPERs which are fragile against interactions.

\subsection{
Topological invariants
}
\label{sec: chiral topoinv}

For a given Fock space, a zero-dimensional $\mathbb{Z}$- ($\mathbb{Z}_{2}$-) invariant can be introduced in non-interacting (interacting) cases. 
This fact indicates the reduction of the topological classification of $\hat{H}$: $\mathbb{Z} \to \mathbb{Z}_2$ (for application to a gapped system, see Appendix~\ref{sec: gapped 0D app}).
In other words, there exist SPERs destroyed by interactions without breaking chiral symmetry. 
As is the case of Sec.~\ref{sec: spin-parity}, the key ingredient is an additional constraint imposed only on the quadratic Hamiltonian $\hat{H}_{0}$ [see Eq.~(\ref{eq: H0=-GHdagG})].

We also note that Eq.~(\ref{eq: XiG=-GXi}) is essential for the above reduction.

\subsubsection{
Non-interacting case
}
\label{sec: Zinv chiral}

In the presence of chiral symmetry, a zero-dimensional $\mathbb{Z}$-invariant can be introduced when the second-quantized Hamiltonian is quadratic.

Firstly, we note that the chiral symmetry~(\ref{eq: chiral mbdy}) imposes the following constraint on the first-quantized Hamiltonian 
\begin{eqnarray}
\label{eq: chiral free}
\xi h^\dagger \xi &=& -h,
\end{eqnarray}
with unitary matrix $\xi$ satisfying~\cite{xi2=1_ftnt} $\xi^2=\1$. 
Here, we have considered that $h$ is a traceless matrix.
We note that the chiral symmetric $\hat{H}_0$ may include $i\gamma_\alpha \left(\hat{n}_{\alpha}-\frac{1}{2}\right)$ with $\gamma_\alpha \in \mathbb{R}$ and $\hat{n}_{\alpha}=\hat{c}^\dagger_\alpha\hat{c}_\alpha$. 
However, this fact does not affect the following argument.
Equation~(\ref{eq: chiral free}) can be seen by noting the relation
\begin{eqnarray}
\hat{\Xi} \hat{\Psi}^\dagger_\alpha \hat{\Xi}^{-1} &=& \xi_{\alpha\beta} \hat{\Psi}_{\beta}.
\end{eqnarray}
Summation is assumed over repeated indices.

As proved below, Eq.~(\ref{eq: chiral free}) results in the following constraint on $\hat{H}_{0}$:
\begin{subequations}
\begin{eqnarray}
\label{eq: H0=-GHdagG}
\hat{H}_0  &=& -\hat{\Gamma}\hat{H}^\dagger_0\hat{\Gamma},
\end{eqnarray}
with 
\begin{eqnarray}
\hat{\Gamma} &=&(-1)^{\hat{N}_-}, \\
\hat{N}_{-}&=& \Psi^\dagger \left( \frac{\1 - \xi}{2} \right) \Psi.
\end{eqnarray}
\end{subequations}

Equation~(\ref{eq: H0=-GHdagG}) can be proven as follows.
Firstly, we rewrite Eq.~(\ref{eq: chiral free}) as
\begin{subequations}
\begin{eqnarray}
\label{eq: h_H xi}
\xi h_{\mathrm{H}} \xi &=& -h_{\mathrm{H}}, \\
\label{eq: h_A xi}
\xi h_{\mathrm{A}} \xi &=& h_{\mathrm{A}},
\end{eqnarray}
\end{subequations}
where we have decomposed $h$ into the Hermitian part $h_{\mathrm{H}}$ and the anti-Hermitian part $h_{\mathrm{A}}$.
Equation~(\ref{eq: h_H xi}) indicates that applying $\hat{H}_{0\mathrm{H}}=\Psi^\dagger h_{\mathrm{H}} \Psi$ increases/decreases the number $N_-$ by one~\cite{hHmat_ftnt}, where $N_{-}$ denotes eigenvalues of $\hat{N}_{-}$.
Thus, $\hat{H}_{0\mathrm{H}}$ anti-commutes with $\hat{\Gamma}=(-1)^{\hat{N}_-}$.
Equation~(\ref{eq: h_A xi}) indicates that $\hat{H}_{0\mathrm{A}}=\Psi^\dagger h_{\mathrm{A}} \Psi$ commutes with $\hat{\Gamma}$.
Noting the relation $\hat{H}_{0}=\hat{H}_{0\mathrm{H}}+\hat{H}_{0\mathrm{A}}$, we obtain Eq.~(\ref{eq: H0=-GHdagG}).

Equation~(\ref{eq: H0=-GHdagG}) allows us to define the zero-th Chern number $N_{0\mathrm{Ch}}$ which is the number of eigenstates with negative eigenvalues of 
\begin{eqnarray}
\label{eq: H0Gamma}
\hat{H}_{0\Gamma}  &=& i\left( \hat{H}_0 -E_{\mathrm{ref}}\1 \right)\hat{\Gamma}.
\end{eqnarray}
Here, we have supposed that $\hat{\Xi}$ and $\hat{\Gamma}$ are anti-commute with each other
\begin{eqnarray}
\label{eq: XiG=-GXi}
\hat{\Xi} \hat{\Gamma} &=& - \hat{\Gamma} \hat{\Xi}.
\end{eqnarray}
In addition, we have supposed that the point-gap opens ($\mathrm{det}[\hat{H}_0-E_{\mathrm{ref}}\1]\neq 0$) for $E_{\mathrm{ref}}\in i\mathbb{R}$.
The above zero-th Chern number is previously introduced for the first-quantized Hamiltonian $h$~\cite{Yoshida_SPERs_PRB19,Kawabata_gapless_PRL19,Yoshida_PtGpZtoZ2_PRB21}.

The anti-commutation relation between $\hat{\Gamma}$ and $\hat{\Xi}$ is essential for the above topological characterization.
For systems where $\hat{\Gamma}$ and $\hat{\Xi}$ commute with each other, $N_{0\mathrm{Ch}}$ does not characterize the topology due to the relation~\cite{PHSofH0_ftnt} $\hat{\Xi} \hat{H}_{0\Gamma}=-\hat{H}_{0\Gamma}\hat{\Xi}$.

In the above we have introduced the zero-dimensional $\mathbb{Z}$-invariant $N_{0\mathrm{Ch}}$ for chiral symmetric systems where $\hat{\Xi}$ satisfies Eq.~(\ref{eq: XiG=-GXi}).

\subsubsection{
Interacting case
}
\label{sec: Z2inv chiral}

In the presence of interactions, the second-quantized Hamiltonian is no longer quadratic, meaning that Eq.~(\ref{eq: H0=-GHdagG}) does not hold.
However, we can still introduce the following $\mathbb{Z}_2$-invariant
\begin{eqnarray}
\label{eq: Z2 mbdyH chiral}
\nu &=& \sgn \left( \mathrm{\det} [\hat{H}-E_{\mathrm{ref}}\1] \right),
\end{eqnarray}
for $E_{\mathrm{ref}}\in \mathbb{R}$
due to the symmetry constraint~(\ref{eq: chiral mbdy}). Here, $\sgn(x)$ takes $1$ ($-1$) for $x>0$ ($x<0$).

In the non-interacting case, the parity of $N_{0\mathrm{Ch}}$ corresponds to $\nu$ for $E_{\mathrm{ref}}=0$;
\begin{eqnarray}
\label{eq: N0Ch and nu}
\nu &=& \sgn( \mathrm{det}[i\hat{\Gamma}]) (-1)^{N_{0\mathrm{Ch}}}.
\end{eqnarray}
The above relation can be seen as follows
\begin{eqnarray}
(-1)^{N_{0\mathrm{Ch}}} &=& \sgn\left( \mathrm{det}[ \hat{H}_{0\Gamma} ] \right) \nonumber \\
                        &=& \nu \, \sgn\left( \mathrm{det}[i\hat{\Gamma}]\right),
\end{eqnarray}
where we have used the relation $\mathrm{det}[ i\hat{H}_{0}\hat{\Gamma} ]= \mathrm{det}[i\hat{\Gamma}] \mathrm{det}[\hat{H}_{0}]$.

Equation~(\ref{eq: N0Ch and nu}) indicates that for the Fock space with $N$, 
the point-gap topological states form the $\mathbb{Z}$ group in the non-interacting case while the point-gap topological states form its subgroup $\mathbb{Z}_2$ in interacting cases.
In particular, it indicates that interactions may destroy SPERs without breaking chiral symmetry if they are characterized by $\nu=0$ and finite $N_{0\mathrm{Ch}}$.

\subsection{
Analysis of a toy model
}
\label{sec: SPER toy}

The SPERs can be fragile against interactions due to the reduction of the non-Hermitian topological classification for systems with chiral symmetry.
In order to demonstrate this fact, let us analyze a system described by the Hamiltonian with
\begin{eqnarray}
\label{eq: chiral H0}
\hat{H}_0 &=& \hat{\Psi}^\dagger h \hat{\Psi}+\sum_{\sigma=1,0,-1} i\gamma_{\sigma}\left( \hat{n}_{a\sigma} -\frac{1}{2} \right), 
\end{eqnarray}
\begin{eqnarray}
\label{eq: chiral Hint}
\hat{H}_{\mathrm{int}} &=& U\sum_{l=a,b}\left( \hat{n}_{l1}-\frac{1}{2} \right) \left( \hat{n}_{l-1}-\frac{1}{2} \right),
\end{eqnarray}
\begin{eqnarray}
\label{eq: chiral h}
h &=&
\left(
\begin{array}{cc|cc|cc}
2i\beta & z^* &   &   &   &   \\
z & -2i\beta &   &   &   &   \\\hline
  &   & \frac{3}{2}i\beta & 2z^* &   &   \\
  &   & 2z & -\frac{3}{2}i\beta &   &   \\\hline
  &   &   &   & i\beta & 3z^* \\
  &   &   &   & 3z & -i\beta \\
\end{array}
\right).
\end{eqnarray}
Here, operators $\hat{\Psi}$ and $\hat{n}_{l\sigma}$ are defined as  $\hat{\Psi}=(\hat{c}_{a1},\hat{c}_{b1}|\hat{c}_{a0},\hat{c}_{b0}|\hat{c}_{a-1},\hat{c}_{b-1})^{T}$ and $\hat{n}_{l\sigma}=\hat{c}^\dagger_{l\sigma}\hat{c}_{l\sigma}$, respectively.
Subscripts $l=a,b$ and $\sigma=1,0,-1$ label orbital and spin degrees of freedom~\cite{pseudo-spin_ftnt}, respectively.
Parameters $\beta$, $U$ and $\gamma_{\sigma}$ are real numbers. A parameter $z$ takes a complex number, $z=x+iy$ with $x,y\in \mathbb{R}$.

The Hamiltonian is chiral symmetric; $\hat{H}$ satisfies Eq.~(\ref{eq: chiral mbdy}) with~\cite{Gurarie_chiral_PRB11,Manmana_Chiral1D_PRB12,Yoshida_SPERs_PRB19}
\begin{eqnarray}
\label{eq: chiral Xi toy}
\hat{\Xi}&=& \prod_{\sigma=1,0,-1} (\hat{c}^\dagger_{a\sigma}+\hat{c}_{a\sigma})(\hat{c}^\dagger_{b\sigma}-\hat{c}_{b\sigma}) \mathcal{K}.
\end{eqnarray}
Here, $\mathcal{K}$ is the complex-conjugation operator.

As well as chiral symmetry, the above Hamiltonian preserves charge $\mathrm{U}(1)$ symmetry and spin $\mathrm{U}(1)$ symmetry~\cite{chiral-NSz_ftnt}; the Hamiltonian $\hat{H}$ satisfies
%
\begin{eqnarray}
{}[\hat{H},\hat{N}]_{\mathrm{c}}&=& 0, \\
{}[\hat{H},\hat{S}_z]_{\mathrm{c}}&=& 0,
\end{eqnarray}
with $\hat{N}=\sum_{l\sigma}\hat{n}_{l\sigma}$ and $\hat{S}_z=\sum_{l\sigma}\sigma \hat{n}_{l\sigma}$.
Thus, the Hamiltonian can be brock-diagonalized with respect to $\hat{N}$ and $\hat{S}_z$. 
By $\hat{H}_{(N,S_z)}$, we denote the second-quantized Hamiltonian for the Fock space with $(N,S_z)$. Here, $N$ and $S_z$ denote eigenvalues of $\hat{N}$ and $\hat{S}_z$, respectively.

\begin{figure}[!h]
\begin{minipage}{1\hsize}
\begin{center}
\includegraphics[width=1\hsize,clip]{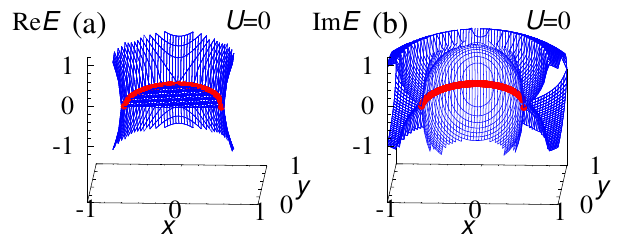}
\end{center}
\end{minipage}
\begin{minipage}{0.9\hsize}
\begin{center}
\includegraphics[width=1\hsize,clip]{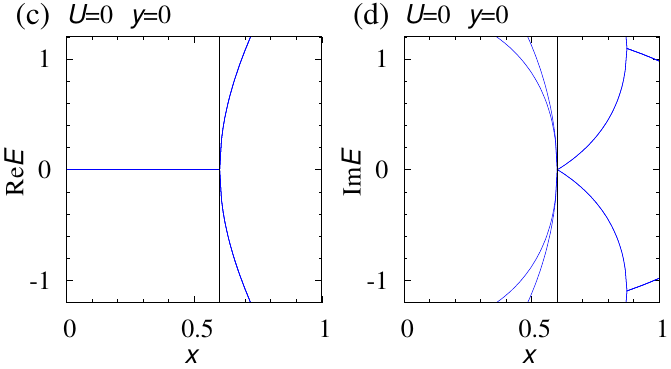}
\end{center}
\end{minipage}
\begin{minipage}{0.48\hsize}
\begin{center}
\includegraphics[width=1\hsize,clip]{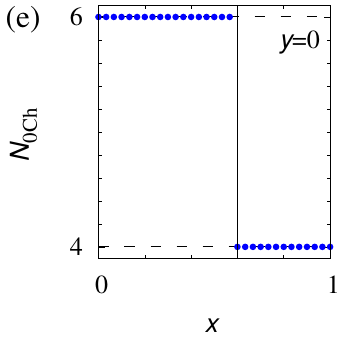}
\end{center}
\end{minipage}
\begin{minipage}{0.48\hsize}
\begin{center}
\includegraphics[width=1\hsize,clip]{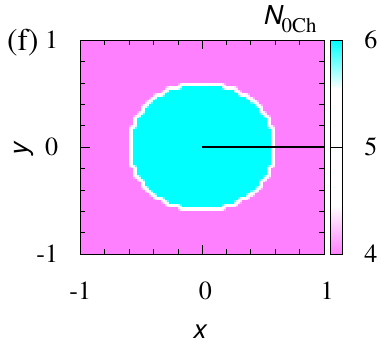}
\end{center}
\end{minipage}
\caption{
Eigenvalues of $\hat{H}_{(3,0)}$ and its point-gap topology for $U=0$. By $\hat{H}_{(3,0)}$, we denote the second-quantized Hamiltonian $\hat{H}$ for the Fock space with $(N,S_z)=(3,0)$.
(a) [(b)] The real- [imaginary-] part of the eigenvalues against $x$ and $y$.
The red lines denote the SPER.
(c) [(d)] The real- [imaginary-] part of the eigenvalues for $y=0$. 
(e) The zero-th Chern number $N_{0\mathrm{Ch}}$ for $y=0$. (f) Color plot of $N_{0\mathrm{Ch}}$.
The vertical lines in panels (c), (d), and (e) denote the critical value $x_c\sim0.6$ where the band touching occurs.
Data in panel (e) correspond to the zero-th Chern number on the black line in panel (f).
These data are obtained for $(\beta,\gamma_{1},\gamma_{0},\gamma_{-1})=(0.8,-3,-2.945,1)$.
}
\label{fig: chiral free}
\end{figure}

We demonstrate that $\hat{H}_{(3,0)}$ hosts an SPER characterized by the zero-th Chern number for $E_{\mathrm{ref}}=0$ in the non-interacting case which is fragile against the interaction $U$.
The emergence of the SPER at zero energy $E=0$ can be observed in Figs.~\ref{fig: chiral free}(a)~and~\ref{fig: chiral free}(b) (see red lines).
On the SPER, four eigenvalues touch for both real- and imaginary-parts [see Figs.~\ref{fig: chiral free}(c)~and~\ref{fig: chiral free}(d)]. 
In addition, Figs.~\ref{fig: chiral free}(e)~and~\ref{fig: chiral free}(f) indicate that the zero-th Chern number for $E_{\mathrm{ref}}=0$ jumps from $N_{0\mathrm{Ch}}=6$ to $N_{0\mathrm{Ch}}=4$ on the SPER with increasing $x$.
The above results indicate that the system exhibits the SPER at zero energy $E=0$ which is characterized by the zero-th Chern number, a $\mathbb{Z}$-invariant.

Here, the $\mathbb{Z}_2$-invariant $\nu$ does not change its value on the SPER;
form Eq.~(\ref{eq: N0Ch and nu})~and Fig.~\ref{fig: chiral free}(e), we can see that the $\mathbb{Z}_2$-invariant remains $\nu=1$ for $E_{\mathrm{ref}}=0$ by noting the relation $\mathrm{det}[i\hat{\Gamma}_{(3,0)}]=1$.

\begin{figure}[!h]
\begin{minipage}{1\hsize}
\begin{center}
\includegraphics[width=1\hsize,clip]{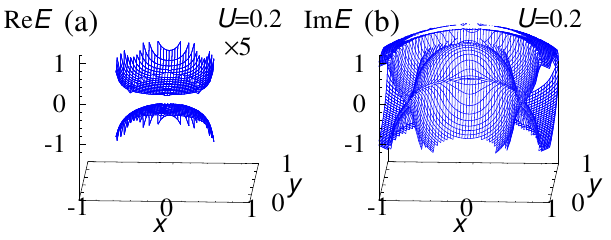}
\end{center}
\end{minipage}
\begin{minipage}{0.9\hsize}
\begin{center}
\includegraphics[width=1\hsize,clip]{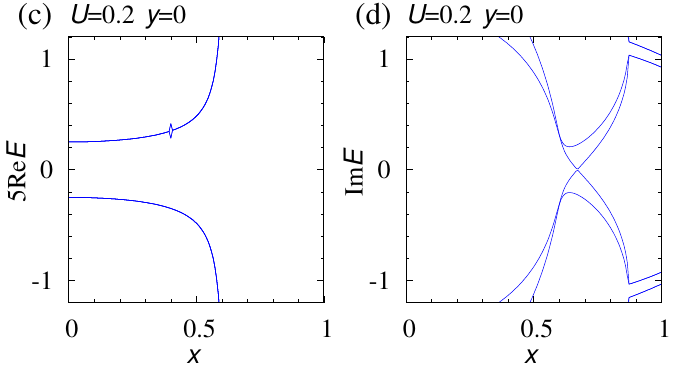}
\end{center}
\end{minipage}
\caption{
Eigenvalues of $\hat{H}_{(3,0)}$ for $U=0.2$.
(a) [(b)] The real- [imaginary-] parts of the eigenvalues against $x$ and $y$.
(c) [(d)] The real- [imaginary-] parts of the eigenvalues for $y=0$. 
Panel~(c) displays the data multiplied by 5 [i.e., $5\mathrm{Re}E_n$ $n=1,2,\ldots,8$].
These data are obtained for $(\beta,\gamma_{1},\gamma_{0},\gamma_{-1})=(0.8,-3,-2.945,1)$.
}
\label{fig: chiral U=02}
\end{figure}

The fact that the $\mathbb{Z}_2$-invariant does not jump on the SPER indicates the fragility of the SPER against the interaction $U$, which is demonstrated below.
Figures~\ref{fig: chiral U=02}(a)~and~\ref{fig: chiral U=02}(b) display the real- and imaginary-parts of the eigenvalues against $x$ and $y$ for $U=0.2$.
In contrast to the non-interacting case, the SPER cannot be observed; the real- and imaginary-parts do not touch simultaneously.
The absence of the SPER can also be confirmed in Figs.~\ref{fig: chiral U=02}(c)~and~\ref{fig: chiral U=02}(d).
These numerical results demonstrate that the interaction $U$ destroys the SPERs on which the zero-th Chern number jumps by an even number.

Putting the above results [Figs.~\ref{fig: chiral free}~and~\ref{fig: chiral U=02} and Eq.~(\ref{eq: N0Ch and nu})] together, we end up with the conclusion that the reduction $\mathbb{Z}\to\mathbb{Z}_2$ results in the fragility of the SPERs against interactions.

\section{
Summary
}
\label{sec: summary}
We have addressed interaction effects on the EPs and the SPERs in the two-dimensional parameter space. 
Our analysis elucidates that interactions may destroy the EPs and the SPERs without breaking relevant symmetry. The fragility of EPs and the SPERs is due to the reduction of the non-Hermitian topological classification. 
Specifically, we have seen that the reduction $\mathbb{Z}^{(N+P'+1)/2}\to\mathbb{Z}$ ($\mathbb{Z}\to\mathbb{Z}_2$) results in the fragility of EPs (SPERs) for systems with charge $\mathrm{U(1)}$ symmetry and spin-parity symmetry (chiral symmetry).
The above results strongly suggest that the reduction of topological classifications results in the fragility of EPs and their variants in generic dimensions and symmetry classes.

We finish this paper with a remark on gapped systems. 
Topological invariants defined in Eqs.~(\ref{eq: W (N Sz) free})~and~(\ref{eq: W [N,P] }) [Eqs.~(\ref{eq: H0Gamma})~and~(\ref{eq: Z2 mbdyH chiral})] are available for the characterization of one- [zero-] dimensional gapped systems. 
In particular, Eqs.~(\ref{eq: W (N Sz) free}),~(\ref{eq: W [N,P] }),~and~(\ref{eq: Wparity=Wsz+Wsz}) indicate the reduction of one-dimensional point-gap topology $\mathbb{Z}^{(N+P'+1)/2}\to\mathbb{Z}$ for gapped systems with charge $\mathrm{U(1)}$ and spin-parity symmetry (see Appendix~\ref{sec: gapped 1D app}). 
In addition, Eqs.~(\ref{eq: H0Gamma}),~(\ref{eq: Z2 mbdyH chiral}),~and~(\ref{eq: N0Ch and nu}) indicate the reduction of zero-dimensional point-gap topology $\mathbb{Z}\to\mathbb{Z}_2$ for gapped systems with chiral symmetry (see Appendix~\ref{sec: gapped 0D app}).

\section*{
Acknowledgements
}
The authors thank Hosho Katsura, Norio Kawakami, and Takuma Isobe for fruitful discussions.
A part of the computation has been done using the facilities of the Supercomputer Center, the Institute for Solid State Physics, the University of Tokyo.
This work is supported by JSPS KAKENHI Grants No.~JP17H06138, No.~JP21K13850 and No.~JP22H05247.
This work is also supported by JST CREST, Grant No.~JPMJCR19T1.


\begin{thebibliography}{129}%
\makeatletter
\providecommand \@ifxundefined [1]{%
 \@ifx{#1\undefined}
}%
\providecommand \@ifnum [1]{%
 \ifnum #1\expandafter \@firstoftwo
 \else \expandafter \@secondoftwo
 \fi
}%
\providecommand \@ifx [1]{%
 \ifx #1\expandafter \@firstoftwo
 \else \expandafter \@secondoftwo
 \fi
}%
\providecommand \natexlab [1]{#1}%
\providecommand \enquote  [1]{``#1''}%
\providecommand \bibnamefont  [1]{#1}%
\providecommand \bibfnamefont [1]{#1}%
\providecommand \citenamefont [1]{#1}%
\providecommand \href@noop [0]{\@secondoftwo}%
\providecommand \href [0]{\begingroup \@sanitize@url \@href}%
\providecommand \@href[1]{\@@startlink{#1}\@@href}%
\providecommand \@@href[1]{\endgroup#1\@@endlink}%
\providecommand \@sanitize@url [0]{\catcode `\\12\catcode `\$12\catcode
  `\&12\catcode `\#12\catcode `\^12\catcode `\_12\catcode `\%12\relax}%
\providecommand \@@startlink[1]{}%
\providecommand \@@endlink[0]{}%
\providecommand \url  [0]{\begingroup\@sanitize@url \@url }%
\providecommand \@url [1]{\endgroup\@href {#1}{\urlprefix }}%
\providecommand \urlprefix  [0]{URL }%
\providecommand \Eprint [0]{\href }%
\providecommand \doibase [0]{http://dx.doi.org/}%
\providecommand \selectlanguage [0]{\@gobble}%
\providecommand \bibinfo  [0]{\@secondoftwo}%
\providecommand \bibfield  [0]{\@secondoftwo}%
\providecommand \translation [1]{[#1]}%
\providecommand \BibitemOpen [0]{}%
\providecommand \bibitemStop [0]{}%
\providecommand \bibitemNoStop [0]{.\EOS\space}%
\providecommand \EOS [0]{\spacefactor3000\relax}%
\providecommand \BibitemShut  [1]{\csname bibitem#1\endcsname}%
\let\auto@bib@innerbib\@empty
\bibitem [{\citenamefont {Tsui}\ \emph {et~al.}(1982)\citenamefont {Tsui},
  \citenamefont {Stormer},\ and\ \citenamefont {Gossard}}]{Tsui_FQHEExp_PRL82}%
  \BibitemOpen
  \bibfield  {author} {\bibinfo {author} {\bibfnamefont {D.~C.}\ \bibnamefont
  {Tsui}}, \bibinfo {author} {\bibfnamefont {H.~L.}\ \bibnamefont {Stormer}}, \
  and\ \bibinfo {author} {\bibfnamefont {A.~C.}\ \bibnamefont {Gossard}},\
  }\href {\doibase 10.1103/PhysRevLett.48.1559} {\bibfield  {journal} {\bibinfo
   {journal} {Phys. Rev. Lett.}\ }\textbf {\bibinfo {volume} {48}},\ \bibinfo
  {pages} {1559} (\bibinfo {year} {1982})}\BibitemShut {NoStop}%
\bibitem [{\citenamefont {Laughlin}(1983)}]{Laughlin_FQHE_PRL83}%
  \BibitemOpen
  \bibfield  {author} {\bibinfo {author} {\bibfnamefont {R.~B.}\ \bibnamefont
  {Laughlin}},\ }\href {\doibase 10.1103/PhysRevLett.50.1395} {\bibfield
  {journal} {\bibinfo  {journal} {Phys. Rev. Lett.}\ }\textbf {\bibinfo
  {volume} {50}},\ \bibinfo {pages} {1395} (\bibinfo {year}
  {1983})}\BibitemShut {NoStop}%
\bibitem [{\citenamefont {Tang}\ \emph {et~al.}(2011)\citenamefont {Tang},
  \citenamefont {Mei},\ and\ \citenamefont {Wen}}]{Tang_FChern_PRL11}%
  \BibitemOpen
  \bibfield  {author} {\bibinfo {author} {\bibfnamefont {E.}~\bibnamefont
  {Tang}}, \bibinfo {author} {\bibfnamefont {J.-W.}\ \bibnamefont {Mei}}, \
  and\ \bibinfo {author} {\bibfnamefont {X.-G.}\ \bibnamefont {Wen}},\ }\href
  {\doibase 10.1103/PhysRevLett.106.236802} {\bibfield  {journal} {\bibinfo
  {journal} {Phys. Rev. Lett.}\ }\textbf {\bibinfo {volume} {106}},\ \bibinfo
  {pages} {236802} (\bibinfo {year} {2011})}\BibitemShut {NoStop}%
\bibitem [{\citenamefont {Sun}\ \emph {et~al.}(2011)\citenamefont {Sun},
  \citenamefont {Gu}, \citenamefont {Katsura},\ and\ \citenamefont
  {Das~Sarma}}]{Sun_FChern_PRL11}%
  \BibitemOpen
  \bibfield  {author} {\bibinfo {author} {\bibfnamefont {K.}~\bibnamefont
  {Sun}}, \bibinfo {author} {\bibfnamefont {Z.}~\bibnamefont {Gu}}, \bibinfo
  {author} {\bibfnamefont {H.}~\bibnamefont {Katsura}}, \ and\ \bibinfo
  {author} {\bibfnamefont {S.}~\bibnamefont {Das~Sarma}},\ }\href {\doibase
  10.1103/PhysRevLett.106.236803} {\bibfield  {journal} {\bibinfo  {journal}
  {Phys. Rev. Lett.}\ }\textbf {\bibinfo {volume} {106}},\ \bibinfo {pages}
  {236803} (\bibinfo {year} {2011})}\BibitemShut {NoStop}%
\bibitem [{\citenamefont {Neupert}\ \emph {et~al.}(2011)\citenamefont
  {Neupert}, \citenamefont {Santos}, \citenamefont {Chamon},\ and\
  \citenamefont {Mudry}}]{Neupert_FChern_PRL11}%
  \BibitemOpen
  \bibfield  {author} {\bibinfo {author} {\bibfnamefont {T.}~\bibnamefont
  {Neupert}}, \bibinfo {author} {\bibfnamefont {L.}~\bibnamefont {Santos}},
  \bibinfo {author} {\bibfnamefont {C.}~\bibnamefont {Chamon}}, \ and\ \bibinfo
  {author} {\bibfnamefont {C.}~\bibnamefont {Mudry}},\ }\href {\doibase
  10.1103/PhysRevLett.106.236804} {\bibfield  {journal} {\bibinfo  {journal}
  {Phys. Rev. Lett.}\ }\textbf {\bibinfo {volume} {106}},\ \bibinfo {pages}
  {236804} (\bibinfo {year} {2011})}\BibitemShut {NoStop}%
\bibitem [{\citenamefont {Regnault}\ and\ \citenamefont
  {Bernevig}(2011)}]{Regnalt_FChen_PRX11}%
  \BibitemOpen
  \bibfield  {author} {\bibinfo {author} {\bibfnamefont {N.}~\bibnamefont
  {Regnault}}\ and\ \bibinfo {author} {\bibfnamefont {B.~A.}\ \bibnamefont
  {Bernevig}},\ }\href {\doibase 10.1103/PhysRevX.1.021014} {\bibfield
  {journal} {\bibinfo  {journal} {Phys. Rev. X}\ }\textbf {\bibinfo {volume}
  {1}},\ \bibinfo {pages} {021014} (\bibinfo {year} {2011})}\BibitemShut
  {NoStop}%
\bibitem [{\citenamefont {Sheng}\ \emph {et~al.}(2011)\citenamefont {Sheng},
  \citenamefont {Gu}, \citenamefont {Sun},\ and\ \citenamefont
  {Sheng}}]{Sheng_FChern_NComm12}%
  \BibitemOpen
  \bibfield  {author} {\bibinfo {author} {\bibfnamefont {D.~N.}\ \bibnamefont
  {Sheng}}, \bibinfo {author} {\bibfnamefont {Z.-C.}\ \bibnamefont {Gu}},
  \bibinfo {author} {\bibfnamefont {K.}~\bibnamefont {Sun}}, \ and\ \bibinfo
  {author} {\bibfnamefont {L.}~\bibnamefont {Sheng}},\ }\href@noop {}
  {\bibfield  {journal} {\bibinfo  {journal} {Nature Communications}\ }\textbf
  {\bibinfo {volume} {2}},\ \bibinfo {pages} {389 EP } (\bibinfo {year}
  {2011})},\ \bibinfo {note} {article}\BibitemShut {NoStop}%
\bibitem [{\citenamefont {Bergholtz}\ and\ \citenamefont
  {Liu}(2013)}]{Bergholtz_FChern_IntJModPhysB13}%
  \BibitemOpen
  \bibfield  {author} {\bibinfo {author} {\bibfnamefont {E.~J.}\ \bibnamefont
  {Bergholtz}}\ and\ \bibinfo {author} {\bibfnamefont {Z.}~\bibnamefont
  {Liu}},\ }\href {\doibase 10.1142/S021797921330017X} {\bibfield  {journal}
  {\bibinfo  {journal} {International Journal of Modern Physics B}\ }\textbf
  {\bibinfo {volume} {27}},\ \bibinfo {pages} {1330017} (\bibinfo {year}
  {2013})}\BibitemShut {NoStop}%
\bibitem [{\citenamefont {Pesin}\ and\ \citenamefont
  {Balents}(2010)}]{Pesin_TMI_NatPhys2010}%
  \BibitemOpen
  \bibfield  {author} {\bibinfo {author} {\bibfnamefont {D.}~\bibnamefont
  {Pesin}}\ and\ \bibinfo {author} {\bibfnamefont {L.}~\bibnamefont
  {Balents}},\ }\href {\doibase 10.1038/nphys1606} {\bibfield  {journal}
  {\bibinfo  {journal} {Nature Physics}\ }\textbf {\bibinfo {volume} {6}},\
  \bibinfo {pages} {376} (\bibinfo {year} {2010})}\BibitemShut {NoStop}%
\bibitem [{\citenamefont {Schnyder}\ \emph {et~al.}(2008)\citenamefont
  {Schnyder}, \citenamefont {Ryu}, \citenamefont {Furusaki},\ and\
  \citenamefont {Ludwig}}]{Schnyder_classification_free_2008}%
  \BibitemOpen
  \bibfield  {author} {\bibinfo {author} {\bibfnamefont {A.~P.}\ \bibnamefont
  {Schnyder}}, \bibinfo {author} {\bibfnamefont {S.}~\bibnamefont {Ryu}},
  \bibinfo {author} {\bibfnamefont {A.}~\bibnamefont {Furusaki}}, \ and\
  \bibinfo {author} {\bibfnamefont {A.~W.~W.}\ \bibnamefont {Ludwig}},\ }\href
  {\doibase 10.1103/PhysRevB.78.195125} {\bibfield  {journal} {\bibinfo
  {journal} {Phys. Rev. B}\ }\textbf {\bibinfo {volume} {78}},\ \bibinfo
  {pages} {195125} (\bibinfo {year} {2008})}\BibitemShut {NoStop}%
\bibitem [{\citenamefont {Kitaev}(2009)}]{Kitaev_classification_free_2009}%
  \BibitemOpen
  \bibfield  {author} {\bibinfo {author} {\bibfnamefont {A.}~\bibnamefont
  {Kitaev}},\ }\href {\doibase 10.1063/1.3149495} {\bibfield  {journal}
  {\bibinfo  {journal} {AIP Conf. Proc.}\ }\textbf {\bibinfo {volume} {1134}},\
  \bibinfo {pages} {22} (\bibinfo {year} {2009})}\BibitemShut {NoStop}%
\bibitem [{\citenamefont {Ryu}\ \emph {et~al.}(2010)\citenamefont {Ryu},
  \citenamefont {Schnyder}, \citenamefont {Furusaki},\ and\ \citenamefont
  {Ludwig}}]{Ryu_classification_free_2010}%
  \BibitemOpen
  \bibfield  {author} {\bibinfo {author} {\bibfnamefont {S.}~\bibnamefont
  {Ryu}}, \bibinfo {author} {\bibfnamefont {A.~P.}\ \bibnamefont {Schnyder}},
  \bibinfo {author} {\bibfnamefont {A.}~\bibnamefont {Furusaki}}, \ and\
  \bibinfo {author} {\bibfnamefont {A.~W.~W.}\ \bibnamefont {Ludwig}},\ }\href
  {http://stacks.iop.org/1367-2630/12/i=6/a=065010} {\bibfield  {journal}
  {\bibinfo  {journal} {New J. Phys.}\ }\textbf {\bibinfo {volume} {12}},\
  \bibinfo {pages} {065010} (\bibinfo {year} {2010})}\BibitemShut {NoStop}%
\bibitem [{\citenamefont {Fidkowski}\ and\ \citenamefont
  {Kitaev}(2010)}]{Z_to_Zn_Fidkowski_PRB10}%
  \BibitemOpen
  \bibfield  {author} {\bibinfo {author} {\bibfnamefont {L.}~\bibnamefont
  {Fidkowski}}\ and\ \bibinfo {author} {\bibfnamefont {A.}~\bibnamefont
  {Kitaev}},\ }\href {\doibase 10.1103/PhysRevB.81.134509} {\bibfield
  {journal} {\bibinfo  {journal} {Phys. Rev. B}\ }\textbf {\bibinfo {volume}
  {81}},\ \bibinfo {pages} {134509} (\bibinfo {year} {2010})}\BibitemShut
  {NoStop}%
\bibitem [{\citenamefont {Pollmann}\ \emph {et~al.}(2010)\citenamefont
  {Pollmann}, \citenamefont {Turner}, \citenamefont {Berg},\ and\ \citenamefont
  {Oshikawa}}]{entanglement_Pollmann10}%
  \BibitemOpen
  \bibfield  {author} {\bibinfo {author} {\bibfnamefont {F.}~\bibnamefont
  {Pollmann}}, \bibinfo {author} {\bibfnamefont {A.~M.}\ \bibnamefont
  {Turner}}, \bibinfo {author} {\bibfnamefont {E.}~\bibnamefont {Berg}}, \ and\
  \bibinfo {author} {\bibfnamefont {M.}~\bibnamefont {Oshikawa}},\ }\href
  {\doibase 10.1103/PhysRevB.81.064439} {\bibfield  {journal} {\bibinfo
  {journal} {Phys. Rev. B}\ }\textbf {\bibinfo {volume} {81}},\ \bibinfo
  {pages} {064439} (\bibinfo {year} {2010})}\BibitemShut {NoStop}%
\bibitem [{\citenamefont {Turner}\ \emph {et~al.}(2011)\citenamefont {Turner},
  \citenamefont {Pollmann},\ and\ \citenamefont {Berg}}]{Turner_ZtoZ8_PRB11}%
  \BibitemOpen
  \bibfield  {author} {\bibinfo {author} {\bibfnamefont {A.~M.}\ \bibnamefont
  {Turner}}, \bibinfo {author} {\bibfnamefont {F.}~\bibnamefont {Pollmann}}, \
  and\ \bibinfo {author} {\bibfnamefont {E.}~\bibnamefont {Berg}},\ }\href
  {\doibase 10.1103/PhysRevB.83.075102} {\bibfield  {journal} {\bibinfo
  {journal} {Phys. Rev. B}\ }\textbf {\bibinfo {volume} {83}},\ \bibinfo
  {pages} {075102} (\bibinfo {year} {2011})}\BibitemShut {NoStop}%
\bibitem [{\citenamefont {Fidkowski}\ and\ \citenamefont
  {Kitaev}(2011)}]{Fidkowski_1Dclassificatin_PRB11}%
  \BibitemOpen
  \bibfield  {author} {\bibinfo {author} {\bibfnamefont {L.}~\bibnamefont
  {Fidkowski}}\ and\ \bibinfo {author} {\bibfnamefont {A.}~\bibnamefont
  {Kitaev}},\ }\href {\doibase 10.1103/PhysRevB.83.075103} {\bibfield
  {journal} {\bibinfo  {journal} {Phys. Rev. B}\ }\textbf {\bibinfo {volume}
  {83}},\ \bibinfo {pages} {075103} (\bibinfo {year} {2011})}\BibitemShut
  {NoStop}%
\bibitem [{\citenamefont {Chen}\ \emph
  {et~al.}(2011{\natexlab{a}})\citenamefont {Chen}, \citenamefont {Gu},\ and\
  \citenamefont {Wen}}]{Chen_classification_1D_1}%
  \BibitemOpen
  \bibfield  {author} {\bibinfo {author} {\bibfnamefont {X.}~\bibnamefont
  {Chen}}, \bibinfo {author} {\bibfnamefont {Z.-C.}\ \bibnamefont {Gu}}, \ and\
  \bibinfo {author} {\bibfnamefont {X.-G.}\ \bibnamefont {Wen}},\ }\href
  {\doibase 10.1103/PhysRevB.83.035107} {\bibfield  {journal} {\bibinfo
  {journal} {Phys. Rev. B}\ }\textbf {\bibinfo {volume} {83}},\ \bibinfo
  {pages} {035107} (\bibinfo {year} {2011}{\natexlab{a}})}\BibitemShut
  {NoStop}%
\bibitem [{\citenamefont {Chen}\ \emph
  {et~al.}(2011{\natexlab{b}})\citenamefont {Chen}, \citenamefont {Gu},\ and\
  \citenamefont {Wen}}]{Chen_classification_1D_2}%
  \BibitemOpen
  \bibfield  {author} {\bibinfo {author} {\bibfnamefont {X.}~\bibnamefont
  {Chen}}, \bibinfo {author} {\bibfnamefont {Z.-C.}\ \bibnamefont {Gu}}, \ and\
  \bibinfo {author} {\bibfnamefont {X.-G.}\ \bibnamefont {Wen}},\ }\href
  {\doibase 10.1103/PhysRevB.84.235128} {\bibfield  {journal} {\bibinfo
  {journal} {Phys. Rev. B}\ }\textbf {\bibinfo {volume} {84}},\ \bibinfo
  {pages} {235128} (\bibinfo {year} {2011}{\natexlab{b}})}\BibitemShut
  {NoStop}%
\bibitem [{\citenamefont {Lu}\ and\ \citenamefont
  {Vishwanath}(2012)}]{Lu_CS_2011}%
  \BibitemOpen
  \bibfield  {author} {\bibinfo {author} {\bibfnamefont {Y.-M.}\ \bibnamefont
  {Lu}}\ and\ \bibinfo {author} {\bibfnamefont {A.}~\bibnamefont
  {Vishwanath}},\ }\href {\doibase 10.1103/PhysRevB.86.125119} {\bibfield
  {journal} {\bibinfo  {journal} {Phys. Rev. B}\ }\textbf {\bibinfo {volume}
  {86}},\ \bibinfo {pages} {125119} (\bibinfo {year} {2012})}\BibitemShut
  {NoStop}%
\bibitem [{\citenamefont {Yao}\ and\ \citenamefont
  {Ryu}(2013)}]{YaoRyu_Z_to_Z8_2013}%
  \BibitemOpen
  \bibfield  {author} {\bibinfo {author} {\bibfnamefont {H.}~\bibnamefont
  {Yao}}\ and\ \bibinfo {author} {\bibfnamefont {S.}~\bibnamefont {Ryu}},\
  }\href {\doibase 10.1103/PhysRevB.88.064507} {\bibfield  {journal} {\bibinfo
  {journal} {Phys. Rev. B}\ }\textbf {\bibinfo {volume} {88}},\ \bibinfo
  {pages} {064507} (\bibinfo {year} {2013})}\BibitemShut {NoStop}%
\bibitem [{\citenamefont {Ryu}\ and\ \citenamefont
  {Zhang}(2012)}]{Ryu_Z_to_Z8_2013}%
  \BibitemOpen
  \bibfield  {author} {\bibinfo {author} {\bibfnamefont {S.}~\bibnamefont
  {Ryu}}\ and\ \bibinfo {author} {\bibfnamefont {S.-C.}\ \bibnamefont
  {Zhang}},\ }\href {\doibase 10.1103/PhysRevB.85.245132} {\bibfield  {journal}
  {\bibinfo  {journal} {Phys. Rev. B}\ }\textbf {\bibinfo {volume} {85}},\
  \bibinfo {pages} {245132} (\bibinfo {year} {2012})}\BibitemShut {NoStop}%
\bibitem [{\citenamefont {Qi}(2013)}]{Qi_Z_to_Z8_2013}%
  \BibitemOpen
  \bibfield  {author} {\bibinfo {author} {\bibfnamefont {X.-L.}\ \bibnamefont
  {Qi}},\ }\href@noop {} {\bibfield  {journal} {\bibinfo  {journal} {New J.
  Phys.}\ }\textbf {\bibinfo {volume} {15}},\ \bibinfo {pages} {065002}
  (\bibinfo {year} {2013})}\BibitemShut {NoStop}%
\bibitem [{\citenamefont {Levin}\ and\ \citenamefont
  {Stern}(2012)}]{Levin_CS_2012}%
  \BibitemOpen
  \bibfield  {author} {\bibinfo {author} {\bibfnamefont {M.}~\bibnamefont
  {Levin}}\ and\ \bibinfo {author} {\bibfnamefont {A.}~\bibnamefont {Stern}},\
  }\href {\doibase 10.1103/PhysRevB.86.115131} {\bibfield  {journal} {\bibinfo
  {journal} {Phys. Rev. B}\ }\textbf {\bibinfo {volume} {86}},\ \bibinfo
  {pages} {115131} (\bibinfo {year} {2012})}\BibitemShut {NoStop}%
\bibitem [{\citenamefont {Hsieh}\ \emph {et~al.}(2014)\citenamefont {Hsieh},
  \citenamefont {Morimoto},\ and\ \citenamefont {Ryu}}]{Hsieh_CS_CPT_2014}%
  \BibitemOpen
  \bibfield  {author} {\bibinfo {author} {\bibfnamefont {C.-T.}\ \bibnamefont
  {Hsieh}}, \bibinfo {author} {\bibfnamefont {T.}~\bibnamefont {Morimoto}}, \
  and\ \bibinfo {author} {\bibfnamefont {S.}~\bibnamefont {Ryu}},\ }\href
  {\doibase 10.1103/PhysRevB.90.245111} {\bibfield  {journal} {\bibinfo
  {journal} {Phys. Rev. B}\ }\textbf {\bibinfo {volume} {90}},\ \bibinfo
  {pages} {245111} (\bibinfo {year} {2014})}\BibitemShut {NoStop}%
\bibitem [{\citenamefont {Isobe}\ and\ \citenamefont
  {Fu}(2015)}]{Isobe_Fu2015}%
  \BibitemOpen
  \bibfield  {author} {\bibinfo {author} {\bibfnamefont {H.}~\bibnamefont
  {Isobe}}\ and\ \bibinfo {author} {\bibfnamefont {L.}~\bibnamefont {Fu}},\
  }\href {\doibase 10.1103/PhysRevB.92.081304} {\bibfield  {journal} {\bibinfo
  {journal} {Phys. Rev. B}\ }\textbf {\bibinfo {volume} {92}},\ \bibinfo
  {pages} {081304} (\bibinfo {year} {2015})}\BibitemShut {NoStop}%
\bibitem [{\citenamefont {Chen}\ \emph {et~al.}(2013)\citenamefont {Chen},
  \citenamefont {Gu}, \citenamefont {Liu},\ and\ \citenamefont
  {Wen}}]{chen_cohomology_3D}%
  \BibitemOpen
  \bibfield  {author} {\bibinfo {author} {\bibfnamefont {X.}~\bibnamefont
  {Chen}}, \bibinfo {author} {\bibfnamefont {Z.-C.}\ \bibnamefont {Gu}},
  \bibinfo {author} {\bibfnamefont {Z.-X.}\ \bibnamefont {Liu}}, \ and\
  \bibinfo {author} {\bibfnamefont {X.-G.}\ \bibnamefont {Wen}},\ }\href
  {\doibase 10.1103/PhysRevB.87.155114} {\bibfield  {journal} {\bibinfo
  {journal} {Phys. Rev. B}\ }\textbf {\bibinfo {volume} {87}},\ \bibinfo
  {pages} {155114} (\bibinfo {year} {2013})}\BibitemShut {NoStop}%
\bibitem [{\citenamefont {Gu}\ and\ \citenamefont
  {Wen}(2014)}]{gu_supercohomology}%
  \BibitemOpen
  \bibfield  {author} {\bibinfo {author} {\bibfnamefont {Z.-C.}\ \bibnamefont
  {Gu}}\ and\ \bibinfo {author} {\bibfnamefont {X.-G.}\ \bibnamefont {Wen}},\
  }\href {\doibase 10.1103/PhysRevB.90.115141} {\bibfield  {journal} {\bibinfo
  {journal} {Phys. Rev. B}\ }\textbf {\bibinfo {volume} {90}},\ \bibinfo
  {pages} {115141} (\bibinfo {year} {2014})}\BibitemShut {NoStop}%
\bibitem [{\citenamefont
  {Kapustin}(2014{\natexlab{a}})}]{kapustin_bosonic_cobordisms2014_1}%
  \BibitemOpen
  \bibfield  {author} {\bibinfo {author} {\bibfnamefont {A.}~\bibnamefont
  {Kapustin}},\ }\href@noop {} {\bibfield  {journal} {\bibinfo  {journal}
  {arXiv:1403.1467}\ } (\bibinfo {year} {2014}{\natexlab{a}})}\BibitemShut
  {NoStop}%
\bibitem [{\citenamefont
  {Kapustin}(2014{\natexlab{b}})}]{kapustin_bosonic_cobordisms2014_2}%
  \BibitemOpen
  \bibfield  {author} {\bibinfo {author} {\bibfnamefont {A.}~\bibnamefont
  {Kapustin}},\ }\href@noop {} {\bibfield  {journal} {\bibinfo  {journal}
  {arXiv:1404.6659}\ } (\bibinfo {year} {2014}{\natexlab{b}})}\BibitemShut
  {NoStop}%
\bibitem [{\citenamefont {Kapustin}\ \emph {et~al.}(2015)\citenamefont
  {Kapustin}, \citenamefont {Thorngren}, \citenamefont {Turzillo},\ and\
  \citenamefont {Wang}}]{kapustin_fermionic_cobordisms2014}%
  \BibitemOpen
  \bibfield  {author} {\bibinfo {author} {\bibfnamefont {A.}~\bibnamefont
  {Kapustin}}, \bibinfo {author} {\bibfnamefont {R.}~\bibnamefont {Thorngren}},
  \bibinfo {author} {\bibfnamefont {A.}~\bibnamefont {Turzillo}}, \ and\
  \bibinfo {author} {\bibfnamefont {Z.}~\bibnamefont {Wang}},\ }\href {\doibase
  10.1007/JHEP12(2015)052} {\bibfield  {journal} {\bibinfo  {journal} {Journal
  of High Energy Physics}\ }\textbf {\bibinfo {volume} {2015}},\ \bibinfo
  {pages} {1} (\bibinfo {year} {2015})}\BibitemShut {NoStop}%
\bibitem [{\citenamefont {Fidkowski}\ \emph {et~al.}(2013)\citenamefont
  {Fidkowski}, \citenamefont {Chen},\ and\ \citenamefont
  {Vishwanath}}]{Fidkowski_Z162013}%
  \BibitemOpen
  \bibfield  {author} {\bibinfo {author} {\bibfnamefont {L.}~\bibnamefont
  {Fidkowski}}, \bibinfo {author} {\bibfnamefont {X.}~\bibnamefont {Chen}}, \
  and\ \bibinfo {author} {\bibfnamefont {A.}~\bibnamefont {Vishwanath}},\
  }\href {\doibase 10.1103/PhysRevX.3.041016} {\bibfield  {journal} {\bibinfo
  {journal} {Phys. Rev. X}\ }\textbf {\bibinfo {volume} {3}},\ \bibinfo {pages}
  {041016} (\bibinfo {year} {2013})}\BibitemShut {NoStop}%
\bibitem [{\citenamefont {Wang}\ \emph {et~al.}(2014)\citenamefont {Wang},
  \citenamefont {Potter},\ and\ \citenamefont
  {Senthil}}]{Wang_Potter_Senthil2014}%
  \BibitemOpen
  \bibfield  {author} {\bibinfo {author} {\bibfnamefont {C.}~\bibnamefont
  {Wang}}, \bibinfo {author} {\bibfnamefont {A.~C.}\ \bibnamefont {Potter}}, \
  and\ \bibinfo {author} {\bibfnamefont {T.}~\bibnamefont {Senthil}},\ }\href
  {\doibase 10.1126/science.1243326} {\bibfield  {journal} {\bibinfo  {journal}
  {Science}\ }\textbf {\bibinfo {volume} {343}},\ \bibinfo {pages} {629}
  (\bibinfo {year} {2014})}\BibitemShut {NoStop}%
\bibitem [{\citenamefont {Metlitski}\ \emph {et~al.}(2014)\citenamefont
  {Metlitski}, \citenamefont {Fidkowski}, \citenamefont {Chen},\ and\
  \citenamefont {Vishwanath}}]{Metlitski_3Dinteraction2014}%
  \BibitemOpen
  \bibfield  {author} {\bibinfo {author} {\bibfnamefont {M.~A.}\ \bibnamefont
  {Metlitski}}, \bibinfo {author} {\bibfnamefont {L.}~\bibnamefont
  {Fidkowski}}, \bibinfo {author} {\bibfnamefont {X.}~\bibnamefont {Chen}}, \
  and\ \bibinfo {author} {\bibfnamefont {A.}~\bibnamefont {Vishwanath}},\
  }\href@noop {} {\bibfield  {journal} {\bibinfo  {journal} {arXiv:1406.3032}\
  } (\bibinfo {year} {2014})}\BibitemShut {NoStop}%
\bibitem [{\citenamefont {Wang}\ and\ \citenamefont
  {Senthil}(2014)}]{Wang_Senthil2014}%
  \BibitemOpen
  \bibfield  {author} {\bibinfo {author} {\bibfnamefont {C.}~\bibnamefont
  {Wang}}\ and\ \bibinfo {author} {\bibfnamefont {T.}~\bibnamefont {Senthil}},\
  }\href {\doibase 10.1103/PhysRevB.89.195124} {\bibfield  {journal} {\bibinfo
  {journal} {Phys. Rev. B}\ }\textbf {\bibinfo {volume} {89}},\ \bibinfo
  {pages} {195124} (\bibinfo {year} {2014})}\BibitemShut {NoStop}%
\bibitem [{\citenamefont {You}\ and\ \citenamefont {Xu}(2014)}]{You_Cenke2014}%
  \BibitemOpen
  \bibfield  {author} {\bibinfo {author} {\bibfnamefont {Y.-Z.}\ \bibnamefont
  {You}}\ and\ \bibinfo {author} {\bibfnamefont {C.}~\bibnamefont {Xu}},\
  }\href {\doibase 10.1103/PhysRevB.90.245120} {\bibfield  {journal} {\bibinfo
  {journal} {Phys. Rev. B}\ }\textbf {\bibinfo {volume} {90}},\ \bibinfo
  {pages} {245120} (\bibinfo {year} {2014})}\BibitemShut {NoStop}%
\bibitem [{\citenamefont {Morimoto}\ \emph {et~al.}(2015)\citenamefont
  {Morimoto}, \citenamefont {Furusaki},\ and\ \citenamefont
  {Mudry}}]{Morimoto_2015}%
  \BibitemOpen
  \bibfield  {author} {\bibinfo {author} {\bibfnamefont {T.}~\bibnamefont
  {Morimoto}}, \bibinfo {author} {\bibfnamefont {A.}~\bibnamefont {Furusaki}},
  \ and\ \bibinfo {author} {\bibfnamefont {C.}~\bibnamefont {Mudry}},\ }\href
  {\doibase 10.1103/PhysRevB.92.125104} {\bibfield  {journal} {\bibinfo
  {journal} {Phys. Rev. B}\ }\textbf {\bibinfo {volume} {92}},\ \bibinfo
  {pages} {125104} (\bibinfo {year} {2015})}\BibitemShut {NoStop}%
\bibitem [{\citenamefont {Yoshida}\ \emph {et~al.}(2017)\citenamefont
  {Yoshida}, \citenamefont {Daido}, \citenamefont {Yanase},\ and\ \citenamefont
  {Kawakami}}]{Superlattice_Yoshida17}%
  \BibitemOpen
  \bibfield  {author} {\bibinfo {author} {\bibfnamefont {T.}~\bibnamefont
  {Yoshida}}, \bibinfo {author} {\bibfnamefont {A.}~\bibnamefont {Daido}},
  \bibinfo {author} {\bibfnamefont {Y.}~\bibnamefont {Yanase}}, \ and\ \bibinfo
  {author} {\bibfnamefont {N.}~\bibnamefont {Kawakami}},\ }\href {\doibase
  10.1103/PhysRevLett.118.147001} {\bibfield  {journal} {\bibinfo  {journal}
  {Phys. Rev. Lett.}\ }\textbf {\bibinfo {volume} {118}},\ \bibinfo {pages}
  {147001} (\bibinfo {year} {2017})}\BibitemShut {NoStop}%
\bibitem [{\citenamefont {Jian}\ and\ \citenamefont
  {Xu}(2018)}]{CMJian_ZtoZnSynthetic_PRX18}%
  \BibitemOpen
  \bibfield  {author} {\bibinfo {author} {\bibfnamefont {C.-M.}\ \bibnamefont
  {Jian}}\ and\ \bibinfo {author} {\bibfnamefont {C.}~\bibnamefont {Xu}},\
  }\href {\doibase 10.1103/PhysRevX.8.041030} {\bibfield  {journal} {\bibinfo
  {journal} {Phys. Rev. X}\ }\textbf {\bibinfo {volume} {8}},\ \bibinfo {pages}
  {041030} (\bibinfo {year} {2018})}\BibitemShut {NoStop}%
\bibitem [{\citenamefont {Hu}\ and\ \citenamefont
  {Hughes}(2011)}]{Hu_nH_PRB11}%
  \BibitemOpen
  \bibfield  {author} {\bibinfo {author} {\bibfnamefont {Y.~C.}\ \bibnamefont
  {Hu}}\ and\ \bibinfo {author} {\bibfnamefont {T.~L.}\ \bibnamefont
  {Hughes}},\ }\href {\doibase 10.1103/PhysRevB.84.153101} {\bibfield
  {journal} {\bibinfo  {journal} {Phys. Rev. B}\ }\textbf {\bibinfo {volume}
  {84}},\ \bibinfo {pages} {153101} (\bibinfo {year} {2011})}\BibitemShut
  {NoStop}%
\bibitem [{\citenamefont {Esaki}\ \emph {et~al.}(2011)\citenamefont {Esaki},
  \citenamefont {Sato}, \citenamefont {Hasebe},\ and\ \citenamefont
  {Kohmoto}}]{Esaki_nH_PRB11}%
  \BibitemOpen
  \bibfield  {author} {\bibinfo {author} {\bibfnamefont {K.}~\bibnamefont
  {Esaki}}, \bibinfo {author} {\bibfnamefont {M.}~\bibnamefont {Sato}},
  \bibinfo {author} {\bibfnamefont {K.}~\bibnamefont {Hasebe}}, \ and\ \bibinfo
  {author} {\bibfnamefont {M.}~\bibnamefont {Kohmoto}},\ }\href {\doibase
  10.1103/PhysRevB.84.205128} {\bibfield  {journal} {\bibinfo  {journal} {Phys.
  Rev. B}\ }\textbf {\bibinfo {volume} {84}},\ \bibinfo {pages} {205128}
  (\bibinfo {year} {2011})}\BibitemShut {NoStop}%
\bibitem [{\citenamefont {Sato}\ \emph {et~al.}(2012)\citenamefont {Sato},
  \citenamefont {Hasebe}, \citenamefont {Esaki},\ and\ \citenamefont
  {Kohmoto}}]{Sato_nHPTEP12}%
  \BibitemOpen
  \bibfield  {author} {\bibinfo {author} {\bibfnamefont {M.}~\bibnamefont
  {Sato}}, \bibinfo {author} {\bibfnamefont {K.}~\bibnamefont {Hasebe}},
  \bibinfo {author} {\bibfnamefont {K.}~\bibnamefont {Esaki}}, \ and\ \bibinfo
  {author} {\bibfnamefont {M.}~\bibnamefont {Kohmoto}},\ }\href {\doibase
  10.1143/PTP.127.937} {\bibfield  {journal} {\bibinfo  {journal} {Progress of
  Theoretical Physics}\ }\textbf {\bibinfo {volume} {127}},\ \bibinfo {pages}
  {937} (\bibinfo {year} {2012})}\BibitemShut {NoStop}%
\bibitem [{\citenamefont {Diehl}\ \emph {et~al.}(2011)\citenamefont {Diehl},
  \citenamefont {Rico}, \citenamefont {Baranov},\ and\ \citenamefont
  {Zoller}}]{Diehl_DissCher_NatPhys11}%
  \BibitemOpen
  \bibfield  {author} {\bibinfo {author} {\bibfnamefont {S.}~\bibnamefont
  {Diehl}}, \bibinfo {author} {\bibfnamefont {E.}~\bibnamefont {Rico}},
  \bibinfo {author} {\bibfnamefont {M.~A.}\ \bibnamefont {Baranov}}, \ and\
  \bibinfo {author} {\bibfnamefont {P.}~\bibnamefont {Zoller}},\ }\href
  {\doibase 10.1038/nphys2106} {\bibfield  {journal} {\bibinfo  {journal}
  {Nature Physics}\ }\textbf {\bibinfo {volume} {7}},\ \bibinfo {pages} {971}
  (\bibinfo {year} {2011})}\BibitemShut {NoStop}%
\bibitem [{\citenamefont {Bardyn}\ \emph {et~al.}(2013)\citenamefont {Bardyn},
  \citenamefont {Baranov}, \citenamefont {Kraus}, \citenamefont {Rico},
  \citenamefont {{\.{I}}mamo{\u{g}}lu}, \citenamefont {Zoller},\ and\
  \citenamefont {Diehl}}]{Bardyn_DissCher_NJP2013}%
  \BibitemOpen
  \bibfield  {author} {\bibinfo {author} {\bibfnamefont {C.-E.}\ \bibnamefont
  {Bardyn}}, \bibinfo {author} {\bibfnamefont {M.~A.}\ \bibnamefont {Baranov}},
  \bibinfo {author} {\bibfnamefont {C.~V.}\ \bibnamefont {Kraus}}, \bibinfo
  {author} {\bibfnamefont {E.}~\bibnamefont {Rico}}, \bibinfo {author}
  {\bibfnamefont {A.}~\bibnamefont {{\.{I}}mamo{\u{g}}lu}}, \bibinfo {author}
  {\bibfnamefont {P.}~\bibnamefont {Zoller}}, \ and\ \bibinfo {author}
  {\bibfnamefont {S.}~\bibnamefont {Diehl}},\ }\href@noop {} {\bibfield
  {journal} {\bibinfo  {journal} {New Journal of Physics}\ }\textbf {\bibinfo
  {volume} {15}},\ \bibinfo {pages} {085001} (\bibinfo {year}
  {2013})}\BibitemShut {NoStop}%
\bibitem [{\citenamefont {Budich}\ \emph {et~al.}(2015)\citenamefont {Budich},
  \citenamefont {Zoller},\ and\ \citenamefont {Diehl}}]{Budich_DissCher_PRA15}%
  \BibitemOpen
  \bibfield  {author} {\bibinfo {author} {\bibfnamefont {J.~C.}\ \bibnamefont
  {Budich}}, \bibinfo {author} {\bibfnamefont {P.}~\bibnamefont {Zoller}}, \
  and\ \bibinfo {author} {\bibfnamefont {S.}~\bibnamefont {Diehl}},\ }\href
  {\doibase 10.1103/PhysRevA.91.042117} {\bibfield  {journal} {\bibinfo
  {journal} {Phys. Rev. A}\ }\textbf {\bibinfo {volume} {91}},\ \bibinfo
  {pages} {042117} (\bibinfo {year} {2015})}\BibitemShut {NoStop}%
\bibitem [{\citenamefont {Lee}(2016)}]{TELeePRL16_Half_quantized}%
  \BibitemOpen
  \bibfield  {author} {\bibinfo {author} {\bibfnamefont {T.~E.}\ \bibnamefont
  {Lee}},\ }\href {\doibase 10.1103/PhysRevLett.116.133903} {\bibfield
  {journal} {\bibinfo  {journal} {Phys. Rev. Lett.}\ }\textbf {\bibinfo
  {volume} {116}},\ \bibinfo {pages} {133903} (\bibinfo {year}
  {2016})}\BibitemShut {NoStop}%
\bibitem [{\citenamefont {Gong}\ \emph {et~al.}(2017)\citenamefont {Gong},
  \citenamefont {Higashikawa},\ and\ \citenamefont {Ueda}}]{ZPGong_PRL17}%
  \BibitemOpen
  \bibfield  {author} {\bibinfo {author} {\bibfnamefont {Z.}~\bibnamefont
  {Gong}}, \bibinfo {author} {\bibfnamefont {S.}~\bibnamefont {Higashikawa}}, \
  and\ \bibinfo {author} {\bibfnamefont {M.}~\bibnamefont {Ueda}},\ }\href
  {\doibase 10.1103/PhysRevLett.118.200401} {\bibfield  {journal} {\bibinfo
  {journal} {Phys. Rev. Lett.}\ }\textbf {\bibinfo {volume} {118}},\ \bibinfo
  {pages} {200401} (\bibinfo {year} {2017})}\BibitemShut {NoStop}%
\bibitem [{\citenamefont {Lieu}(2018)}]{Lieu_nHSSH_PRB2018}%
  \BibitemOpen
  \bibfield  {author} {\bibinfo {author} {\bibfnamefont {S.}~\bibnamefont
  {Lieu}},\ }\href {\doibase 10.1103/PhysRevB.97.045106} {\bibfield  {journal}
  {\bibinfo  {journal} {Phys. Rev. B}\ }\textbf {\bibinfo {volume} {97}},\
  \bibinfo {pages} {045106} (\bibinfo {year} {2018})}\BibitemShut {NoStop}%
\bibitem [{\citenamefont {Gong}\ \emph {et~al.}(2018)\citenamefont {Gong},
  \citenamefont {Ashida}, \citenamefont {Kawabata}, \citenamefont {Takasan},
  \citenamefont {Higashikawa},\ and\ \citenamefont {Ueda}}]{Gong_class_PRX18}%
  \BibitemOpen
  \bibfield  {author} {\bibinfo {author} {\bibfnamefont {Z.}~\bibnamefont
  {Gong}}, \bibinfo {author} {\bibfnamefont {Y.}~\bibnamefont {Ashida}},
  \bibinfo {author} {\bibfnamefont {K.}~\bibnamefont {Kawabata}}, \bibinfo
  {author} {\bibfnamefont {K.}~\bibnamefont {Takasan}}, \bibinfo {author}
  {\bibfnamefont {S.}~\bibnamefont {Higashikawa}}, \ and\ \bibinfo {author}
  {\bibfnamefont {M.}~\bibnamefont {Ueda}},\ }\href {\doibase
  10.1103/PhysRevX.8.031079} {\bibfield  {journal} {\bibinfo  {journal} {Phys.
  Rev. X}\ }\textbf {\bibinfo {volume} {8}},\ \bibinfo {pages} {031079}
  (\bibinfo {year} {2018})}\BibitemShut {NoStop}%
\bibitem [{\citenamefont {Kawabata}\ \emph
  {et~al.}(2019{\natexlab{a}})\citenamefont {Kawabata}, \citenamefont
  {Shiozaki}, \citenamefont {Ueda},\ and\ \citenamefont
  {Sato}}]{Kawabata_gapped_PRX19}%
  \BibitemOpen
  \bibfield  {author} {\bibinfo {author} {\bibfnamefont {K.}~\bibnamefont
  {Kawabata}}, \bibinfo {author} {\bibfnamefont {K.}~\bibnamefont {Shiozaki}},
  \bibinfo {author} {\bibfnamefont {M.}~\bibnamefont {Ueda}}, \ and\ \bibinfo
  {author} {\bibfnamefont {M.}~\bibnamefont {Sato}},\ }\href {\doibase
  10.1103/PhysRevX.9.041015} {\bibfield  {journal} {\bibinfo  {journal} {Phys.
  Rev. X}\ }\textbf {\bibinfo {volume} {9}},\ \bibinfo {pages} {041015}
  (\bibinfo {year} {2019}{\natexlab{a}})}\BibitemShut {NoStop}%
\bibitem [{\citenamefont {Lieu}\ \emph {et~al.}(2020)\citenamefont {Lieu},
  \citenamefont {McGinley},\ and\ \citenamefont
  {Cooper}}]{Lieu_Liouclass_PRL20}%
  \BibitemOpen
  \bibfield  {author} {\bibinfo {author} {\bibfnamefont {S.}~\bibnamefont
  {Lieu}}, \bibinfo {author} {\bibfnamefont {M.}~\bibnamefont {McGinley}}, \
  and\ \bibinfo {author} {\bibfnamefont {N.~R.}\ \bibnamefont {Cooper}},\
  }\href {\doibase 10.1103/PhysRevLett.124.040401} {\bibfield  {journal}
  {\bibinfo  {journal} {Phys. Rev. Lett.}\ }\textbf {\bibinfo {volume} {124}},\
  \bibinfo {pages} {040401} (\bibinfo {year} {2020})}\BibitemShut {NoStop}%
\bibitem [{\citenamefont {Yang}\ \emph {et~al.}(2020)\citenamefont {Yang},
  \citenamefont {Chiu}, \citenamefont {Fang},\ and\ \citenamefont
  {Hu}}]{Yang_nHJPoly_PRL20}%
  \BibitemOpen
  \bibfield  {author} {\bibinfo {author} {\bibfnamefont {Z.}~\bibnamefont
  {Yang}}, \bibinfo {author} {\bibfnamefont {C.-K.}\ \bibnamefont {Chiu}},
  \bibinfo {author} {\bibfnamefont {C.}~\bibnamefont {Fang}}, \ and\ \bibinfo
  {author} {\bibfnamefont {J.}~\bibnamefont {Hu}},\ }\href {\doibase
  10.1103/PhysRevLett.124.186402} {\bibfield  {journal} {\bibinfo  {journal}
  {Phys. Rev. Lett.}\ }\textbf {\bibinfo {volume} {124}},\ \bibinfo {pages}
  {186402} (\bibinfo {year} {2020})}\BibitemShut {NoStop}%
\bibitem [{\citenamefont {Chang}\ \emph {et~al.}(2020)\citenamefont {Chang},
  \citenamefont {You}, \citenamefont {Wen},\ and\ \citenamefont
  {Ryu}}]{Chang_nHES_PRR20}%
  \BibitemOpen
  \bibfield  {author} {\bibinfo {author} {\bibfnamefont {P.-Y.}\ \bibnamefont
  {Chang}}, \bibinfo {author} {\bibfnamefont {J.-S.}\ \bibnamefont {You}},
  \bibinfo {author} {\bibfnamefont {X.}~\bibnamefont {Wen}}, \ and\ \bibinfo
  {author} {\bibfnamefont {S.}~\bibnamefont {Ryu}},\ }\href {\doibase
  10.1103/PhysRevResearch.2.033069} {\bibfield  {journal} {\bibinfo  {journal}
  {Phys. Rev. Research}\ }\textbf {\bibinfo {volume} {2}},\ \bibinfo {pages}
  {033069} (\bibinfo {year} {2020})}\BibitemShut {NoStop}%
\bibitem [{\citenamefont {Ghosh}\ and\ \citenamefont
  {Nag}(2022)}]{Kumer_nHTopo_PRB22}%
  \BibitemOpen
  \bibfield  {author} {\bibinfo {author} {\bibfnamefont {A.~K.}\ \bibnamefont
  {Ghosh}}\ and\ \bibinfo {author} {\bibfnamefont {T.}~\bibnamefont {Nag}},\
  }\href {\doibase 10.1103/PhysRevB.106.L140303} {\bibfield  {journal}
  {\bibinfo  {journal} {Phys. Rev. B}\ }\textbf {\bibinfo {volume} {106}},\
  \bibinfo {pages} {L140303} (\bibinfo {year} {2022})}\BibitemShut {NoStop}%
\bibitem [{\citenamefont {Arouca}\ \emph {et~al.}(2022)\citenamefont {Arouca},
  \citenamefont {Cayao},\ and\ \citenamefont
  {Black-Schaffer}}]{Arouca_nHTopo_arXiv22}%
  \BibitemOpen
  \bibfield  {author} {\bibinfo {author} {\bibfnamefont {R.}~\bibnamefont
  {Arouca}}, \bibinfo {author} {\bibfnamefont {J.}~\bibnamefont {Cayao}}, \
  and\ \bibinfo {author} {\bibfnamefont {A.~M.}\ \bibnamefont
  {Black-Schaffer}},\ }\href@noop {} {\bibfield  {journal} {\bibinfo  {journal}
  {arXiv preprint arXiv:2206.15324}\ } (\bibinfo {year} {2022})}\BibitemShut
  {NoStop}%
\bibitem [{\citenamefont {Cayao}\ and\ \citenamefont
  {Black-Schaffer}(2022)}]{Cayao_nHTopo_arXiv22}%
  \BibitemOpen
  \bibfield  {author} {\bibinfo {author} {\bibfnamefont {J.}~\bibnamefont
  {Cayao}}\ and\ \bibinfo {author} {\bibfnamefont {A.~M.}\ \bibnamefont
  {Black-Schaffer}},\ }\href@noop {} {\bibfield  {journal} {\bibinfo  {journal}
  {arXiv preprint arXiv:2208.05372}\ } (\bibinfo {year} {2022})}\BibitemShut
  {NoStop}%
\bibitem [{\citenamefont {Bergholtz}\ \emph {et~al.}(2021)\citenamefont
  {Bergholtz}, \citenamefont {Budich},\ and\ \citenamefont
  {Kunst}}]{Bergholtz_Review19}%
  \BibitemOpen
  \bibfield  {author} {\bibinfo {author} {\bibfnamefont {E.~J.}\ \bibnamefont
  {Bergholtz}}, \bibinfo {author} {\bibfnamefont {J.~C.}\ \bibnamefont
  {Budich}}, \ and\ \bibinfo {author} {\bibfnamefont {F.~K.}\ \bibnamefont
  {Kunst}},\ }\href@noop {} {\bibfield  {journal} {\bibinfo  {journal} {Rev.
  Mod. Phys.}\ }\textbf {\bibinfo {volume} {93}},\ \bibinfo {pages} {015005}
  (\bibinfo {year} {2021})}\BibitemShut {NoStop}%
\bibitem [{\citenamefont {Ashida}\ \emph {et~al.}(2020)\citenamefont {Ashida},
  \citenamefont {Gong},\ and\ \citenamefont
  {Ueda}}]{Ashida_nHReview_AdvPhys20}%
  \BibitemOpen
  \bibfield  {author} {\bibinfo {author} {\bibfnamefont {Y.}~\bibnamefont
  {Ashida}}, \bibinfo {author} {\bibfnamefont {Z.}~\bibnamefont {Gong}}, \ and\
  \bibinfo {author} {\bibfnamefont {M.}~\bibnamefont {Ueda}},\ }\href {\doibase
  10.1080/00018732.2021.1876991} {\bibfield  {journal} {\bibinfo  {journal}
  {Advances in Physics}\ }\textbf {\bibinfo {volume} {69}},\ \bibinfo {pages}
  {249} (\bibinfo {year} {2020})}\BibitemShut {NoStop}%
\bibitem [{\citenamefont {Yoshida}\ \emph
  {et~al.}(2020{\natexlab{a}})\citenamefont {Yoshida}, \citenamefont {Peters},
  \citenamefont {Kawakami},\ and\ \citenamefont
  {Hatsugai}}]{Yoshida_nHReview_PTEP20}%
  \BibitemOpen
  \bibfield  {author} {\bibinfo {author} {\bibfnamefont {T.}~\bibnamefont
  {Yoshida}}, \bibinfo {author} {\bibfnamefont {R.}~\bibnamefont {Peters}},
  \bibinfo {author} {\bibfnamefont {N.}~\bibnamefont {Kawakami}}, \ and\
  \bibinfo {author} {\bibfnamefont {Y.}~\bibnamefont {Hatsugai}},\ }\href@noop
  {} {\bibfield  {journal} {\bibinfo  {journal} {Progress of Theoretical and
  Experimental Physics}\ }\textbf {\bibinfo {volume} {2020}},\ \bibinfo {pages}
  {12A109} (\bibinfo {year} {2020}{\natexlab{a}})}\BibitemShut {NoStop}%
\bibitem [{\citenamefont {Martinez~Alvarez}\ \emph {et~al.}(2018)\citenamefont
  {Martinez~Alvarez}, \citenamefont {Barrios~Vargas},\ and\ \citenamefont
  {Foa~Torres}}]{Alvarez_nHSkin_PRB18}%
  \BibitemOpen
  \bibfield  {author} {\bibinfo {author} {\bibfnamefont {V.~M.}\ \bibnamefont
  {Martinez~Alvarez}}, \bibinfo {author} {\bibfnamefont {J.~E.}\ \bibnamefont
  {Barrios~Vargas}}, \ and\ \bibinfo {author} {\bibfnamefont {L.~E.~F.}\
  \bibnamefont {Foa~Torres}},\ }\href {\doibase 10.1103/PhysRevB.97.121401}
  {\bibfield  {journal} {\bibinfo  {journal} {Phys. Rev. B}\ }\textbf {\bibinfo
  {volume} {97}},\ \bibinfo {pages} {121401} (\bibinfo {year}
  {2018})}\BibitemShut {NoStop}%
\bibitem [{\citenamefont {Yao}\ and\ \citenamefont
  {Wang}(2018)}]{SYao_nHSkin-1D_PRL18}%
  \BibitemOpen
  \bibfield  {author} {\bibinfo {author} {\bibfnamefont {S.}~\bibnamefont
  {Yao}}\ and\ \bibinfo {author} {\bibfnamefont {Z.}~\bibnamefont {Wang}},\
  }\href {\doibase 10.1103/PhysRevLett.121.086803} {\bibfield  {journal}
  {\bibinfo  {journal} {Phys. Rev. Lett.}\ }\textbf {\bibinfo {volume} {121}},\
  \bibinfo {pages} {086803} (\bibinfo {year} {2018})}\BibitemShut {NoStop}%
\bibitem [{\citenamefont {Kunst}\ \emph {et~al.}(2018)\citenamefont {Kunst},
  \citenamefont {Edvardsson}, \citenamefont {Budich},\ and\ \citenamefont
  {Bergholtz}}]{KFlore_nHSkin_PRL18}%
  \BibitemOpen
  \bibfield  {author} {\bibinfo {author} {\bibfnamefont {F.~K.}\ \bibnamefont
  {Kunst}}, \bibinfo {author} {\bibfnamefont {E.}~\bibnamefont {Edvardsson}},
  \bibinfo {author} {\bibfnamefont {J.~C.}\ \bibnamefont {Budich}}, \ and\
  \bibinfo {author} {\bibfnamefont {E.~J.}\ \bibnamefont {Bergholtz}},\ }\href
  {\doibase 10.1103/PhysRevLett.121.026808} {\bibfield  {journal} {\bibinfo
  {journal} {Phys. Rev. Lett.}\ }\textbf {\bibinfo {volume} {121}},\ \bibinfo
  {pages} {026808} (\bibinfo {year} {2018})}\BibitemShut {NoStop}%
\bibitem [{\citenamefont {Lee}\ and\ \citenamefont
  {Thomale}(2019)}]{Lee_Skin19}%
  \BibitemOpen
  \bibfield  {author} {\bibinfo {author} {\bibfnamefont {C.~H.}\ \bibnamefont
  {Lee}}\ and\ \bibinfo {author} {\bibfnamefont {R.}~\bibnamefont {Thomale}},\
  }\href {\doibase 10.1103/PhysRevB.99.201103} {\bibfield  {journal} {\bibinfo
  {journal} {Phys. Rev. B}\ }\textbf {\bibinfo {volume} {99}},\ \bibinfo
  {pages} {201103} (\bibinfo {year} {2019})}\BibitemShut {NoStop}%
\bibitem [{\citenamefont {Lee}\ \emph {et~al.}(2019)\citenamefont {Lee},
  \citenamefont {Ahn}, \citenamefont {Zhou},\ and\ \citenamefont
  {Vishwanath}}]{Lee_nHSkin_PRL19}%
  \BibitemOpen
  \bibfield  {author} {\bibinfo {author} {\bibfnamefont {J.~Y.}\ \bibnamefont
  {Lee}}, \bibinfo {author} {\bibfnamefont {J.}~\bibnamefont {Ahn}}, \bibinfo
  {author} {\bibfnamefont {H.}~\bibnamefont {Zhou}}, \ and\ \bibinfo {author}
  {\bibfnamefont {A.}~\bibnamefont {Vishwanath}},\ }\href {\doibase
  10.1103/PhysRevLett.123.206404} {\bibfield  {journal} {\bibinfo  {journal}
  {Phys. Rev. Lett.}\ }\textbf {\bibinfo {volume} {123}},\ \bibinfo {pages}
  {206404} (\bibinfo {year} {2019})}\BibitemShut {NoStop}%
\bibitem [{\citenamefont {Borgnia}\ \emph {et~al.}(2020)\citenamefont
  {Borgnia}, \citenamefont {Kruchkov},\ and\ \citenamefont
  {Slager}}]{Borgnia_ptGapPRL2020}%
  \BibitemOpen
  \bibfield  {author} {\bibinfo {author} {\bibfnamefont {D.~S.}\ \bibnamefont
  {Borgnia}}, \bibinfo {author} {\bibfnamefont {A.~J.}\ \bibnamefont
  {Kruchkov}}, \ and\ \bibinfo {author} {\bibfnamefont {R.-J.}\ \bibnamefont
  {Slager}},\ }\href {\doibase 10.1103/PhysRevLett.124.056802} {\bibfield
  {journal} {\bibinfo  {journal} {Phys. Rev. Lett.}\ }\textbf {\bibinfo
  {volume} {124}},\ \bibinfo {pages} {056802} (\bibinfo {year}
  {2020})}\BibitemShut {NoStop}%
\bibitem [{\citenamefont {Zhang}\ \emph
  {et~al.}(2020{\natexlab{a}})\citenamefont {Zhang}, \citenamefont {Yang},\
  and\ \citenamefont {Fang}}]{Zhang_BECskin19}%
  \BibitemOpen
  \bibfield  {author} {\bibinfo {author} {\bibfnamefont {K.}~\bibnamefont
  {Zhang}}, \bibinfo {author} {\bibfnamefont {Z.}~\bibnamefont {Yang}}, \ and\
  \bibinfo {author} {\bibfnamefont {C.}~\bibnamefont {Fang}},\ }\href@noop {}
  {\bibfield  {journal} {\bibinfo  {journal} {Phys. Rev. Lett.}\ }\textbf
  {\bibinfo {volume} {125}},\ \bibinfo {pages} {126402} (\bibinfo {year}
  {2020}{\natexlab{a}})}\BibitemShut {NoStop}%
\bibitem [{\citenamefont {Okuma}\ \emph {et~al.}(2020)\citenamefont {Okuma},
  \citenamefont {Kawabata}, \citenamefont {Shiozaki},\ and\ \citenamefont
  {Sato}}]{Okuma_BECskin19}%
  \BibitemOpen
  \bibfield  {author} {\bibinfo {author} {\bibfnamefont {N.}~\bibnamefont
  {Okuma}}, \bibinfo {author} {\bibfnamefont {K.}~\bibnamefont {Kawabata}},
  \bibinfo {author} {\bibfnamefont {K.}~\bibnamefont {Shiozaki}}, \ and\
  \bibinfo {author} {\bibfnamefont {M.}~\bibnamefont {Sato}},\ }\href {\doibase
  10.1103/PhysRevLett.124.086801} {\bibfield  {journal} {\bibinfo  {journal}
  {Phys. Rev. Lett.}\ }\textbf {\bibinfo {volume} {124}},\ \bibinfo {pages}
  {086801} (\bibinfo {year} {2020})}\BibitemShut {NoStop}%
\bibitem [{\citenamefont {Hofmann}\ \emph {et~al.}(2020)\citenamefont
  {Hofmann}, \citenamefont {Helbig}, \citenamefont {Schindler}, \citenamefont
  {Salgo}, \citenamefont {Brzezi\ifmmode~\acute{n}\else \'{n}\fi{}ska},
  \citenamefont {Greiter}, \citenamefont {Kiessling}, \citenamefont {Wolf},
  \citenamefont {Vollhardt}, \citenamefont {Kaba\ifmmode~\check{s}\else
  \v{s}\fi{}i}, \citenamefont {Lee}, \citenamefont {Bilu\ifmmode \check{s}\else
  \v{s}\fi{}i\ifmmode~\acute{c}\else \'{c}\fi{}}, \citenamefont {Thomale},\
  and\ \citenamefont {Neupert}}]{Hofmann_ExpRecipSkin_19}%
  \BibitemOpen
  \bibfield  {author} {\bibinfo {author} {\bibfnamefont {T.}~\bibnamefont
  {Hofmann}}, \bibinfo {author} {\bibfnamefont {T.}~\bibnamefont {Helbig}},
  \bibinfo {author} {\bibfnamefont {F.}~\bibnamefont {Schindler}}, \bibinfo
  {author} {\bibfnamefont {N.}~\bibnamefont {Salgo}}, \bibinfo {author}
  {\bibfnamefont {M.}~\bibnamefont {Brzezi\ifmmode~\acute{n}\else
  \'{n}\fi{}ska}}, \bibinfo {author} {\bibfnamefont {M.}~\bibnamefont
  {Greiter}}, \bibinfo {author} {\bibfnamefont {T.}~\bibnamefont {Kiessling}},
  \bibinfo {author} {\bibfnamefont {D.}~\bibnamefont {Wolf}}, \bibinfo {author}
  {\bibfnamefont {A.}~\bibnamefont {Vollhardt}}, \bibinfo {author}
  {\bibfnamefont {A.}~\bibnamefont {Kaba\ifmmode~\check{s}\else \v{s}\fi{}i}},
  \bibinfo {author} {\bibfnamefont {C.~H.}\ \bibnamefont {Lee}}, \bibinfo
  {author} {\bibfnamefont {A.}~\bibnamefont {Bilu\ifmmode \check{s}\else
  \v{s}\fi{}i\ifmmode~\acute{c}\else \'{c}\fi{}}}, \bibinfo {author}
  {\bibfnamefont {R.}~\bibnamefont {Thomale}}, \ and\ \bibinfo {author}
  {\bibfnamefont {T.}~\bibnamefont {Neupert}},\ }\href@noop {} {\bibfield
  {journal} {\bibinfo  {journal} {Phys. Rev. Research}\ }\textbf {\bibinfo
  {volume} {2}},\ \bibinfo {pages} {023265} (\bibinfo {year}
  {2020})}\BibitemShut {NoStop}%
\bibitem [{\citenamefont {Yoshida}\ \emph
  {et~al.}(2020{\natexlab{b}})\citenamefont {Yoshida}, \citenamefont
  {Mizoguchi},\ and\ \citenamefont {Hatsugai}}]{Yoshida_MSkinPRR20}%
  \BibitemOpen
  \bibfield  {author} {\bibinfo {author} {\bibfnamefont {T.}~\bibnamefont
  {Yoshida}}, \bibinfo {author} {\bibfnamefont {T.}~\bibnamefont {Mizoguchi}},
  \ and\ \bibinfo {author} {\bibfnamefont {Y.}~\bibnamefont {Hatsugai}},\
  }\href {\doibase 10.1103/PhysRevResearch.2.022062} {\bibfield  {journal}
  {\bibinfo  {journal} {Phys. Rev. Research}\ }\textbf {\bibinfo {volume}
  {2}},\ \bibinfo {pages} {022062} (\bibinfo {year}
  {2020}{\natexlab{b}})}\BibitemShut {NoStop}%
\bibitem [{\citenamefont {Bessho}\ and\ \citenamefont
  {Sato}(2021)}]{Bessho_nHNN_PRL21}%
  \BibitemOpen
  \bibfield  {author} {\bibinfo {author} {\bibfnamefont {T.}~\bibnamefont
  {Bessho}}\ and\ \bibinfo {author} {\bibfnamefont {M.}~\bibnamefont {Sato}},\
  }\href {\doibase 10.1103/PhysRevLett.127.196404} {\bibfield  {journal}
  {\bibinfo  {journal} {Phys. Rev. Lett.}\ }\textbf {\bibinfo {volume} {127}},\
  \bibinfo {pages} {196404} (\bibinfo {year} {2021})}\BibitemShut {NoStop}%
\bibitem [{\citenamefont {Kawabata}\ \emph {et~al.}(2021)\citenamefont
  {Kawabata}, \citenamefont {Shiozaki},\ and\ \citenamefont
  {Ryu}}]{Kawabata_TQFTSkin_PRL21}%
  \BibitemOpen
  \bibfield  {author} {\bibinfo {author} {\bibfnamefont {K.}~\bibnamefont
  {Kawabata}}, \bibinfo {author} {\bibfnamefont {K.}~\bibnamefont {Shiozaki}},
  \ and\ \bibinfo {author} {\bibfnamefont {S.}~\bibnamefont {Ryu}},\ }\href
  {\doibase 10.1103/PhysRevLett.126.216405} {\bibfield  {journal} {\bibinfo
  {journal} {Phys. Rev. Lett.}\ }\textbf {\bibinfo {volume} {126}},\ \bibinfo
  {pages} {216405} (\bibinfo {year} {2021})}\BibitemShut {NoStop}%
\bibitem [{\citenamefont {Okuma}\ and\ \citenamefont
  {Sato}(2023)}]{Okuma_nHSkinRev_arXiv22}%
  \BibitemOpen
  \bibfield  {author} {\bibinfo {author} {\bibfnamefont {N.}~\bibnamefont
  {Okuma}}\ and\ \bibinfo {author} {\bibfnamefont {M.}~\bibnamefont {Sato}},\
  }\href@noop {} {\bibfield  {journal} {\bibinfo  {journal} {Annual Review of
  Condensed Matter Physics}\ }\textbf {\bibinfo {volume} {14}},\ \bibinfo
  {pages} {null} (\bibinfo {year} {2023})}\BibitemShut {NoStop}%
\bibitem [{\citenamefont {Shen}\ \emph {et~al.}(2018)\citenamefont {Shen},
  \citenamefont {Zhen},\ and\ \citenamefont {Fu}}]{HShen2017_non-Hermi}%
  \BibitemOpen
  \bibfield  {author} {\bibinfo {author} {\bibfnamefont {H.}~\bibnamefont
  {Shen}}, \bibinfo {author} {\bibfnamefont {B.}~\bibnamefont {Zhen}}, \ and\
  \bibinfo {author} {\bibfnamefont {L.}~\bibnamefont {Fu}},\ }\href {\doibase
  10.1103/PhysRevLett.120.146402} {\bibfield  {journal} {\bibinfo  {journal}
  {Phys. Rev. Lett.}\ }\textbf {\bibinfo {volume} {120}},\ \bibinfo {pages}
  {146402} (\bibinfo {year} {2018})}\BibitemShut {NoStop}%
\bibitem [{\citenamefont {Xu}\ \emph {et~al.}(2017)\citenamefont {Xu},
  \citenamefont {Wang},\ and\ \citenamefont
  {Duan}}]{YXuPRL17_exceptional_ring}%
  \BibitemOpen
  \bibfield  {author} {\bibinfo {author} {\bibfnamefont {Y.}~\bibnamefont
  {Xu}}, \bibinfo {author} {\bibfnamefont {S.-T.}\ \bibnamefont {Wang}}, \ and\
  \bibinfo {author} {\bibfnamefont {L.-M.}\ \bibnamefont {Duan}},\ }\href
  {\doibase 10.1103/PhysRevLett.118.045701} {\bibfield  {journal} {\bibinfo
  {journal} {Phys. Rev. Lett.}\ }\textbf {\bibinfo {volume} {118}},\ \bibinfo
  {pages} {045701} (\bibinfo {year} {2017})}\BibitemShut {NoStop}%
\bibitem [{\citenamefont {Hassan}\ \emph {et~al.}(2017)\citenamefont {Hassan},
  \citenamefont {Zhen}, \citenamefont {Solja\ifmmode \check{c}\else
  \v{c}\fi{}i\ifmmode~\acute{c}\else \'{c}\fi{}}, \citenamefont {Khajavikhan},\
  and\ \citenamefont {Christodoulides}}]{Hassan_EP_PRL17}%
  \BibitemOpen
  \bibfield  {author} {\bibinfo {author} {\bibfnamefont {A.~U.}\ \bibnamefont
  {Hassan}}, \bibinfo {author} {\bibfnamefont {B.}~\bibnamefont {Zhen}},
  \bibinfo {author} {\bibfnamefont {M.}~\bibnamefont {Solja\ifmmode
  \check{c}\else \v{c}\fi{}i\ifmmode~\acute{c}\else \'{c}\fi{}}}, \bibinfo
  {author} {\bibfnamefont {M.}~\bibnamefont {Khajavikhan}}, \ and\ \bibinfo
  {author} {\bibfnamefont {D.~N.}\ \bibnamefont {Christodoulides}},\ }\href
  {\doibase 10.1103/PhysRevLett.118.093002} {\bibfield  {journal} {\bibinfo
  {journal} {Phys. Rev. Lett.}\ }\textbf {\bibinfo {volume} {118}},\ \bibinfo
  {pages} {093002} (\bibinfo {year} {2017})}\BibitemShut {NoStop}%
\bibitem [{\citenamefont {Zhou}\ \emph {et~al.}(2018)\citenamefont {Zhou},
  \citenamefont {Peng}, \citenamefont {Yoon}, \citenamefont {Hsu},
  \citenamefont {Nelson}, \citenamefont {Fu}, \citenamefont {Joannopoulos},
  \citenamefont {Solja{\v c}i{\'c}},\ and\ \citenamefont
  {Zhen}}]{Zhou_ObEP_Science18}%
  \BibitemOpen
  \bibfield  {author} {\bibinfo {author} {\bibfnamefont {H.}~\bibnamefont
  {Zhou}}, \bibinfo {author} {\bibfnamefont {C.}~\bibnamefont {Peng}}, \bibinfo
  {author} {\bibfnamefont {Y.}~\bibnamefont {Yoon}}, \bibinfo {author}
  {\bibfnamefont {C.~W.}\ \bibnamefont {Hsu}}, \bibinfo {author} {\bibfnamefont
  {K.~A.}\ \bibnamefont {Nelson}}, \bibinfo {author} {\bibfnamefont
  {L.}~\bibnamefont {Fu}}, \bibinfo {author} {\bibfnamefont {J.~D.}\
  \bibnamefont {Joannopoulos}}, \bibinfo {author} {\bibfnamefont
  {M.}~\bibnamefont {Solja{\v c}i{\'c}}}, \ and\ \bibinfo {author}
  {\bibfnamefont {B.}~\bibnamefont {Zhen}},\ }\href {\doibase
  10.1126/science.aap9859} {\bibfield  {journal} {\bibinfo  {journal}
  {Science}\ }\textbf {\bibinfo {volume} {359}},\ \bibinfo {pages} {1009}
  (\bibinfo {year} {2018})}\BibitemShut {NoStop}%
\bibitem [{\citenamefont {Kozii}\ and\ \citenamefont
  {Fu}(2017)}]{VKozii_nH_arXiv17}%
  \BibitemOpen
  \bibfield  {author} {\bibinfo {author} {\bibfnamefont {V.}~\bibnamefont
  {Kozii}}\ and\ \bibinfo {author} {\bibfnamefont {L.}~\bibnamefont {Fu}},\
  }\href@noop {} {\bibfield  {journal} {\bibinfo  {journal} {arXiv preprint
  arXiv:1708.05841}\ } (\bibinfo {year} {2017})}\BibitemShut {NoStop}%
\bibitem [{\citenamefont {Yoshida}\ \emph {et~al.}(2018)\citenamefont
  {Yoshida}, \citenamefont {Peters},\ and\ \citenamefont
  {Kawakami}}]{Yoshida_EP_DMFT_PRB18}%
  \BibitemOpen
  \bibfield  {author} {\bibinfo {author} {\bibfnamefont {T.}~\bibnamefont
  {Yoshida}}, \bibinfo {author} {\bibfnamefont {R.}~\bibnamefont {Peters}}, \
  and\ \bibinfo {author} {\bibfnamefont {N.}~\bibnamefont {Kawakami}},\ }\href
  {\doibase 10.1103/PhysRevB.98.035141} {\bibfield  {journal} {\bibinfo
  {journal} {Phys. Rev. B}\ }\textbf {\bibinfo {volume} {98}},\ \bibinfo
  {pages} {035141} (\bibinfo {year} {2018})}\BibitemShut {NoStop}%
\bibitem [{\citenamefont {Wojcik}\ \emph {et~al.}(2020)\citenamefont {Wojcik},
  \citenamefont {Sun}, \citenamefont {Bzdu\ifmmode~\check{s}\else
  \v{s}\fi{}ek},\ and\ \citenamefont {Fan}}]{Wojcik_DiscEP_PRB20}%
  \BibitemOpen
  \bibfield  {author} {\bibinfo {author} {\bibfnamefont {C.~C.}\ \bibnamefont
  {Wojcik}}, \bibinfo {author} {\bibfnamefont {X.-Q.}\ \bibnamefont {Sun}},
  \bibinfo {author} {\bibfnamefont {T.~c.~v.}\ \bibnamefont
  {Bzdu\ifmmode~\check{s}\else \v{s}\fi{}ek}}, \ and\ \bibinfo {author}
  {\bibfnamefont {S.}~\bibnamefont {Fan}},\ }\href {\doibase
  10.1103/PhysRevB.101.205417} {\bibfield  {journal} {\bibinfo  {journal}
  {Phys. Rev. B}\ }\textbf {\bibinfo {volume} {101}},\ \bibinfo {pages}
  {205417} (\bibinfo {year} {2020})}\BibitemShut {NoStop}%
\bibitem [{\citenamefont {Yang}\ \emph
  {et~al.}(2021{\natexlab{a}})\citenamefont {Yang}, \citenamefont {Schnyder},
  \citenamefont {Hu},\ and\ \citenamefont {Chiu}}]{Yang_DiscEP_PRL21}%
  \BibitemOpen
  \bibfield  {author} {\bibinfo {author} {\bibfnamefont {Z.}~\bibnamefont
  {Yang}}, \bibinfo {author} {\bibfnamefont {A.~P.}\ \bibnamefont {Schnyder}},
  \bibinfo {author} {\bibfnamefont {J.}~\bibnamefont {Hu}}, \ and\ \bibinfo
  {author} {\bibfnamefont {C.-K.}\ \bibnamefont {Chiu}},\ }\href {\doibase
  10.1103/PhysRevLett.126.086401} {\bibfield  {journal} {\bibinfo  {journal}
  {Phys. Rev. Lett.}\ }\textbf {\bibinfo {volume} {126}},\ \bibinfo {pages}
  {086401} (\bibinfo {year} {2021}{\natexlab{a}})}\BibitemShut {NoStop}%
\bibitem [{\citenamefont {Budich}\ \emph {et~al.}(2019)\citenamefont {Budich},
  \citenamefont {Carlstr\"om}, \citenamefont {Kunst},\ and\ \citenamefont
  {Bergholtz}}]{Budich_SPERs_PRB19}%
  \BibitemOpen
  \bibfield  {author} {\bibinfo {author} {\bibfnamefont {J.~C.}\ \bibnamefont
  {Budich}}, \bibinfo {author} {\bibfnamefont {J.}~\bibnamefont {Carlstr\"om}},
  \bibinfo {author} {\bibfnamefont {F.~K.}\ \bibnamefont {Kunst}}, \ and\
  \bibinfo {author} {\bibfnamefont {E.~J.}\ \bibnamefont {Bergholtz}},\ }\href
  {\doibase 10.1103/PhysRevB.99.041406} {\bibfield  {journal} {\bibinfo
  {journal} {Phys. Rev. B}\ }\textbf {\bibinfo {volume} {99}},\ \bibinfo
  {pages} {041406} (\bibinfo {year} {2019})}\BibitemShut {NoStop}%
\bibitem [{\citenamefont {Okugawa}\ and\ \citenamefont
  {Yokoyama}(2019)}]{Okugawa_SPERs_PRB19}%
  \BibitemOpen
  \bibfield  {author} {\bibinfo {author} {\bibfnamefont {R.}~\bibnamefont
  {Okugawa}}\ and\ \bibinfo {author} {\bibfnamefont {T.}~\bibnamefont
  {Yokoyama}},\ }\href {\doibase 10.1103/PhysRevB.99.041202} {\bibfield
  {journal} {\bibinfo  {journal} {Phys. Rev. B}\ }\textbf {\bibinfo {volume}
  {99}},\ \bibinfo {pages} {041202} (\bibinfo {year} {2019})}\BibitemShut
  {NoStop}%
\bibitem [{\citenamefont {Yoshida}\ \emph
  {et~al.}(2019{\natexlab{a}})\citenamefont {Yoshida}, \citenamefont {Peters},
  \citenamefont {Kawakami},\ and\ \citenamefont
  {Hatsugai}}]{Yoshida_SPERs_PRB19}%
  \BibitemOpen
  \bibfield  {author} {\bibinfo {author} {\bibfnamefont {T.}~\bibnamefont
  {Yoshida}}, \bibinfo {author} {\bibfnamefont {R.}~\bibnamefont {Peters}},
  \bibinfo {author} {\bibfnamefont {N.}~\bibnamefont {Kawakami}}, \ and\
  \bibinfo {author} {\bibfnamefont {Y.}~\bibnamefont {Hatsugai}},\ }\href
  {\doibase 10.1103/PhysRevB.99.121101} {\bibfield  {journal} {\bibinfo
  {journal} {Phys. Rev. B}\ }\textbf {\bibinfo {volume} {99}},\ \bibinfo
  {pages} {121101} (\bibinfo {year} {2019}{\natexlab{a}})}\BibitemShut
  {NoStop}%
\bibitem [{\citenamefont {Zhou}\ \emph {et~al.}(2019)\citenamefont {Zhou},
  \citenamefont {Lee}, \citenamefont {Liu},\ and\ \citenamefont
  {Zhen}}]{Zhou_SPERs_Optica19}%
  \BibitemOpen
  \bibfield  {author} {\bibinfo {author} {\bibfnamefont {H.}~\bibnamefont
  {Zhou}}, \bibinfo {author} {\bibfnamefont {J.~Y.}\ \bibnamefont {Lee}},
  \bibinfo {author} {\bibfnamefont {S.}~\bibnamefont {Liu}}, \ and\ \bibinfo
  {author} {\bibfnamefont {B.}~\bibnamefont {Zhen}},\ }\href@noop {} {\bibfield
   {journal} {\bibinfo  {journal} {Optica}\ }\textbf {\bibinfo {volume} {6}},\
  \bibinfo {pages} {190} (\bibinfo {year} {2019})}\BibitemShut {NoStop}%
\bibitem [{\citenamefont {Kawabata}\ \emph
  {et~al.}(2019{\natexlab{b}})\citenamefont {Kawabata}, \citenamefont
  {Bessho},\ and\ \citenamefont {Sato}}]{Kawabata_gapless_PRL19}%
  \BibitemOpen
  \bibfield  {author} {\bibinfo {author} {\bibfnamefont {K.}~\bibnamefont
  {Kawabata}}, \bibinfo {author} {\bibfnamefont {T.}~\bibnamefont {Bessho}}, \
  and\ \bibinfo {author} {\bibfnamefont {M.}~\bibnamefont {Sato}},\ }\href
  {\doibase 10.1103/PhysRevLett.123.066405} {\bibfield  {journal} {\bibinfo
  {journal} {Phys. Rev. Lett.}\ }\textbf {\bibinfo {volume} {123}},\ \bibinfo
  {pages} {066405} (\bibinfo {year} {2019}{\natexlab{b}})}\BibitemShut
  {NoStop}%
\bibitem [{\citenamefont {Delplace}\ \emph {et~al.}(2021)\citenamefont
  {Delplace}, \citenamefont {Yoshida},\ and\ \citenamefont
  {Hatsugai}}]{Delplace_Resul_PRL21}%
  \BibitemOpen
  \bibfield  {author} {\bibinfo {author} {\bibfnamefont {P.}~\bibnamefont
  {Delplace}}, \bibinfo {author} {\bibfnamefont {T.}~\bibnamefont {Yoshida}}, \
  and\ \bibinfo {author} {\bibfnamefont {Y.}~\bibnamefont {Hatsugai}},\ }\href
  {\doibase 10.1103/PhysRevLett.127.186602} {\bibfield  {journal} {\bibinfo
  {journal} {Phys. Rev. Lett.}\ }\textbf {\bibinfo {volume} {127}},\ \bibinfo
  {pages} {186602} (\bibinfo {year} {2021})}\BibitemShut {NoStop}%
\bibitem [{\citenamefont {Mandal}\ and\ \citenamefont
  {Bergholtz}(2021)}]{Mandal_EP3_PRL21}%
  \BibitemOpen
  \bibfield  {author} {\bibinfo {author} {\bibfnamefont {I.}~\bibnamefont
  {Mandal}}\ and\ \bibinfo {author} {\bibfnamefont {E.~J.}\ \bibnamefont
  {Bergholtz}},\ }\href {\doibase 10.1103/PhysRevLett.127.186601} {\bibfield
  {journal} {\bibinfo  {journal} {Phys. Rev. Lett.}\ }\textbf {\bibinfo
  {volume} {127}},\ \bibinfo {pages} {186601} (\bibinfo {year}
  {2021})}\BibitemShut {NoStop}%
\bibitem [{\citenamefont {San-Jose}\ \emph {et~al.}(2016)\citenamefont
  {San-Jose}, \citenamefont {Cayao}, \citenamefont {Prada},\ and\ \citenamefont
  {Aguado}}]{Jose_DissSuperEP_SciRep16}%
  \BibitemOpen
  \bibfield  {author} {\bibinfo {author} {\bibfnamefont {P.}~\bibnamefont
  {San-Jose}}, \bibinfo {author} {\bibfnamefont {J.}~\bibnamefont {Cayao}},
  \bibinfo {author} {\bibfnamefont {E.}~\bibnamefont {Prada}}, \ and\ \bibinfo
  {author} {\bibfnamefont {R.}~\bibnamefont {Aguado}},\ }\href {\doibase
  10.1038/srep21427} {\bibfield  {journal} {\bibinfo  {journal} {Scientific
  Reports}\ }\textbf {\bibinfo {volume} {6}},\ \bibinfo {pages} {21427}
  (\bibinfo {year} {2016})}\BibitemShut {NoStop}%
\bibitem [{\citenamefont {Ozawa}\ \emph {et~al.}(2019)\citenamefont {Ozawa},
  \citenamefont {Price}, \citenamefont {Amo}, \citenamefont {Goldman},
  \citenamefont {Hafezi}, \citenamefont {Lu}, \citenamefont {Rechtsman},
  \citenamefont {Schuster}, \citenamefont {Simon}, \citenamefont {Zilberberg},\
  and\ \citenamefont {Carusotto}}]{Ozawa_TopoPhoto_RMP19}%
  \BibitemOpen
  \bibfield  {author} {\bibinfo {author} {\bibfnamefont {T.}~\bibnamefont
  {Ozawa}}, \bibinfo {author} {\bibfnamefont {H.~M.}\ \bibnamefont {Price}},
  \bibinfo {author} {\bibfnamefont {A.}~\bibnamefont {Amo}}, \bibinfo {author}
  {\bibfnamefont {N.}~\bibnamefont {Goldman}}, \bibinfo {author} {\bibfnamefont
  {M.}~\bibnamefont {Hafezi}}, \bibinfo {author} {\bibfnamefont
  {L.}~\bibnamefont {Lu}}, \bibinfo {author} {\bibfnamefont {M.~C.}\
  \bibnamefont {Rechtsman}}, \bibinfo {author} {\bibfnamefont {D.}~\bibnamefont
  {Schuster}}, \bibinfo {author} {\bibfnamefont {J.}~\bibnamefont {Simon}},
  \bibinfo {author} {\bibfnamefont {O.}~\bibnamefont {Zilberberg}}, \ and\
  \bibinfo {author} {\bibfnamefont {I.}~\bibnamefont {Carusotto}},\ }\href
  {\doibase 10.1103/RevModPhys.91.015006} {\bibfield  {journal} {\bibinfo
  {journal} {Rev. Mod. Phys.}\ }\textbf {\bibinfo {volume} {91}},\ \bibinfo
  {pages} {015006} (\bibinfo {year} {2019})}\BibitemShut {NoStop}%
\bibitem [{\citenamefont {Yoshida}\ and\ \citenamefont
  {Hatsugai}(2019)}]{Yoshida_SPERs_mech19}%
  \BibitemOpen
  \bibfield  {author} {\bibinfo {author} {\bibfnamefont {T.}~\bibnamefont
  {Yoshida}}\ and\ \bibinfo {author} {\bibfnamefont {Y.}~\bibnamefont
  {Hatsugai}},\ }\href {\doibase 10.1103/PhysRevB.100.054109} {\bibfield
  {journal} {\bibinfo  {journal} {Phys. Rev. B}\ }\textbf {\bibinfo {volume}
  {100}},\ \bibinfo {pages} {054109} (\bibinfo {year} {2019})}\BibitemShut
  {NoStop}%
\bibitem [{\citenamefont {Li}\ \emph {et~al.}(2019)\citenamefont {Li},
  \citenamefont {Peng}, \citenamefont {Han}, \citenamefont {Miri},
  \citenamefont {Li}, \citenamefont {Xiao}, \citenamefont {Zhu}, \citenamefont
  {Zhao}, \citenamefont {Al{\`u}}, \citenamefont {Fan},\ and\ \citenamefont
  {Qiu}}]{LiSciencePT19}%
  \BibitemOpen
  \bibfield  {author} {\bibinfo {author} {\bibfnamefont {Y.}~\bibnamefont
  {Li}}, \bibinfo {author} {\bibfnamefont {Y.-G.}\ \bibnamefont {Peng}},
  \bibinfo {author} {\bibfnamefont {L.}~\bibnamefont {Han}}, \bibinfo {author}
  {\bibfnamefont {M.-A.}\ \bibnamefont {Miri}}, \bibinfo {author}
  {\bibfnamefont {W.}~\bibnamefont {Li}}, \bibinfo {author} {\bibfnamefont
  {M.}~\bibnamefont {Xiao}}, \bibinfo {author} {\bibfnamefont {X.-F.}\
  \bibnamefont {Zhu}}, \bibinfo {author} {\bibfnamefont {J.}~\bibnamefont
  {Zhao}}, \bibinfo {author} {\bibfnamefont {A.}~\bibnamefont {Al{\`u}}},
  \bibinfo {author} {\bibfnamefont {S.}~\bibnamefont {Fan}}, \ and\ \bibinfo
  {author} {\bibfnamefont {C.-W.}\ \bibnamefont {Qiu}},\ }\href {\doibase
  10.1126/science.aaw6259} {\bibfield  {journal} {\bibinfo  {journal}
  {Science}\ }\textbf {\bibinfo {volume} {364}},\ \bibinfo {pages} {170}
  (\bibinfo {year} {2019})}\BibitemShut {NoStop}%
\bibitem [{\citenamefont {Partanen}\ \emph {et~al.}(2019)\citenamefont
  {Partanen}, \citenamefont {Goetz}, \citenamefont {Tan}, \citenamefont
  {Kohvakka}, \citenamefont {Sevriuk}, \citenamefont {Lake}, \citenamefont
  {Kokkoniemi}, \citenamefont {Ikonen}, \citenamefont {Hazra}, \citenamefont
  {M\"akinen}, \citenamefont {Hyypp\"a}, \citenamefont {Gr\"onberg},
  \citenamefont {Vesterinen}, \citenamefont {Silveri},\ and\ \citenamefont
  {M\"ott\"onen}}]{Partanen_EPQbit_PRB2019}%
  \BibitemOpen
  \bibfield  {author} {\bibinfo {author} {\bibfnamefont {M.}~\bibnamefont
  {Partanen}}, \bibinfo {author} {\bibfnamefont {J.}~\bibnamefont {Goetz}},
  \bibinfo {author} {\bibfnamefont {K.~Y.}\ \bibnamefont {Tan}}, \bibinfo
  {author} {\bibfnamefont {K.}~\bibnamefont {Kohvakka}}, \bibinfo {author}
  {\bibfnamefont {V.}~\bibnamefont {Sevriuk}}, \bibinfo {author} {\bibfnamefont
  {R.~E.}\ \bibnamefont {Lake}}, \bibinfo {author} {\bibfnamefont
  {R.}~\bibnamefont {Kokkoniemi}}, \bibinfo {author} {\bibfnamefont
  {J.}~\bibnamefont {Ikonen}}, \bibinfo {author} {\bibfnamefont
  {D.}~\bibnamefont {Hazra}}, \bibinfo {author} {\bibfnamefont
  {A.}~\bibnamefont {M\"akinen}}, \bibinfo {author} {\bibfnamefont
  {E.}~\bibnamefont {Hyypp\"a}}, \bibinfo {author} {\bibfnamefont
  {L.}~\bibnamefont {Gr\"onberg}}, \bibinfo {author} {\bibfnamefont
  {V.}~\bibnamefont {Vesterinen}}, \bibinfo {author} {\bibfnamefont
  {M.}~\bibnamefont {Silveri}}, \ and\ \bibinfo {author} {\bibfnamefont
  {M.}~\bibnamefont {M\"ott\"onen}},\ }\href {\doibase
  10.1103/PhysRevB.100.134505} {\bibfield  {journal} {\bibinfo  {journal}
  {Phys. Rev. B}\ }\textbf {\bibinfo {volume} {100}},\ \bibinfo {pages}
  {134505} (\bibinfo {year} {2019})}\BibitemShut {NoStop}%
\bibitem [{\citenamefont {Naghiloo}\ \emph {et~al.}(2019)\citenamefont
  {Naghiloo}, \citenamefont {Abbasi}, \citenamefont {Joglekar},\ and\
  \citenamefont {Murch}}]{Naghiloo_EPQbit_NatPhys2019}%
  \BibitemOpen
  \bibfield  {author} {\bibinfo {author} {\bibfnamefont {M.}~\bibnamefont
  {Naghiloo}}, \bibinfo {author} {\bibfnamefont {M.}~\bibnamefont {Abbasi}},
  \bibinfo {author} {\bibfnamefont {Y.~N.}\ \bibnamefont {Joglekar}}, \ and\
  \bibinfo {author} {\bibfnamefont {K.~W.}\ \bibnamefont {Murch}},\ }\href
  {\doibase 10.1038/s41567-019-0652-z} {\bibfield  {journal} {\bibinfo
  {journal} {Nature Physics}\ }\textbf {\bibinfo {volume} {15}},\ \bibinfo
  {pages} {1232} (\bibinfo {year} {2019})}\BibitemShut {NoStop}%
\bibitem [{\citenamefont {Yoshida}\ \emph {et~al.}(2022)\citenamefont
  {Yoshida}, \citenamefont {Mizoguchi},\ and\ \citenamefont
  {Hatsugai}}]{Yoshida_nHgame_SciRep2022}%
  \BibitemOpen
  \bibfield  {author} {\bibinfo {author} {\bibfnamefont {T.}~\bibnamefont
  {Yoshida}}, \bibinfo {author} {\bibfnamefont {T.}~\bibnamefont {Mizoguchi}},
  \ and\ \bibinfo {author} {\bibfnamefont {Y.}~\bibnamefont {Hatsugai}},\
  }\href {\doibase 10.1038/s41598-021-04178-8} {\bibfield  {journal} {\bibinfo
  {journal} {Scientific Reports}\ }\textbf {\bibinfo {volume} {12}},\ \bibinfo
  {pages} {560} (\bibinfo {year} {2022})}\BibitemShut {NoStop}%
\bibitem [{\citenamefont {Luitz}\ and\ \citenamefont
  {Piazza}(2019)}]{Luitz_EPcorrPRR2019}%
  \BibitemOpen
  \bibfield  {author} {\bibinfo {author} {\bibfnamefont {D.~J.}\ \bibnamefont
  {Luitz}}\ and\ \bibinfo {author} {\bibfnamefont {F.}~\bibnamefont {Piazza}},\
  }\href {\doibase 10.1103/PhysRevResearch.1.033051} {\bibfield  {journal}
  {\bibinfo  {journal} {Phys. Rev. Research}\ }\textbf {\bibinfo {volume}
  {1}},\ \bibinfo {pages} {033051} (\bibinfo {year} {2019})}\BibitemShut
  {NoStop}%
\bibitem [{\citenamefont {Yoshida}\ \emph
  {et~al.}(2019{\natexlab{b}})\citenamefont {Yoshida}, \citenamefont {Kudo},\
  and\ \citenamefont {Hatsugai}}]{Yoshida_nHFQH19}%
  \BibitemOpen
  \bibfield  {author} {\bibinfo {author} {\bibfnamefont {T.}~\bibnamefont
  {Yoshida}}, \bibinfo {author} {\bibfnamefont {K.}~\bibnamefont {Kudo}}, \
  and\ \bibinfo {author} {\bibfnamefont {Y.}~\bibnamefont {Hatsugai}},\ }\href
  {\doibase 10.1038/s41598-019-53253-8} {\bibfield  {journal} {\bibinfo
  {journal} {Scientific Reports}\ }\textbf {\bibinfo {volume} {9}},\ \bibinfo
  {pages} {16895} (\bibinfo {year} {2019}{\natexlab{b}})}\BibitemShut {NoStop}%
\bibitem [{\citenamefont {Yoshida}\ \emph
  {et~al.}(2020{\natexlab{c}})\citenamefont {Yoshida}, \citenamefont {Kudo},
  \citenamefont {Katsura},\ and\ \citenamefont
  {Hatsugai}}]{Yoshida_nHFQHJ_PRR20}%
  \BibitemOpen
  \bibfield  {author} {\bibinfo {author} {\bibfnamefont {T.}~\bibnamefont
  {Yoshida}}, \bibinfo {author} {\bibfnamefont {K.}~\bibnamefont {Kudo}},
  \bibinfo {author} {\bibfnamefont {H.}~\bibnamefont {Katsura}}, \ and\
  \bibinfo {author} {\bibfnamefont {Y.}~\bibnamefont {Hatsugai}},\ }\href
  {\doibase 10.1103/PhysRevResearch.2.033428} {\bibfield  {journal} {\bibinfo
  {journal} {Phys. Rev. Research}\ }\textbf {\bibinfo {volume} {2}},\ \bibinfo
  {pages} {033428} (\bibinfo {year} {2020}{\natexlab{c}})}\BibitemShut
  {NoStop}%
\bibitem [{\citenamefont {Guo}\ \emph {et~al.}(2020)\citenamefont {Guo},
  \citenamefont {Wang}, \citenamefont {Wang},\ and\ \citenamefont
  {Kou}}]{Guo_nHToric_PRB20}%
  \BibitemOpen
  \bibfield  {author} {\bibinfo {author} {\bibfnamefont {C.-X.}\ \bibnamefont
  {Guo}}, \bibinfo {author} {\bibfnamefont {X.-R.}\ \bibnamefont {Wang}},
  \bibinfo {author} {\bibfnamefont {C.}~\bibnamefont {Wang}}, \ and\ \bibinfo
  {author} {\bibfnamefont {S.-P.}\ \bibnamefont {Kou}},\ }\href {\doibase
  10.1103/PhysRevB.101.144439} {\bibfield  {journal} {\bibinfo  {journal}
  {Phys. Rev. B}\ }\textbf {\bibinfo {volume} {101}},\ \bibinfo {pages}
  {144439} (\bibinfo {year} {2020})}\BibitemShut {NoStop}%
\bibitem [{\citenamefont {Matsumoto}\ \emph {et~al.}(2020)\citenamefont
  {Matsumoto}, \citenamefont {Kawabata}, \citenamefont {Ashida}, \citenamefont
  {Furukawa},\ and\ \citenamefont {Ueda}}]{Matsumoto_nHToric_PRL20}%
  \BibitemOpen
  \bibfield  {author} {\bibinfo {author} {\bibfnamefont {N.}~\bibnamefont
  {Matsumoto}}, \bibinfo {author} {\bibfnamefont {K.}~\bibnamefont {Kawabata}},
  \bibinfo {author} {\bibfnamefont {Y.}~\bibnamefont {Ashida}}, \bibinfo
  {author} {\bibfnamefont {S.}~\bibnamefont {Furukawa}}, \ and\ \bibinfo
  {author} {\bibfnamefont {M.}~\bibnamefont {Ueda}},\ }\href {\doibase
  10.1103/PhysRevLett.125.260601} {\bibfield  {journal} {\bibinfo  {journal}
  {Phys. Rev. Lett.}\ }\textbf {\bibinfo {volume} {125}},\ \bibinfo {pages}
  {260601} (\bibinfo {year} {2020})}\BibitemShut {NoStop}%
\bibitem [{\citenamefont {Zhang}\ \emph
  {et~al.}(2020{\natexlab{b}})\citenamefont {Zhang}, \citenamefont {Chen},
  \citenamefont {Zhang}, \citenamefont {Lang}, \citenamefont {Li},\ and\
  \citenamefont {Zhu}}]{Zhang_nHTMI_PRB20}%
  \BibitemOpen
  \bibfield  {author} {\bibinfo {author} {\bibfnamefont {D.-W.}\ \bibnamefont
  {Zhang}}, \bibinfo {author} {\bibfnamefont {Y.-L.}\ \bibnamefont {Chen}},
  \bibinfo {author} {\bibfnamefont {G.-Q.}\ \bibnamefont {Zhang}}, \bibinfo
  {author} {\bibfnamefont {L.-J.}\ \bibnamefont {Lang}}, \bibinfo {author}
  {\bibfnamefont {Z.}~\bibnamefont {Li}}, \ and\ \bibinfo {author}
  {\bibfnamefont {S.-L.}\ \bibnamefont {Zhu}},\ }\href@noop {} {\bibfield
  {journal} {\bibinfo  {journal} {Phys. Rev. B}\ }\textbf {\bibinfo {volume}
  {101}},\ \bibinfo {pages} {235150} (\bibinfo {year}
  {2020}{\natexlab{b}})}\BibitemShut {NoStop}%
\bibitem [{\citenamefont {Liu}\ \emph {et~al.}(2020)\citenamefont {Liu},
  \citenamefont {He}, \citenamefont {Yoshida}, \citenamefont {Xiang},\ and\
  \citenamefont {Nori}}]{Liu_nHTMI_RPB20}%
  \BibitemOpen
  \bibfield  {author} {\bibinfo {author} {\bibfnamefont {T.}~\bibnamefont
  {Liu}}, \bibinfo {author} {\bibfnamefont {J.~J.}\ \bibnamefont {He}},
  \bibinfo {author} {\bibfnamefont {T.}~\bibnamefont {Yoshida}}, \bibinfo
  {author} {\bibfnamefont {Z.-L.}\ \bibnamefont {Xiang}}, \ and\ \bibinfo
  {author} {\bibfnamefont {F.}~\bibnamefont {Nori}},\ }\href {\doibase
  10.1103/PhysRevB.102.235151} {\bibfield  {journal} {\bibinfo  {journal}
  {Phys. Rev. B}\ }\textbf {\bibinfo {volume} {102}},\ \bibinfo {pages}
  {235151} (\bibinfo {year} {2020})}\BibitemShut {NoStop}%
\bibitem [{\citenamefont {Xu}\ and\ \citenamefont
  {Chen}(2020)}]{Xu_nHBM_PRB20}%
  \BibitemOpen
  \bibfield  {author} {\bibinfo {author} {\bibfnamefont {Z.}~\bibnamefont
  {Xu}}\ and\ \bibinfo {author} {\bibfnamefont {S.}~\bibnamefont {Chen}},\
  }\href@noop {} {\bibfield  {journal} {\bibinfo  {journal} {Phys. Rev. B}\
  }\textbf {\bibinfo {volume} {102}},\ \bibinfo {pages} {035153} (\bibinfo
  {year} {2020})}\BibitemShut {NoStop}%
\bibitem [{\citenamefont {Pan}\ \emph {et~al.}(2020)\citenamefont {Pan},
  \citenamefont {Wang}, \citenamefont {Cui},\ and\ \citenamefont
  {Chen}}]{Pan_PTHubb_oQS_PRA20}%
  \BibitemOpen
  \bibfield  {author} {\bibinfo {author} {\bibfnamefont {L.}~\bibnamefont
  {Pan}}, \bibinfo {author} {\bibfnamefont {X.}~\bibnamefont {Wang}}, \bibinfo
  {author} {\bibfnamefont {X.}~\bibnamefont {Cui}}, \ and\ \bibinfo {author}
  {\bibfnamefont {S.}~\bibnamefont {Chen}},\ }\href {\doibase
  10.1103/PhysRevA.102.023306} {\bibfield  {journal} {\bibinfo  {journal}
  {Phys. Rev. A}\ }\textbf {\bibinfo {volume} {102}},\ \bibinfo {pages}
  {023306} (\bibinfo {year} {2020})}\BibitemShut {NoStop}%
\bibitem [{\citenamefont {Mu}\ \emph {et~al.}(2020)\citenamefont {Mu},
  \citenamefont {Lee}, \citenamefont {Li},\ and\ \citenamefont
  {Gong}}]{Mu_MbdySkin_PRB20}%
  \BibitemOpen
  \bibfield  {author} {\bibinfo {author} {\bibfnamefont {S.}~\bibnamefont
  {Mu}}, \bibinfo {author} {\bibfnamefont {C.~H.}\ \bibnamefont {Lee}},
  \bibinfo {author} {\bibfnamefont {L.}~\bibnamefont {Li}}, \ and\ \bibinfo
  {author} {\bibfnamefont {J.}~\bibnamefont {Gong}},\ }\href {\doibase
  10.1103/PhysRevB.102.081115} {\bibfield  {journal} {\bibinfo  {journal}
  {Phys. Rev. B}\ }\textbf {\bibinfo {volume} {102}},\ \bibinfo {pages}
  {081115} (\bibinfo {year} {2020})}\BibitemShut {NoStop}%
\bibitem [{\citenamefont {Yoshida}\ and\ \citenamefont
  {Hatsugai}(2021)}]{Yoshida_PtGpZtoZ2_PRB21}%
  \BibitemOpen
  \bibfield  {author} {\bibinfo {author} {\bibfnamefont {T.}~\bibnamefont
  {Yoshida}}\ and\ \bibinfo {author} {\bibfnamefont {Y.}~\bibnamefont
  {Hatsugai}},\ }\href {\doibase 10.1103/PhysRevB.104.075106} {\bibfield
  {journal} {\bibinfo  {journal} {Phys. Rev. B}\ }\textbf {\bibinfo {volume}
  {104}},\ \bibinfo {pages} {075106} (\bibinfo {year} {2021})}\BibitemShut
  {NoStop}%
\bibitem [{\citenamefont {Yang}\ \emph
  {et~al.}(2021{\natexlab{b}})\citenamefont {Yang}, \citenamefont {Morampudi},\
  and\ \citenamefont {Bergholtz}}]{Yang_EPKitaev_PRL21}%
  \BibitemOpen
  \bibfield  {author} {\bibinfo {author} {\bibfnamefont {K.}~\bibnamefont
  {Yang}}, \bibinfo {author} {\bibfnamefont {S.~C.}\ \bibnamefont {Morampudi}},
  \ and\ \bibinfo {author} {\bibfnamefont {E.~J.}\ \bibnamefont {Bergholtz}},\
  }\href {\doibase 10.1103/PhysRevLett.126.077201} {\bibfield  {journal}
  {\bibinfo  {journal} {Phys. Rev. Lett.}\ }\textbf {\bibinfo {volume} {126}},\
  \bibinfo {pages} {077201} (\bibinfo {year} {2021}{\natexlab{b}})}\BibitemShut
  {NoStop}%
\bibitem [{\citenamefont {Shen}\ and\ \citenamefont
  {Lee}(2022)}]{Shen_CorrSkin_arXiv21}%
  \BibitemOpen
  \bibfield  {author} {\bibinfo {author} {\bibfnamefont {R.}~\bibnamefont
  {Shen}}\ and\ \bibinfo {author} {\bibfnamefont {C.~H.}\ \bibnamefont {Lee}},\
  }\href {\doibase 10.1038/s42005-022-01015-w} {\bibfield  {journal} {\bibinfo
  {journal} {Communications Physics}\ }\textbf {\bibinfo {volume} {5}},\
  \bibinfo {pages} {238} (\bibinfo {year} {2022})}\BibitemShut {NoStop}%
\bibitem [{\citenamefont {Lee}(2021)}]{Lee_MbdySkin_PRB21}%
  \BibitemOpen
  \bibfield  {author} {\bibinfo {author} {\bibfnamefont {C.~H.}\ \bibnamefont
  {Lee}},\ }\href {\doibase 10.1103/PhysRevB.104.195102} {\bibfield  {journal}
  {\bibinfo  {journal} {Phys. Rev. B}\ }\textbf {\bibinfo {volume} {104}},\
  \bibinfo {pages} {195102} (\bibinfo {year} {2021})}\BibitemShut {NoStop}%
\bibitem [{\citenamefont {Zhang}\ \emph {et~al.}(2022)\citenamefont {Zhang},
  \citenamefont {Denner}, \citenamefont {Bzdu\ifmmode~\check{s}\else
  \v{s}\fi{}ek}, \citenamefont {Sentef},\ and\ \citenamefont
  {Neupert}}]{Zhang_CorrSkin_arXiv22}%
  \BibitemOpen
  \bibfield  {author} {\bibinfo {author} {\bibfnamefont {S.-B.}\ \bibnamefont
  {Zhang}}, \bibinfo {author} {\bibfnamefont {M.~M.}\ \bibnamefont {Denner}},
  \bibinfo {author} {\bibfnamefont {T.~c.~v.}\ \bibnamefont
  {Bzdu\ifmmode~\check{s}\else \v{s}\fi{}ek}}, \bibinfo {author} {\bibfnamefont
  {M.~A.}\ \bibnamefont {Sentef}}, \ and\ \bibinfo {author} {\bibfnamefont
  {T.}~\bibnamefont {Neupert}},\ }\href {\doibase 10.1103/PhysRevB.106.L121102}
  {\bibfield  {journal} {\bibinfo  {journal} {Phys. Rev. B}\ }\textbf {\bibinfo
  {volume} {106}},\ \bibinfo {pages} {L121102} (\bibinfo {year}
  {2022})}\BibitemShut {NoStop}%
\bibitem [{\citenamefont {Kawabata}\ \emph {et~al.}(2022)\citenamefont
  {Kawabata}, \citenamefont {Shiozaki},\ and\ \citenamefont
  {Ryu}}]{Kawabata_CorrSkin_PRB22}%
  \BibitemOpen
  \bibfield  {author} {\bibinfo {author} {\bibfnamefont {K.}~\bibnamefont
  {Kawabata}}, \bibinfo {author} {\bibfnamefont {K.}~\bibnamefont {Shiozaki}},
  \ and\ \bibinfo {author} {\bibfnamefont {S.}~\bibnamefont {Ryu}},\ }\href
  {\doibase 10.1103/PhysRevB.105.165137} {\bibfield  {journal} {\bibinfo
  {journal} {Phys. Rev. B}\ }\textbf {\bibinfo {volume} {105}},\ \bibinfo
  {pages} {165137} (\bibinfo {year} {2022})}\BibitemShut {NoStop}%
\bibitem [{\citenamefont {Sch\"afer}\ \emph {et~al.}(2022)\citenamefont
  {Sch\"afer}, \citenamefont {Budich},\ and\ \citenamefont
  {Luitz}}]{Schafers_EPcorr_arXiv2022}%
  \BibitemOpen
  \bibfield  {author} {\bibinfo {author} {\bibfnamefont {R.}~\bibnamefont
  {Sch\"afer}}, \bibinfo {author} {\bibfnamefont {J.~C.}\ \bibnamefont
  {Budich}}, \ and\ \bibinfo {author} {\bibfnamefont {D.~J.}\ \bibnamefont
  {Luitz}},\ }\href {\doibase 10.1103/PhysRevResearch.4.033181} {\bibfield
  {journal} {\bibinfo  {journal} {Phys. Rev. Res.}\ }\textbf {\bibinfo {volume}
  {4}},\ \bibinfo {pages} {033181} (\bibinfo {year} {2022})}\BibitemShut
  {NoStop}%
\bibitem [{\citenamefont {Orito}\ and\ \citenamefont
  {Imura}(2022)}]{Orito_CorrSkin_PRB22}%
  \BibitemOpen
  \bibfield  {author} {\bibinfo {author} {\bibfnamefont {T.}~\bibnamefont
  {Orito}}\ and\ \bibinfo {author} {\bibfnamefont {K.-I.}\ \bibnamefont
  {Imura}},\ }\href {\doibase 10.1103/PhysRevB.105.024303} {\bibfield
  {journal} {\bibinfo  {journal} {Phys. Rev. B}\ }\textbf {\bibinfo {volume}
  {105}},\ \bibinfo {pages} {024303} (\bibinfo {year} {2022})}\BibitemShut
  {NoStop}%
\bibitem [{\citenamefont {Tsubota}\ \emph {et~al.}(2022)\citenamefont
  {Tsubota}, \citenamefont {Yang}, \citenamefont {Akagi},\ and\ \citenamefont
  {Katsura}}]{Tsubota_CorrInv_PRB22}%
  \BibitemOpen
  \bibfield  {author} {\bibinfo {author} {\bibfnamefont {S.}~\bibnamefont
  {Tsubota}}, \bibinfo {author} {\bibfnamefont {H.}~\bibnamefont {Yang}},
  \bibinfo {author} {\bibfnamefont {Y.}~\bibnamefont {Akagi}}, \ and\ \bibinfo
  {author} {\bibfnamefont {H.}~\bibnamefont {Katsura}},\ }\href {\doibase
  10.1103/PhysRevB.105.L201113} {\bibfield  {journal} {\bibinfo  {journal}
  {Phys. Rev. B}\ }\textbf {\bibinfo {volume} {105}},\ \bibinfo {pages}
  {L201113} (\bibinfo {year} {2022})}\BibitemShut {NoStop}%
\bibitem [{\citenamefont {Faugno}\ and\ \citenamefont
  {Ozawa}(2022)}]{Gaugno_corrnHSkin_arXiv2022}%
  \BibitemOpen
  \bibfield  {author} {\bibinfo {author} {\bibfnamefont {W.~N.}\ \bibnamefont
  {Faugno}}\ and\ \bibinfo {author} {\bibfnamefont {T.}~\bibnamefont {Ozawa}},\
  }\href {\doibase 10.1103/PhysRevLett.129.180401} {\bibfield  {journal}
  {\bibinfo  {journal} {Phys. Rev. Lett.}\ }\textbf {\bibinfo {volume} {129}},\
  \bibinfo {pages} {180401} (\bibinfo {year} {2022})}\BibitemShut {NoStop}%
\bibitem [{\citenamefont {Yoshida}\ and\ \citenamefont
  {Hatsugai}(2022)}]{Yoshida_reduction1Dptgp_arXiv2022}%
  \BibitemOpen
  \bibfield  {author} {\bibinfo {author} {\bibfnamefont {T.}~\bibnamefont
  {Yoshida}}\ and\ \bibinfo {author} {\bibfnamefont {Y.}~\bibnamefont
  {Hatsugai}},\ }\href {\doibase 10.1103/PhysRevB.106.205147} {\bibfield
  {journal} {\bibinfo  {journal} {Phys. Rev. B}\ }\textbf {\bibinfo {volume}
  {106}},\ \bibinfo {pages} {205147} (\bibinfo {year} {2022})}\BibitemShut
  {NoStop}%
\bibitem [{\citenamefont {Qin}\ \emph {et~al.}(2023)\citenamefont {Qin},
  \citenamefont {Shen},\ and\ \citenamefont {Lee}}]{Qin_CorrPolarons_arXiv22}%
  \BibitemOpen
  \bibfield  {author} {\bibinfo {author} {\bibfnamefont {F.}~\bibnamefont
  {Qin}}, \bibinfo {author} {\bibfnamefont {R.}~\bibnamefont {Shen}}, \ and\
  \bibinfo {author} {\bibfnamefont {C.~H.}\ \bibnamefont {Lee}},\ }\href
  {\doibase 10.1103/PhysRevA.107.L010202} {\bibfield  {journal} {\bibinfo
  {journal} {Phys. Rev. A}\ }\textbf {\bibinfo {volume} {107}},\ \bibinfo
  {pages} {L010202} (\bibinfo {year} {2023})}\BibitemShut {NoStop}%
\bibitem [{\citenamefont {Tomita}\ \emph {et~al.}(2017)\citenamefont {Tomita},
  \citenamefont {Nakajima}, \citenamefont {Danshita}, \citenamefont {Takasu},\
  and\ \citenamefont {Takahashi}}]{Tomita_Zeno_SciAdv17}%
  \BibitemOpen
  \bibfield  {author} {\bibinfo {author} {\bibfnamefont {T.}~\bibnamefont
  {Tomita}}, \bibinfo {author} {\bibfnamefont {S.}~\bibnamefont {Nakajima}},
  \bibinfo {author} {\bibfnamefont {I.}~\bibnamefont {Danshita}}, \bibinfo
  {author} {\bibfnamefont {Y.}~\bibnamefont {Takasu}}, \ and\ \bibinfo {author}
  {\bibfnamefont {Y.}~\bibnamefont {Takahashi}},\ }\href@noop {} {\bibfield
  {journal} {\bibinfo  {journal} {Science Advances}\ }\textbf {\bibinfo
  {volume} {3}},\ \bibinfo {pages} {e1701513} (\bibinfo {year}
  {2017})}\BibitemShut {NoStop}%
\bibitem [{\citenamefont {Tomita}\ \emph {et~al.}(2019)\citenamefont {Tomita},
  \citenamefont {Nakajima}, \citenamefont {Takasu},\ and\ \citenamefont
  {Takahashi}}]{Tomita_2BdyLoss_PRA19}%
  \BibitemOpen
  \bibfield  {author} {\bibinfo {author} {\bibfnamefont {T.}~\bibnamefont
  {Tomita}}, \bibinfo {author} {\bibfnamefont {S.}~\bibnamefont {Nakajima}},
  \bibinfo {author} {\bibfnamefont {Y.}~\bibnamefont {Takasu}}, \ and\ \bibinfo
  {author} {\bibfnamefont {Y.}~\bibnamefont {Takahashi}},\ }\href {\doibase
  10.1103/PhysRevA.99.031601} {\bibfield  {journal} {\bibinfo  {journal} {Phys.
  Rev. A}\ }\textbf {\bibinfo {volume} {99}},\ \bibinfo {pages} {031601}
  (\bibinfo {year} {2019})}\BibitemShut {NoStop}%
\bibitem [{\citenamefont {Takasu}\ \emph {et~al.}(2020)\citenamefont {Takasu},
  \citenamefont {Yagami}, \citenamefont {Ashida}, \citenamefont {Hamazaki},
  \citenamefont {Kuno},\ and\ \citenamefont
  {Takahashi}}]{Takasu_nHPTcoldAtom_PTEP2020}%
  \BibitemOpen
  \bibfield  {author} {\bibinfo {author} {\bibfnamefont {Y.}~\bibnamefont
  {Takasu}}, \bibinfo {author} {\bibfnamefont {T.}~\bibnamefont {Yagami}},
  \bibinfo {author} {\bibfnamefont {Y.}~\bibnamefont {Ashida}}, \bibinfo
  {author} {\bibfnamefont {R.}~\bibnamefont {Hamazaki}}, \bibinfo {author}
  {\bibfnamefont {Y.}~\bibnamefont {Kuno}}, \ and\ \bibinfo {author}
  {\bibfnamefont {Y.}~\bibnamefont {Takahashi}},\ }\href@noop {} {\bibfield
  {journal} {\bibinfo  {journal} {Progress of Theoretical and Experimental
  Physics}\ }\textbf {\bibinfo {volume} {2020}},\ \bibinfo {pages} {12A110}
  (\bibinfo {year} {2020})}\BibitemShut {NoStop}%
\bibitem [{\citenamefont {Ma}\ \emph {et~al.}(2019)\citenamefont {Ma},
  \citenamefont {Saxberg}, \citenamefont {Owens}, \citenamefont {Leung},
  \citenamefont {Lu}, \citenamefont {Simon},\ and\ \citenamefont
  {Schuster}}]{Ma_LossQuantumCircuits_Nature2019}%
  \BibitemOpen
  \bibfield  {author} {\bibinfo {author} {\bibfnamefont {R.}~\bibnamefont
  {Ma}}, \bibinfo {author} {\bibfnamefont {B.}~\bibnamefont {Saxberg}},
  \bibinfo {author} {\bibfnamefont {C.}~\bibnamefont {Owens}}, \bibinfo
  {author} {\bibfnamefont {N.}~\bibnamefont {Leung}}, \bibinfo {author}
  {\bibfnamefont {Y.}~\bibnamefont {Lu}}, \bibinfo {author} {\bibfnamefont
  {J.}~\bibnamefont {Simon}}, \ and\ \bibinfo {author} {\bibfnamefont {D.~I.}\
  \bibnamefont {Schuster}},\ }\href {\doibase 10.1038/s41586-019-0897-9}
  {\bibfield  {journal} {\bibinfo  {journal} {Nature}\ }\textbf {\bibinfo
  {volume} {566}},\ \bibinfo {pages} {51} (\bibinfo {year} {2019})}\BibitemShut
  {NoStop}%
\bibitem [{U1-()}]{U1-nc_ftnt}%
  \BibitemOpen
  \href@noop {} {}\bibinfo {note} {We have imposed the additional constraint on
  the $\hat{H}$. However, this does not affect the discussion provided in
  Sec.~\ref{sec: spin-parity topoinv} because the argument in Sec.~\ref{sec:
  spin-parity topoinv} is directly available by replacing the $\hat{H}$ to the
  block-diagonalized Hamiltonian with $\sum_\sigma \hat{n}_{c\sigma}$
  \hspace{-2mm}}\BibitemShut {NoStop}%
\bibitem [{Com()}]{CommentN2P-1_ftnt}%
  \BibitemOpen
  \href@noop {} {}\bibinfo {note} {Specifically, the winding numbers take
  $W_{(1,1)}=W_{[2,-1]}=0$ for $V=0$ \hspace{-2mm}}\BibitemShut {NoStop}%
\bibitem [{4x4()}]{4x4EPdiag_ftnt}%
  \BibitemOpen
  \href@noop {} {}\bibinfo {note} {This $4 \times 4$-matrix is diagonalized as
  follows. The Hamiltonian can be rewritten as \begin{eqnarray}
  \hat{H}_{(2,1)}&=& \left( \begin{array}{cc} e^{i\frac{\theta}{2}}M_{\theta} &
  iV\tau_0 \\ iV\tau_0 & e^{-i\frac{\theta}{2}} M_{\theta} \end{array} \right).
  \end{eqnarray} Here, the $2\times 2$ identity matrix is denoted by $\tau_0$.
  The matrix $M_{\theta}$ is defined as $ M_{\theta} = \left( \begin{array}{cc}
  0 & e^{i\frac{\theta}{2}} \\ e^{-i\frac{\theta}{2}} & 0 \end{array} \right).
  $ Thus, firstly diagonazling the matrix $M_{\theta}$, we obtain the
  eigenvalues shown in Eq.~(\ref{eq: Ep Em (N,P)=(2,1) s=-1})
  \hspace{-2mm}}\BibitemShut {NoStop}%
\bibitem [{xi2()}]{xi2=1_ftnt}%
  \BibitemOpen
  \href@noop {} {}\bibinfo {note} {We note that $\xi^2 = \1$ holds due to the
  relations $ u_{\mathrm{T}} u^*_{\mathrm{T}} =\pm \1$ and $ u_{\mathrm{C}}
  u^*_{\mathrm{C}} =\pm \1$. Here $u_{\mathrm{T}}$ ($u_{\mathrm{C}}$) denotes a
  unitary matrix describing time-reversal (particle-hole) symmetry. In order to
  obtain the relation $\xi^2 = \1$, we note that $u_{\mathrm{T}}
  u^*_{\mathrm{C}}$ squares to be $\pm \1$; $(u_{\mathrm{T}} u^*_{\mathrm{C}})
  (u_{\mathrm{T}} u^*_{\mathrm{C}})= u_{\mathrm{T}} u^*_{\mathrm{T}}
  u_{\mathrm{C}} u^*_{\mathrm{C}}$. Here we have used the fact that without
  loss of generality, $ u_{\mathrm{T}}$ and $u_{\mathrm{C}}$ can be chosen so
  that $ u_{\mathrm{T}} u^*_{\mathrm{C}} = u_{\mathrm{C}} u^*_{\mathrm{T}}$ is
  satisfied. For $(u_{\mathrm{T}} u^*_{\mathrm{C}})^2=\1$, we can define $\xi$
  as $\xi= u_{\mathrm{T}} u^*_{\mathrm{C}}$. For $(u_{\mathrm{T}}
  u^*_{\mathrm{C}})^2=-\1$, we can define $\xi$ as $\xi= i u_{\mathrm{T}}
  u^*_{\mathrm{C}}$. Therefore, we can see that $\xi^2 = \1$ holds}\BibitemShut
  {NoStop}%
\bibitem [{hHm()}]{hHmat_ftnt}%
  \BibitemOpen
  \href@noop {} {}\bibinfo {note} {This fact can be directly seen by noting
  that $ h_{\mathrm{H}}$ is written as $h_{\mathrm{H}}= \left(
  \begin{array}{c|c} 0 & q \\ \hline q^\dagger & 0 \end{array} \right) $ with
  the basis where $\xi$ is written as $ \xi= \left( \begin{array}{c|c} \1 & 0
  \\ \hline 0 & -\1 \end{array} \right) $. Here, $q$ is a matrix. We also note
  that $h_{\mathrm{A}}$ is written as $h_{\mathrm{A}}= \left(
  \begin{array}{c|c} q_{\mathrm{A}+} & 0 \\ \hline 0 & q_{\mathrm{A}-}
  \end{array} \right) $ with the above basis. Here, $q_{A+}$ and $q_{A-}$ are
  anti-Hermitian matrices \hspace{-2mm}}\BibitemShut {NoStop}%
\bibitem [{PHS()}]{PHSofH0_ftnt}%
  \BibitemOpen
  \href@noop {} {}\bibinfo {note} {For systems where $\hat{\Gamma}$ and
  $\hat{\Xi}$ commute each other, consider an eigenstate $|\tilde{E}\rangle$ of
  the Hermitian operator $\hat{H}_{0\Gamma}$ with eigenvalue $\tilde{E} \in
  \mathbb{R}$. Then, we have $\hat{H}_{0\Gamma} \hat{\Xi} |\tilde{E}\rangle =
  -\tilde{E} \hat{\Xi}|\tilde{E}\rangle $, meaning that the state $\hat{\Xi}
  |\tilde{E}\rangle$ is an eigenstate with eigenvalue $-\tilde{E}$. Thus,
  tuning parameters does not change the zero-th Chern number $N_{0\mathrm{Ch}}$
  when $\hat{\Gamma}$ and $\hat{\Xi}$ commute each other. In contrast, for
  systems where $\hat{\Gamma}$ and $\hat{\Xi}$ anti-commute each other [see
  Eq.~(\ref{eq: XiG=-GXi})], the zero-th Chern number $N_{0\mathrm{Ch}}$ can
  change its value; in this case, the relation $\hat{\Xi}
  \hat{H}_{0\Gamma}=\hat{H}_{0\Gamma}\hat{\Xi}$ holds
  \hspace{-2mm}}\BibitemShut {NoStop}%
\bibitem [{pse()}]{pseudo-spin_ftnt}%
  \BibitemOpen
  \href@noop {} {}\bibinfo {note} {To be strict, the subscript $\sigma=1,0,-1$
  labels pseudo-spin because fermions should have odd half-integer spin
  ($1/2,3/2,5/2\ldots$) \hspace{-2mm}}\BibitemShut {NoStop}%
\bibitem [{\citenamefont {Gurarie}(2011)}]{Gurarie_chiral_PRB11}%
  \BibitemOpen
  \bibfield  {author} {\bibinfo {author} {\bibfnamefont {V.}~\bibnamefont
  {Gurarie}},\ }\href {\doibase 10.1103/PhysRevB.83.085426} {\bibfield
  {journal} {\bibinfo  {journal} {Phys. Rev. B}\ }\textbf {\bibinfo {volume}
  {83}},\ \bibinfo {pages} {085426} (\bibinfo {year} {2011})}\BibitemShut
  {NoStop}%
\bibitem [{\citenamefont {Manmana}\ \emph {et~al.}(2012)\citenamefont
  {Manmana}, \citenamefont {Essin}, \citenamefont {Noack},\ and\ \citenamefont
  {Gurarie}}]{Manmana_Chiral1D_PRB12}%
  \BibitemOpen
  \bibfield  {author} {\bibinfo {author} {\bibfnamefont {S.~R.}\ \bibnamefont
  {Manmana}}, \bibinfo {author} {\bibfnamefont {A.~M.}\ \bibnamefont {Essin}},
  \bibinfo {author} {\bibfnamefont {R.~M.}\ \bibnamefont {Noack}}, \ and\
  \bibinfo {author} {\bibfnamefont {V.}~\bibnamefont {Gurarie}},\ }\href
  {\doibase 10.1103/PhysRevB.86.205119} {\bibfield  {journal} {\bibinfo
  {journal} {Phys. Rev. B}\ }\textbf {\bibinfo {volume} {86}},\ \bibinfo
  {pages} {205119} (\bibinfo {year} {2012})}\BibitemShut {NoStop}%
\bibitem [{chi()}]{chiral-NSz_ftnt}%
  \BibitemOpen
  \href@noop {} {}\bibinfo {note} {We have imposed the additional constraints
  on the $\hat{H}$. However, these constraints do not affect the discussion
  provided in Sec.~\ref{sec: chiral topoinv} because the block-diagonalization
  is not effected by the presence/absence of interactions in contrast to the
  case of Sec.~\ref{sec: spin-parity topoinv}. Namely, the argument in
  Sec.~\ref{sec: chiral topoinv} is directly available by replacing the
  $\hat{H}$ to the block-diagonalized Hamiltonian with $\hat{N}$ and
  $\hat{S}_z$, although an explicit analysis of a toy model without additional
  symmetry constraints is left as a future work \hspace{-2mm}}\BibitemShut
  {NoStop}%
\end{thebibliography}
%


\appendix

\section{
The number of the subspaces for the Fock space with $[N,P]$
}
\label{sec: num Fock app}
For a given set of $[N,P]$, we count how many sets of $(N_\uparrow,N_\downarrow)$ are allowed, which elucidates the number of the subspaces for the given Fock space with $[N,P]$.
We count the number of the sets $(N_\uparrow,N_\downarrow)$ for the following four cases.

(i) For even $N$ and $P=1$, the following sets of $(N_\uparrow, N_\downarrow)$ are allowed: 
\begin{eqnarray}
\left\{ (2M,0), (2M-2,2), (2M-4,4), \ldots,  (0,2M) \right\}, \nonumber \\
\end{eqnarray}
with $N=2M$ and $M$ being a non-negative integer.
Thus, there exist the number $N/2+1$ of the sets $(N_\uparrow,N_\downarrow)$ for the given set of $[N,P]=[2M,1]$.

(ii) For even $N$ and $P=-1$, the following sets of $(N_\uparrow, N_\downarrow)$ are allowed: 
\begin{eqnarray}
\left\{ (2M-1,1), (2M-3,3), (2M-5,5), \ldots,  (1,2M-1) \right\}, \nonumber \\
\end{eqnarray}
with $N=2M$ and $M$ being a non-negative integer.
Thus, there exist the number $N/2$ of the sets $(N_\uparrow,N_\downarrow)$ for the given set of $[N,P]=[2M,-1]$.

(iii) For odd $N$ and $P=1$, the following sets of $(N_\uparrow, N_\downarrow)$ are allowed: 
\begin{eqnarray}
\left\{ (2M+1,0), (2M-1,2), (2M-3,4), \ldots,  (1,2M) \right\}, \nonumber \\
\end{eqnarray}
with $N=2M+1$ and $M$ being a non-negative integer.
Thus, there exist the number $(N+1)/2$ of the sets $(N_\uparrow,N_\downarrow)$ for the given set of $[N,P]=[2M+1,1]$.

(iv) For odd $N$ and $P=-1$, the following sets of $(N_\uparrow, N_\downarrow)$ are allowed: 
\begin{eqnarray}
\left\{ (2M,1), (2M-2,3), (2M-4,5), \ldots,  (0,2M+1) \right\}, \nonumber \\
\end{eqnarray}
with $N=2M+1$ and $M$ being a non-negative integer.
Thus, there exist the number $(N+1)/2$ of the sets $(N_\uparrow,N_\downarrow)$ for the given set of $[N,P]=[2M+1,-1]$.

Taking into account the above results, we end up with the fact that there exist the number $(N+P'+1)/2$ of the subspaces with $(N_\uparrow,N_\downarrow)$ for the given Fock space with $[N,P]$.
Here, $P'$ takes $P$ ($0$) for even (odd) $N$.

\section{
Robustness of EPs for $s_\downarrow=1$
}
\label{sec: EP for s=-1 app}

EPs characterized by a finite value of the winding number $W_{[N,P]}$ are robust against interactions.
In order to demonstrate this fact, let us analyze the Hamiltonian discussed in Sec.~\ref{sec: spin-parity toy} for $s_\downarrow=1$ [i.e., the Hamiltonian~(\ref{eq: generic H}) specified by Eqs.~(\ref{eq: spin-parity h})~and~(\ref{eq: parity intVonly})].

As is the case of $s_\downarrow=-1$ (see Sec.~\ref{sec: spin-parity toy}), we focus on the Fock space with $[N,P]=[2,1]$.
\begin{figure}[!h]
\begin{minipage}{1\hsize}
\begin{center}
\includegraphics[width=1\hsize,clip]{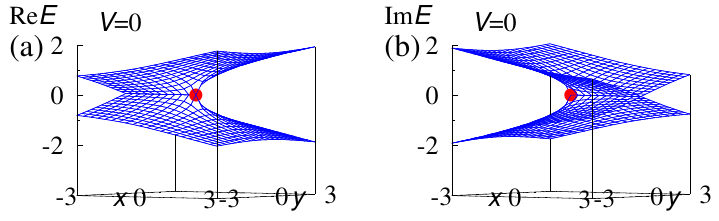}
\end{center}
\end{minipage}
\begin{minipage}{1\hsize}
\begin{center}
\includegraphics[width=1\hsize,clip]{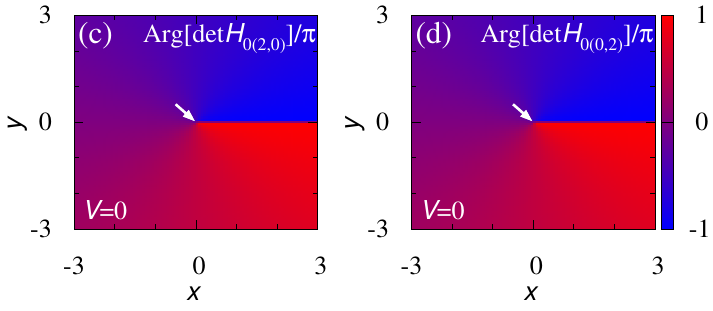}
\end{center}
\end{minipage}
\caption{
(a) [(b)] The real- [imaginary-] part of eigenvalues of $\hat{H}_{[2,1]}$ for $V=0$ and $s_\downarrow=1$.
The red dots denote EPs.
(c) [(d)] The argument of $\mathrm{det}\hat{H}_{0(2,0)}$ [$\mathrm{det}\hat{H}_{0(0,2)}$]. 
The data are plotted in a similar way to Fig.~\ref{fig: MbdyEfermi free (N,P)=(2,1)}.
}
\label{fig: MbdyEfermi free (N,P)=(2,1) app}
\end{figure}
Figures~\ref{fig: MbdyEfermi free (N,P)=(2,1) app}(a)~and~\ref{fig: MbdyEfermi free (N,P)=(2,1) app}(b) display eigenvalues of $\hat{H}_{[2,1]}$ for $V=0$ and $s_\downarrow=1$.
For $V=0$, the Fock space can be divided into subspaces with $(N_{\uparrow}, N_{\downarrow})=(2,0)$ and $(0,2)$. 
For both subspaces, EPs emerge which are characterized by the winding numbers $(W_{(2,0)},W_{(0,2)})=(1,1)$ for $E_{\mathrm{ref}}=0$ [see Figs.~\ref{fig: MbdyEfermi free (N,P)=(2,1) app}(c)~and~\ref{fig: MbdyEfermi free (N,P)=(2,1) app}(d)].
Thus, recalling Eq.~(\ref{eq: Wparity=Wsz+Wsz}), we have $W_{[N,P]}=2$ for $E_{\mathrm{ref}}=0$. 
This result is consistent with Fig.~\ref{fig: MbdyEfermi ArgH (N,P)=(2,1)}(a); $W_{[N,P]}$ takes two which is computed along a path winding around the singular point (denoted by a white arrow) in the counterclockwise direction.

\begin{figure}[!h]
\begin{minipage}{1\hsize}
\begin{center}
\includegraphics[width=1\hsize,clip]{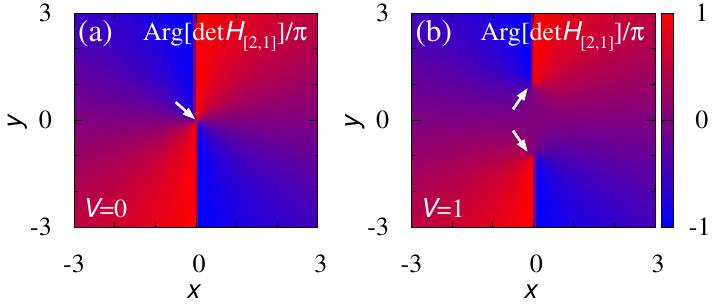}
\end{center}
\end{minipage}
\caption{
(a) [(b)] The argument of $\mathrm{det}\hat{H}_{[2,1]}$ for $V=0$ [$V=1$] and $s_{\downarrow}=1$.
The white arrows indicate the points where EPs emerge.
}
\label{fig: MbdyEfermi ArgH (N,P)=(2,1)}
\end{figure}
The finite value of $W_{[N,P]}$ in the non-interacting case indicates the robustness of EPs against interactions.
Figure~\ref{fig: MbdyEfermi ArgH (N,P)=(2,1)}(b) displays color map of $\mathrm{det}\hat{H}_{[2,1]}$ for $V=1$.
In this figure, we can see that $W_{[N,P]}$ takes one which is computed along a path winding around the singular point (denoted by a white arrow) in the counterclockwise direction.
\begin{figure}[!h]
\begin{minipage}{1\hsize}
\begin{center}
\includegraphics[width=1\hsize,clip]{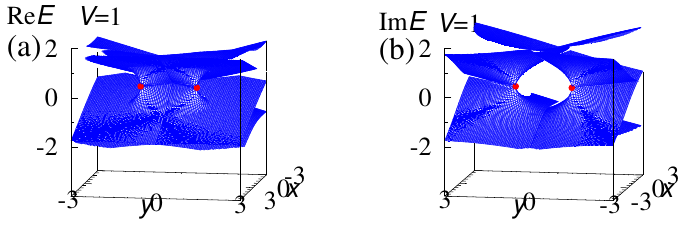}
\end{center}
\end{minipage}
\caption{
(a) [(b)] The real- [imaginary-] part of eigenvalues of $\hat{H}_{[2,1]}$ for $V=1$ and $s_\downarrow=1$.
Red dots denote the EPs at zero energy $E=0$.
The data are plotted in a similar way to Fig.~\ref{fig: MbdyEfermi corr (N,P)=(2,1)}.
}
\label{fig: EPU1 s1 V1 app}
\end{figure}
Correspondingly, the EPs emerges even in the interacting cases. Figure~\ref{fig: EPU1 s1 V1 app} displays eigenvalues of $\hat{H}_{[2,1]}$ for $V=1$ and $s_\downarrow=1$.
The above results indicate that EPs characterized by a finite value of $W_{[N,P]}$ are robust against interactions.

\section{
Reduction of one-dimensional topology for gapped systems
}
\label{sec: gapped 1D app}

Computing the topological invariants [Eqs.~(\ref{eq: W (N Sz) free})~and~(\ref{eq: W [N,P] })], we analyze the one-dimensional topology of a gapped system, which justifies the reduction $\mathbb{Z}^{(N+P'+1)/2}\to \mathbb{Z}$ for the one-dimensional point-gap topology with charge $\mathrm{U(1)}$ symmetry and spin-parity symmetry.
In the following, the topology in a one-dimensional parameter space is mainly analyzed, although a similar analysis can be done for the topology in one spatial dimension as briefly explained around the end of this section.

Consider the Hamiltonian~(\ref{eq: generic H}) specified with $\hat{\Psi}=(\hat{c}_{a\uparrow},\hat{c}_{b\uparrow},\hat{c}_{a\downarrow},\hat{c}_{b\downarrow})^{T}$,
\begin{eqnarray}
h(\theta) &=&  \mathrm{diag}(e^{i\theta},0,e^{-i\theta},0), \\
\hat{H}_{\mathrm{int}}&=&iV[\hat{S}^+_a\hat{S}^+_b+\hat{S}^-_a\hat{S}^-_b].
\end{eqnarray}
Here, $\mathrm{diag}(\ldots)$ denotes a diagonal matrix whose elements are specified by the numbers enclosed in the parentheses.
The coefficient $V$ is a real number. 
The one-dimensional parameter space is described by $\theta$ ($0 \leq \theta < 2\pi$).

The above Hamiltonian preserves charge $\mathrm{U(1)}$ symmetry and spin-parity symmetry [see Eqs.~(\ref{eq: U1 symm mbdyH})~and~(\ref{eq: spin-parity symm mbdyH})].
We also note that $\hat{H}$ commutes with $\hat{n}_{b\sigma}=\hat{c}^\dagger_{b\sigma}\hat{c}_{b\sigma}$ for $\sigma=\uparrow,\downarrow$.
Thus, we suppose that a fermion occupies orbital $b$.

For the Fock space with $[N,P]=[2,1]$, the Hamiltonian is written as
\begin{eqnarray}
\hat{H}_{[2,1]}&=& 
\left(
\begin{array}{cc}
e^{i\theta} & iV \\
iV & e^{-i\theta}
\end{array}
\right),
\end{eqnarray}
with the basis
\begin{eqnarray}
\left(
\hat{c}^{\dagger}_{a\uparrow}\hat{c}^{\dagger}_{b\uparrow} |0\rangle, \,
\hat{c}^{\dagger}_{a\downarrow}\hat{c}^{\dagger}_{b\downarrow} |0\rangle
\right).
\end{eqnarray}

\begin{figure}[!h]
\begin{minipage}{1\hsize}
\begin{center}
\includegraphics[width=1\hsize,clip]{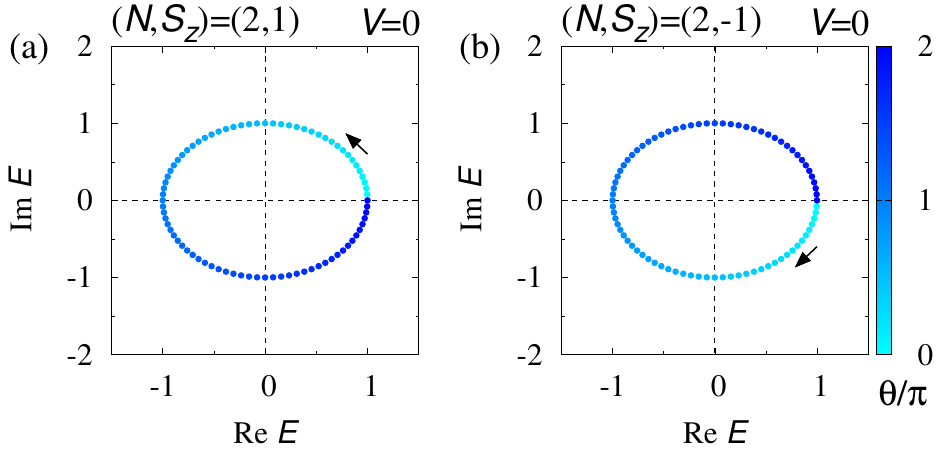}
\end{center}
\end{minipage}
\begin{minipage}{1\hsize}
\begin{center}
\includegraphics[width=0.5\hsize,clip]{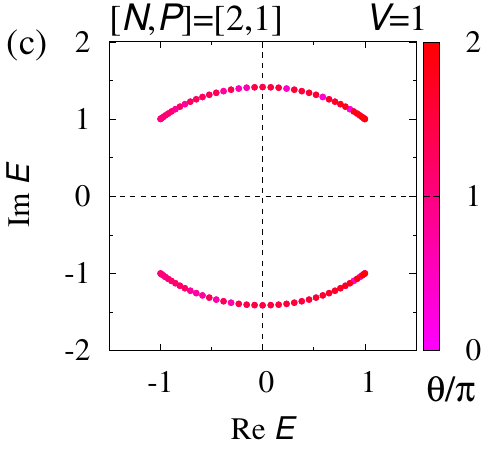}
\end{center}
\end{minipage}
\caption{
Spectral flow of the Hamiltonian $\hat{H}$ for the Fock space with $[N,P]=[2,1]$.
(a) [(b)] Spectral flow for the subspaces with $(N,S_z)=(2,1)$ [$(2,-1)$] at $V=0$.
(c) Spectral flow for the Fock space with $[N,P]=[2,1]$ at $V=1$.
}
\label{fig: Gapped1D}
\end{figure}
For $V=0$, the above representation indicates that the Fock space is decomposed into subspaces with $(N,S_z)=(2,1)$ and $(2,-1)$, and that the winding numbers for these subspaces take $W_{(2,1)}=1$ and $W_{(2,-1)}=-1$ at $E_{\mathrm{ref}}=0$, respectively [see Figs.~\ref{fig: Gapped1D}(a)~and~\ref{fig: Gapped1D}(b)].
Therefore, the loop structures observed in Figs.~\ref{fig: Gapped1D}(a)~and~\ref{fig: Gapped1D}(b) are robust against perturbations at the non-interacting level because they are protected by the non-trivial point-gap topology.

Here, Eq.~(\ref{eq: Wparity=Wsz+Wsz}) indicates that the above loop structure is no longer protected by the non-trivial topology in the presence of interactions; because of the winding numbers $W_{(2,1)}=1$ and $W_{(2,1)}=-1$ for $E_{\mathrm{ref}}=0$, Eq.~(\ref{eq: Wparity=Wsz+Wsz}) results in $W_{[2,1]}=0$.
Indeed, the interaction $V$ destroys the loop structure observed in Figs.~\ref{fig: Gapped1D}(a)~and~\ref{fig: Gapped1D}(b).
In order to see this, we compute the eigenvalues which are written as
\begin{eqnarray}
E_{\pm}(\theta)&=& \cos\theta \pm i\sqrt{\sin^2\theta+V^2}.
\end{eqnarray}
This result elucidates that the loop structure is destroyed by interactions mixing the subspaces with $(N,S_z)=(2,1)$ and $(2,-1)$; $|\mathrm{Im}E_{\pm}(\theta)|>0$ holds for an arbitrary $\theta$.
This fact is explicitly presented in Fig.~\ref{fig: Gapped1D}(c).

The above results demonstrate that the reduction of the one-dimensional point-gap topology for the gapped systems: $\mathbb{Z}^{2}\to \mathbb{Z}$ for $[N,P]=[2,1]$.
Namely, in the non-interacting case, the second quantized Hamiltonian $\hat{H}_0$ possesses the non-trivial properties characterized by $W_{(2,1)}=1$ and $W_{(2,-1)}=-1$ for $E_{\mathrm{ref}}=0$.
However, the non-trivial topology is not maintained in the presence of interactions ($W_{[2,1]}=0$), which is supported by the fragility of the loop structure against the interactions [see Fig.~\ref{fig: Gapped1D}].

We finish this section with two remarks.
The previous work~\cite{Yoshida_reduction1Dptgp_arXiv2022} has also addressed the reduction phenomena. 
We note, however, that Ref.~\onlinecite{Yoshida_reduction1Dptgp_arXiv2022} has compared the topology of the first-quantized Hamiltonian $h$ and the second-quantized Hamiltonian $\hat{H}$.
In contrast, the analysis provided in this section compares the topology of the non-interacting second-quantized Hamiltonian $\hat{H}_0$ with that of the interacting second-quantized Hamiltonian $\hat{H}$ which clearly elucidates that 
the $\mathbb{Z}^{(N+P'+1)/2}$ group formed by the point-gap topological states in non-interacting cases reduces to its subgroup $\mathbb{Z}$ due to the interactions [see Eq.~(\ref{eq: Wparity=Wsz+Wsz})].

We also note that a similar argument is applied to the point-gap topology in one spatial dimension. 
In Ref.~\onlinecite{Yoshida_reduction1Dptgp_arXiv2022}, the point-gap topology in one spatial dimension is analyzed for an interacting non-Hermitian chain with charge $\mathrm{U(1)}$ symmetry and spin-parity symmetry [see Eq.~(9) of Ref.~\onlinecite{Yoshida_reduction1Dptgp_arXiv2022}]. 
Because the Fock space with $[N,P]=[3,-1]$ is divided into subspaces with $(N,S_z)=(3,3/2)$ and $(3,-1/2)$ in the non-interacting case, the topology of non-interacting Hamiltonian is characterized by the two winding numbers taking $W_{(3,3/2)}=1$ and $W_{(3,-1/2)}=-1$ at $E_{\mathrm{ref}}=0$ for the Fock space with $[N,P]= [3,-1]$.
Thus, the loop structure is observed for the non-interacting non-Hermitian chain [see Fig.~3(b) of Ref.~\onlinecite{Yoshida_reduction1Dptgp_arXiv2022}]. 
This loop structure is fragile against interactions because Eq.~(\ref{eq: Wparity=Wsz+Wsz}) results in the vanishing winding number $W_{[N,P]}=0$. 
Indeed, interactions destroy the loop structure with keeping the point-gap for $E_{\mathrm{ref}}=0$ and the relevant symmetry [see Fig.~3(d) of Ref.~\onlinecite{Yoshida_reduction1Dptgp_arXiv2022}]. 
Correspondingly, the non-Hermitian skin effect observed at the non-interacting level is also destroyed by the interactions [see Figs.~3(c)~and~3(e) of Ref.~\onlinecite{Yoshida_reduction1Dptgp_arXiv2022}]. 

\section{
Reduction of zero-dimensional topology for a gapped system
}
\label{sec: gapped 0D app}

We analyze the zero-dimensional topology of a gapped system, which justifies the reduction $\mathbb{Z}\to \mathbb{Z}_2$ for the zero-dimensional point-gap topology with chiral symmetry.
Although the following results are essentially the same as the ones in Sec.~\ref{sec: SPER toy}, we discuss the details by focusing on a zero-dimensional system with the point-gap.

Consider the Hamiltonian $\hat{H}$ specified with Eqs.~(\ref{eq: chiral H0}),~(\ref{eq: chiral Hint}),~and~(\ref{eq: chiral h}) for given real parameters $x$ and $y=0$ (i.e., $z=x$, $0 \leq x \leq 1$).
The Hamiltonian $\hat{H}$ preserves the chiral symmetry~(\ref{eq: chiral mbdy}) with $\hat{\Xi}$ defined in Eq.~(\ref{eq: chiral Xi toy}). Thus, in the absence of interactions, the topology of $\hat{H}$ is characterized by the zero-th Chern number $N_{0\mathrm{Ch}}$, a $\mathbb{Z}$-invariant [see Sec.~\ref{sec: Zinv chiral}].
In the presence of interactions, the topology of $\hat{H}$ is characterized by the $\mathbb{Z}_2$-invariant $\nu$.

\begin{figure}[!h]
\vspace{5mm}
\begin{minipage}{1\hsize}
\begin{center}
\includegraphics[width=0.8\hsize,clip]{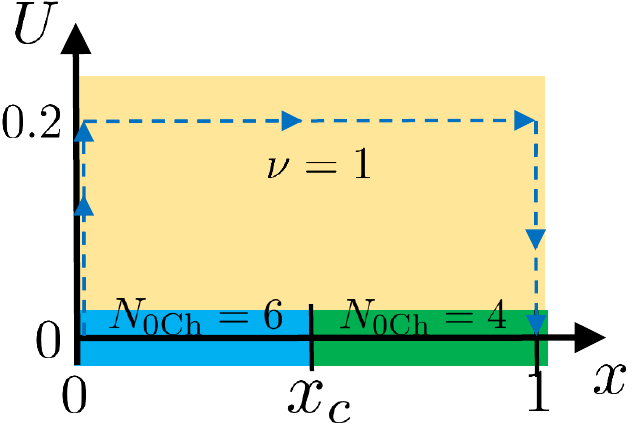}
\end{center}
\end{minipage}
\caption{
Phase diagram of the Hamiltonian $\hat{H}_{(3,0)}$.
In the absence of interactions, the zero-th Chern number takes $N_{0\mathrm{Ch}}=4$ (6) at $E_{\mathrm{ref}}=0$ for $0\leq x< x_c$ ($x_c< x\leq 1$) with $x_c\sim 0.6$.
In the presence of interaction, the $\mathbb{Z}_2$-invariant takes $\nu=1$ for $0\leq x \leq 1$.
Dashed allows illustrate a path parameterized by $\lambda$.
}
\label{fig: Gapped0D phase}
\end{figure}

Let us focus on the Fock space with $(N,S_z)=(3,0)$.
Figure~\ref{fig: Gapped0D phase} displays topological invariants against $x$ and $U$. 
In the non-interacting case, the point-gap at $E_{\mathrm{ref}}=0$ closes ($\mathrm{\det}\hat{H}_{(3,0)}=0$) at the point $(x,U)=(x_c,0)$ with $x_c\sim 0.6$ which separates two regions of distinct point-gap topology with $N_{0\mathrm{Ch}}=6$ and $N_{0\mathrm{Ch}}=4$ for $E_{\mathrm{ref}}=0$.
In the presence of interactions $U$, the above point-gap closing does not occur. 
Therefore, one can identify the topology of $N_{0\mathrm{Ch}}=6$ and that of $N_{0\mathrm{Ch}}=4$ in the presence of interactions.
Correspondingly the topology is characterized by $\nu$ which takes $\nu=1$ in the entire region.

\begin{figure}[!h]
\begin{minipage}{1\hsize}
\begin{center}
\includegraphics[width=1\hsize,clip]{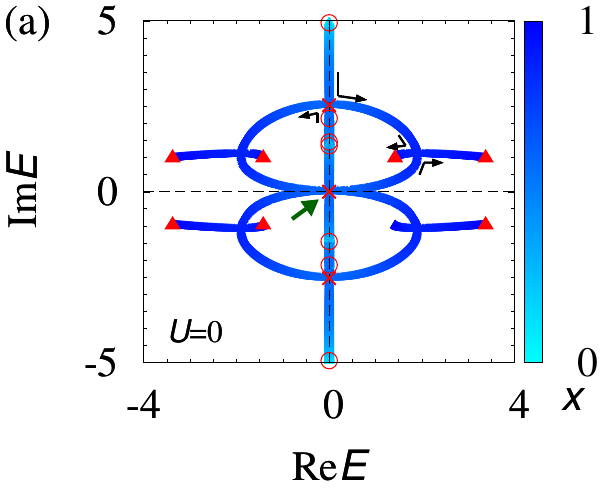}
\end{center}
\end{minipage}
\begin{minipage}{1\hsize}
\begin{center}
\includegraphics[width=1\hsize,clip]{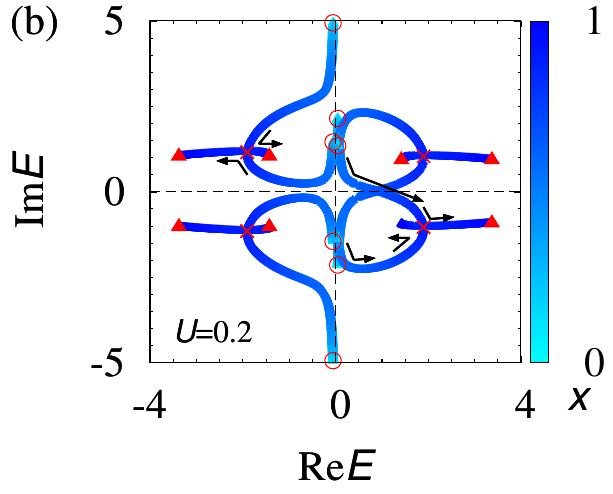}
\end{center}
\end{minipage}
\caption{
(a) [(b)] Spectral flow of $\hat{H}_{(3,0)}$ for $U=0$ [$0.2$]. 
With increasing $x$ from $0$ to $1$, the eigenvalues flow along blue curves as indicated by black arrows.
In panel (a) [(b)], eigenvalues for $x=0$, $0.6$, and $1$  [$x=0$, $0.8725$, and $1$] are denoted by open circles, crosses, and closed triangles, respectively.
In panel (a), the point-gap at $E_{\mathrm{ref}}=0$ closes as denoted by the green arrow.
The flow in panel (a) is symmetric about the real- and imaginary-axes due to Eqs.~(\ref{eq: chiral mbdy})~and~(\ref{eq: H0=-GHdagG}).
The flow in panel (b) is symmetric about the real-axis due to Eqs.~(\ref{eq: chiral mbdy}).
These data are obtained for $(\beta,\gamma_1,\gamma_0,\gamma_{-1})=(0.8,-3,-2.945,1)$.
}
\label{fig: Gapped0D flow}
\end{figure}

\begin{figure}[!h]
\begin{minipage}{1\hsize}
\begin{center}
\includegraphics[width=1\hsize,clip]{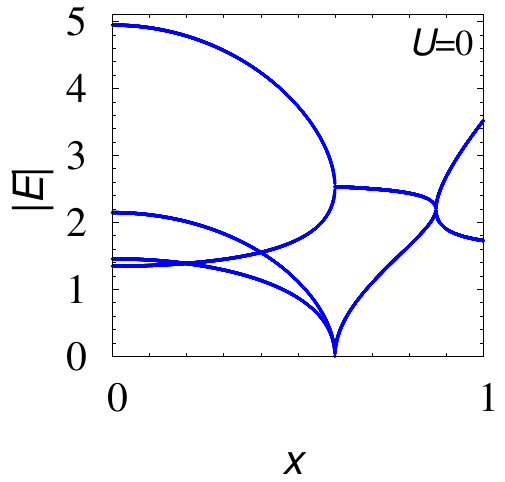}
\end{center}
\end{minipage}
\caption{
The absolute value of eigenvalues $|E_n|$ ($n=1,\ldots,8$) as functions of $x$ for $U=0$.
At $x=x_c\sim 0.6$, the eigenvalues become zero.
}
\label{fig: Gapped0D AbsE U0}
\end{figure}

In the following, we numerically demonstrate that interactions allow the smooth deformation of $\hat{H}$ characterized by $N_{0\mathrm{Ch}}=6$ to the $\hat{H}$ characterized by $N_{0\mathrm{Ch}}=4$ which keeps the point-gap and chiral symmetry.
Figures~\ref{fig: Gapped0D flow}(a)~and~\ref{fig: Gapped0D flow}(b) display the spectral flow for $U=0$ and $U=0.2$, respectively.
As shown in Fig.~\ref{fig: Gapped0D flow}(a), the point-gap closes at $x=x_c\sim0.6$ in the non-interacting case [see also Fig.~\ref{fig: Gapped0D AbsE U0}]. 
In contrast, the point-gap remains open in the interacting case [see Fig.~\ref{fig: Gapped0D flow}(b)], which is consistent with the phase diagram [see Fig.~\ref{fig: Gapped0D phase}].

\begin{figure}[!h]
\begin{minipage}{1\hsize}
\begin{center}
\includegraphics[width=1\hsize,clip]{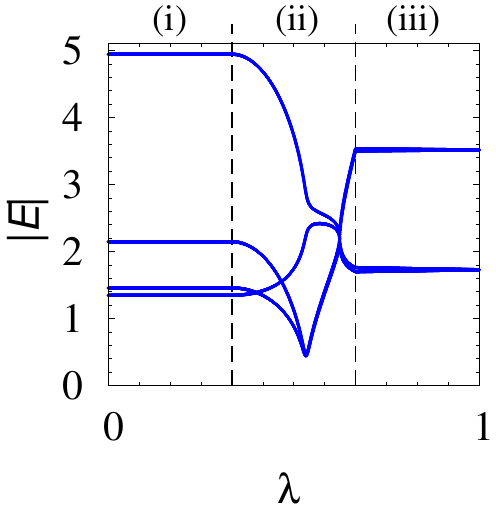}
\end{center}
\end{minipage}
\caption{
The absolute value of eigenvalues $|E_n|$ ($n=1,\ldots,8$) as functions of $\lambda$ which parameterizes the path illustrated in Fig.~\ref{fig: Gapped0D phase}.
On the dashed vertical lines $\lambda$ takes $\lambda=1/3$ and $2/3$, respectively.
Here, $\lambda$ parameterizes $(x,U)$ as follows: [region (i)] for $0 \leq \lambda <1/3$, it parameterizes as $(x,U)=(0,0.6\lambda)$; [region (ii)] for $1/3 \leq \lambda <2/3$ , it parameterizes as $(x,U)=(3\lambda-1,0.2)$; [region (iii)] $2/3 \leq \lambda \leq 1$, it parameterizes as  $(x,U)=(1,0.2-0.6\lambda)$.
}
\label{fig: Gapped0D AbsE}
\end{figure}

In a similar way, we can see that the gap remains open along the path parameterized by $\lambda$ ($0 \leq \lambda \leq 1$) which is illustrated by dashed arrows in Fig.~\ref{fig: Gapped0D phase}. 
Figure~\ref{fig: Gapped0D AbsE} indicates that interactions allow the smooth deformation of $\hat{H}$ characterized by $N_{0\mathrm{Ch}}=6$ to the $\hat{H}$ characterized by $N_{0\mathrm{Ch}}=4$ which keeps the point-gap and chiral symmetry.

The above results justifies the reduction of the point-gap topology $\mathbb{Z} \to \mathbb{Z}_2$ for zero-dimensional systems with chiral symmetry.

The previous work~\cite{Yoshida_PtGpZtoZ2_PRB21} has also addressed the reduction phenomenon. 
We note, however, that Ref.~\onlinecite{Yoshida_PtGpZtoZ2_PRB21} has compared the topology of the first-quantized Hamiltonian $h$ and the second-quantized Hamiltonian $\hat{H}$.
In contrast, the analysis provided in this section compares the topology of the non-interacting second-quantized Hamiltonian $\hat{H}_0$ with that of the interacting second-quantized Hamiltonian $\hat{H}$ which clearly elucidates that 
the $\mathbb{Z}$ group formed by the point-gap topological states in non-interacting cases reduces to its subgroup $\mathbb{Z}_2$ due to the interactions [see Eq.~(\ref{eq: N0Ch and nu})].

\end{document}